\documentclass{tlp}

\usepackage{aopmath, latexsym, amsfonts, url}

\newcommand{\defAs}{\stackrel{\mbox{\tiny\it ${\cal D}$\hspace*{-.4ex}ef}}{=}}
\newcommand{\pfin}[1]{\wp_{\mbox{\tiny fin}}\hspace*{-.3ex}\left({#1}\right)}
\newcommand{\with}{\mathbin{\tt with}}
\newcommand{\dime}[1]{||{#1}||}
\newcommand{\bis}{\mathbin{b}}

\newcommand{\fp}{\tt} 
\newcommand{\vars}{\mathit{vars}}
\newcommand{\consts}{\mathit{consts}}

\newcommand{\stir}[2]{{#1 \atopwithdelims \{\} #2}}
\newcommand{\e}{\emptyset}

\newcommand{\dif}{\bigtriangleup}
\newcommand{\w}{\{\cdot\,|\,\cdot\}}
\newcommand{\mo}{\{\![}
\newcommand{\mc}{]\!\}}
\newcommand{\lev}{{\it lev}}
\newcommand{\nat}{\mathbb{N}}

\newtheorem{definition}{Definition} 
\newtheorem{example}{Example} 
\newtheorem{remark}{Remark} 

\title[Set Unification]
    {Set Unification}

\author[Agostino Dovier,
Enrico Pontelli, and Gianfranco Rossi] {AGOSTINO
DOVIER\thanks{A.~Dovier is partially supported by MIUR project
Sybilla, and by GNCS 2005 project on constraints and their
applications.
}\\
Universit\`a di Udine, Dip. di Matematica e Informatica\\
Via delle Scienze 206, 33100 Udine (Italy)\\
\email{dovier@dimi.uniud.it}\\
\and ENRICO PONTELLI\thanks{E.~Pontelli is partially supported by
NSF Grants CNS-0220590, CNS-0454066, and HRD-0420407.
}\\
New Mexico State University, Dept. Computer Science\\
Box 30001, MSC  CS, Las Cruces, NM 88003 (USA)\\
\email{epontell@cs.nmsu.edu}\\
\and GIANFRANCO ROSSI\thanks{G.~Rossi is partially supported by
MIUR project AIDA, and by GNCS 2005 project on constraints and
their applications.
}\\
Universit\`a di Parma, Dip. di Matematica\\
Via M. D'Azeglio, 85/A, 43100 Parma (Italy)\\
\email{gianfranco.rossi@unipr.it}
}

\begin{document}

\maketitle

\begin{abstract}
The unification problem in algebras capable of describing \emph{sets}
has been tackled, directly or indirectly, by many
researchers and it finds important applications in various
research areas---e.g., deductive databases, theorem proving,
static analysis, rapid software prototyping. The various
solutions proposed are spread across a large literature.
In this paper we provide a uniform presentation of unification of sets,
formalizing it at the level of set theory. We address
the problem of deciding existence of solutions at an abstract
level. This provides also the ability to classify different types
of set unification problems. Unification algorithms are
uniformly proposed to solve the unification problem  in each of
such classes.

The algorithms presented are partly drawn from the
literature---and properly revisited and analyzed---and partly
novel proposals. In particular, we present a new goal-driven
algorithm for general $ACI1$ unification and a new simpler
algorithm for general $(Ab)(C\ell)$ unification.
\end{abstract}

\begin{keywords}
Unification Theory, Set Theory, $ACI1$ Unification.
\end{keywords}


\section{Introduction}\label{introduction}

Sets are familiar mathematical objects, and they are often used as
an high-level abstraction to represent complex data structures,
whenever the order and repetitions of elements are immaterial. A
key operation when dealing with set data structures is comparing
two sets. According to the traditional \emph{extensionality
axiom}~\cite{Kun80}, two sets are equal if and only if they
contain the same elements. The problem of set equality is usually
formally addressed within first-order logic. In this context, a
set is represented by  a first-order term, called a \emph{set
term}, built from symbols of a suitable alphabet, using selected
function symbols as set constructors. Since, in general, variables
can occur within a set term in place of either individuals or
sets, solving equations between set terms amounts to solving a
\emph{set unification} or a \emph{set matching} problem.
Intuitively, the set unification problem is the problem of
computing (or simply testing the existence of) an assignment of
values to the variables occurring in two set terms which makes
them denote the same set. Set matching can be seen as a special
case of set unification, where variables are allowed to occur in
only one of the two set terms which are compared. Set unification
can be thought of as an instance  of $E$-unification \cite{SIEK},
i.e., unification modulo an equational theory $E$\/, where $E$
describes the (semantic) properties of  the interpreted symbols
used to represent sets.

Two main approaches for representing sets as terms have been
presented in the literature. The \emph{union-based
representation} makes use of the union operator ($\cup$\/) to
construct sets, while the \emph{list-like representation} builds
sets using an \emph{element insertion} constructor (typically
denoted by $\w$\/). The list-like representation has been
frequently used in the context of logic languages embedding sets.
It is used for instance in~\cite{Kup90}, in~\cite{Jay92},
in~\cite{BNST91}---where $\w$ is called {\bf scons}---in the
language \{log\}~\cite{JLP1}, and  in the G\"{o}del
language~\cite{HL94}. In various papers dealing with computable
set theory, $\w$ is used and called $\with$ \cite{CST2}.

The union-based representation, on the contrary, has been often
used when dealing with the problem of set unification on its
own~\cite{But86,LS76}, where set unification is dealt with as an
\emph{Associative-Commutative-Idempotent ($ACI$) unification}
problem---i.e., unification in presence of operators satisfying
the \emph{Associativity}, \emph{Commutativity}, and
\emph{Idempotence} properties. In~\cite{LAL91} sets are
represented using the union-based approach; however, since set
operations are evaluated only when applied to \emph{ground sets},
set unification is not required at all.

The computational complexity properties of the
set unification and set
matching problems have been investigated by Kapur
and Narendran~\cite{KN86,KN92}, who
established that these decision problems are NP-complete.
Complexity of the set unification/matching operation, however,
depends on which forms of set terms (e.g.,  flat or
nested sets, with zero, one, or more set variables)
are allowed. The form of
set terms in turn is influenced by the set constructors used to
build them. Thus, different complexity results can be obtained for
different classes of set terms.

\medskip

In this paper we present a uniform survey of the problem of
unification in presence of sets, across different set
representations and different admissible classes of set terms. We
provide a uniform presentation of a number of different approaches
and compare them. Unification algorithms for each considered
unification problem are presented and analyzed. These algorithms
are either drawn from the literature or they represent \emph{novel
solutions} proposed by the authors. In particular a goal-driven
algorithm for general $ACI$ unification is proposed, together with
a new algorithm (with a simple termination proof) for general
$(Ab)(C\ell)$ unification.

\subsection{Application Domains of Set Unification}

Various forms of set unification have been proposed by many
authors, in different application frameworks:

\begin{description}
\item[Declarative programming languages with sets:]
Various declarative programming languages relying on sets as
\emph{first-class  objects} have been proposed, which provide
different forms of set unification. Most of these languages are
instances of the \emph{Constraint Logic Programming}
paradigm~\cite{JLP1,TOPLAS2000,russia} or of the
\emph{Functional-Logic} paradigm~\cite{Jay92,SETA}. The
specification language Z \cite{z} makes use of sets as data
abstraction; attempts have been made to produce executable
versions of Z, such as the ZAP compiler \cite{Grieskamp} (whose
implementation, however, does not embed a set unification
algorithm).

\item[Deductive databases:]
Various proposals have been put forward for
embedding sets as primitive data structures in deductive database
languages, providing
 set unification or set matching
as a built-in mechanism for set
manipulation~\cite{relationlog,AG91,ldl,STZ92,ng,flogic}. In these
frameworks, it is common to deal with sets involving unions of
variables.

\item[AI and Automated deduction:]
Set abstraction and operations have been shown to be fundamental
in various subfields of Artificial Intelligence. They have been
used as tools for the description of linguistic theories in
Natural Language Processing \cite{mana}. In
particular, unification based grammars augmented with set
descriptions (e.g.,~\cite{Rou88,PM90}) require
set unification. Set unification has  been used in discovery
procedures for determining categorial grammars from linguistic
data (e.g.,~\cite{Mar97}). Set data structures have also
been used in pattern matching and pattern directed invocation in
various AI languages \cite{LS76}. Proposals dealing with
computable properties and algorithmic manipulation of set
structures have appeared also in the area of automated
deduction, e.g., to reduce the length of proofs~\cite{PS95}.

\item[Program analysis and Security:]
Codish and Lagoon~\cite{cl2000} described an application of
elementary $ACI1$ unification to the problem of \emph{sharing
analysis} of logic programs.
Wang et al.\ show how a system based on $CLP(\mathcal{SET})$ (hence,
on set unification) can be used to model
access control~\cite{Wang04}.
\end{description}

\subsection{Unification Algorithms}

The problem of solving set unification has been mostly
tackled in the form of  $ACI$ unification, and
unification algorithms, returning the set of all the solutions to
a given problem, have been proposed. The first work proposing a
viable solution to $ACI$ unification is~\cite{LS76}. This work
mostly deals with $AC$ unification---by reducing it to
the solution of Diophantine equations---and only in the end
it suggests a solution of the $ACI$ problem, by replacing
arithmetic equations with Boolean equations. Direct solutions of
the $ACI$ problem have been proposed by B\"{u}ttner~\cite{But86}
and Baader and B\"{u}ttner~\cite{BB88}. More recently, Baader and
Schulz~\cite{BS96} provided a general methodology allowing the
unification with constants algorithms proposed for $ACI$ to be
extended to general $ACI1$ unification algorithms.

In recent years, a number of efforts have emerged that propose
set unification algorithms for the list-like representation of
sets, hence for a different equational theory (called
$(Ab)(C\ell)$ in~\cite{JLP1}). A first proposal in this direction
is the algorithm sketched by Jayaraman and Plaisted
in~\cite{JP89}. A more general and complete algorithm is the one
in~\cite{JLP1}. The problem of set unification in this context
has been tackled by different
authors~\cite{AD97,DPR98funda,Sto96,Sto99,voronkov}.
In particular, the algorithms presented
in~\cite{AD97} and~\cite{Sto99} provide solutions which are
optimal, in terms of number of unifiers, for large classes of
unification problems. The algorithms in~\cite{AD97,voronkov}
ensure polynomial time complexity in each non-deterministic
branch of the computation.

Various authors have considered simplified versions of the
$(Ab)(C\ell)$ problem obtained by imposing restrictions on the
form of the set terms. In particular, various works have been
proposed to study the simpler cases of matching (e.g.,
\cite{STZ92}) and unification of \emph{Bound Simple} set terms,
i.e., bound set terms of the form $\{s_1, \dots, s_n\}$, where
each $s_i$ is either a constant or a
variable~\cite{arni,greco1,greco2}. A parallel algorithm for such
restricted $(Ab)(C\ell)$ unification has been presented
in~\cite{ng}. Set matching is also discussed in~\cite{KN86}.

All these algorithms, however, have been developed in separate
contexts, without considering any relationship among them. They
have never been formally compared and related.
A contribution of this paper  is to provide a
uniform presentation of the problem, covering most of its
different instances, and  surveying  the different solutions
developed.

\subsection{Overall Structure of the Paper}

The paper is organized as follows. In Section~\ref{set universe}
we define the universe of sets we are dealing with, along with a
suitable abstract syntax for representing them and a syntactical
classification of the set unification problems. In
Section~\ref{examples} we present a number of examples of
unification problems which provide motivations for set
unification. In Section~\ref{SUD} we discuss the complexity of the
set unification decision problem for each syntactic class of set
terms listed in Section~\ref{set universe}. In
Section~\ref{notation} we introduce the basic notions and notation
concerning $E$-unification and the equational theories used in
$E$-unification with sets. In Section~\ref{aci1-costanti} we
describe the problem of \emph{$ACI1$ unification with constants}
and its impact on set unification. In Section~\ref{setlog} we
extend the discussion to the \emph{$(Ab)(C\ell)$ unification}
problem, i.e., the problem of set unification in presence of set
terms based on the element insertion constructor $\w$\/, and we
present a new algorithm for this case. In Section~\ref{ACI_gen} we
tackle the most general problem of unification of terms containing
both $ACI1$ and free (uninterpreted)
function symbols. A new general $ACI1$
unification algorithm is presented. Some related topics are
discussed in Section~\ref{related}, and concluding remarks are
presented in Section~\ref{conclusioni}.
In~\ref{auxproofs} the proofs of the main results of the paper are reported.

\section{Sets and the Set Unification Problem}\label{set universe}

In this section we characterize the universe of sets we deal with,
and we discuss some well-known operations on sets. Finally, we
formally introduce the set unification problem.

\subsection{A Universe of Sets} \label{universe}

A set is an arbitrary, unordered, collection of elements.
Typically, a set is specified either \emph{intensionally}, by
means of a  property that characterizes membership to the set, or
\emph{extensionally}, by explicit enumeration of all its elements.
In this paper we restrict our attention to extensional sets. For
instance, $\{a,b,c\}$ is the (extensional) set which contains
exactly the elements $a$, $b$, and $c$. We denote mathematically
the fact: ``$a$ belongs to the set $\{a,b,c\}$'' using the
membership relation: $a \in \{a,b,c\}$. We assume the
\emph{extensionality} axiom~\cite{Kun80} that states that two sets
are equal if and only if they contain the same elements. Thus,
$\{a,b,c\}$  is the unique set containing exactly $a$, $b$, and
$c$. $\{a,c,b\}$\/, $\{b,a,c\}$\/, etc. are alternative ways to
describe the same set. A particular set is the empty set $\e$,
that contains no elements. A set containing only one element is
said to be a \emph{singleton}. If $s$ is a set, then we will
denote with $|s|$ its cardinality.

\medskip

A set is \emph{finite} if it contains a finite number of
elements.\footnote{A precise, formal, characterization of the
notion of finiteness is outside the scope of this work. For a
theoretical analysis of this topic see~\cite{Tar24}.} For instance
$\e, \{ \e \}$, and $ \{a,b,c\}$ are finite sets. However, this
definition does not remove all possible cases leading to infinity.
The singleton set $\{ \nat \}$ is a finite set, but its unique
element $\nat$ is an infinite set. A set is said to be
\emph{hereditarily finite} if it is finite and all its elements
are hereditarily finite. This definition leaves still a further
possibility for infinity. Let us consider the sets $x$ and $y$ that
satisfy the equations $x = \{\e,y\}, y = \{x\}$. They are
hereditarily finite, but they hide an infinite descending chain $x
\ni y \ni x \ni y \ni   \cdots$. These sets, where the membership
relation is allowed to be not well-founded,  are called \emph{non
well-founded sets} (or \emph{hypersets})~\cite{Acz88,BM96}.
Hypersets are very important in some areas, such as concurrency
theory, but they are not accepted in traditional set theory, where
sets are expected to be well-founded.

\medskip

Let us focus on \emph{hereditarily finite and well-founded sets}.
We can consider two approaches to set theory:
\begin{itemize}
\item \emph{pure sets}, in which the only entity
that does not contain elements is the empty set $\e$, and
\item \emph{sets with individuals}, in which there exists a
collection $\cal U$ of individuals, where each
element of $\cal U$ is not a set
and does not contain elements. Since the elements of
$\cal U$ are not sets, we also have that $\e \notin {\cal U}$\/.
\end{itemize}
In the second approach, the extensionality axiom has to be revised
for the elements of $\cal U$\/, since
\begin{enumerate}
\item[$(i)$] two individuals are different even if
they contain the same elements (namely, none), and
\item[$(ii)$] all the elements in $\cal U$ are different from $\e$.
\end{enumerate}
In this paper we will focus on the
approach based on sets with individuals, as it generalizes the
pure sets approach (by taking ${\cal U} = \e$).

Let us introduce the universe of sets we are interested in (see
also~\cite[pg.~88]{CST2}).
As usual, the subset relation $x \subseteq y$ denotes the
formula $\forall z\,(z\in x \rightarrow z \in y)$.
If $s$ is a set, with $\pfin{s} = \{ x
\,:\, x \subseteq s \wedge\, x \mbox{ is finite} \}$
we denote the set of all its finite subsets.

\begin{definition}
The \emph{Universe $\mathbb{HF}^{\cal U}$ of
hereditarily finite sets based on $\cal U$}
is obtained as follows:
$$\left\{
  \begin{array}{lcl}
          \mathbb{HF}^{\cal U}_0 & \defAs & \pfin{ {\cal U} }  \\
          \mathbb{HF}^{\cal U}_{i+1} & \defAs  & \mathbb{HF}^{\cal U}_{i} \cup
                      \pfin{\mathbb{HF}^{\cal U}_{i}}\\
          \mathbb{HF}^{\cal U} & \defAs & \bigcup_{i \in \nat}
          \mathbb{HF}^{\cal U}_{i}
  \end{array}
  \right.$$
\end{definition}
The sets in $\mathbb{HF}^{\cal U}_{0}$ contain the finite subsets
of the set  of individuals: these particular sets are called
\emph{flat} sets. The sets introduced in $\mathbb{HF}^{\cal
U}_{i}$, with $i > 0$\/, may contain elements that are  sets
themselves. We refer to such sets as \emph{nested sets}. For
instance, if ${\cal U} = \{a,b,c\}$, then $\mathbb{HF}^{\cal
U}_{0}$ consists of the flat sets:
$$\e, \{a\}, \{b\},\{c\},\{a,b\},\{a,c\},\{b,c\},\{a,b,c\}$$
Some nested sets are the following:
$$\{\e\},\, \{\{a\}\},\, \{ \e, \{ \{a\}, b\}, \{\{\{c\}\}\} \}$$

\subsection{Abstract Set Terms}\label{representation}

So far we have represented sets by exploiting the usual intuitive
notation based on braces
and commas. In order to deal with sets as primitive data objects
in a first-order language, however, we need to precisely
represent them as first-order terms of the language. For this
reason, one or more function symbols are selected to be used as
\emph{set constructors}. Set constructors will allow complex sets
to be built from simpler ones. Many different approaches are
possible. The two approaches that, to our knowledge, have
received more attention in the literature are the following.

\begin{enumerate}
\item \label{first}
\emph{Union-based representation}. This solution is based on the
use of the \emph{union} constructor $\cup$\/ and, possibly, the
\emph{singleton} constructor $\{\cdot\}$\/. $s \cup t$ represents
the set which contains the elements of the sets $s$ and $t$\/,
that is,
$$s \cup t   = \{
x \,:\, x \in s \vee x \in t \}\,,$$ while  $\{t\}$ represents the
set containing the single element $t$\/. With this approach, the
finite set $\{t_0,\dots,t_n\}$ is represented by a union of
singletons: $\{t_0\}\cup\cdots\cup\{t_n\}$\/, where $t_0, \dots,
t_n$ are either sets or individuals. The empty set is represented
by a distinguished  constant $\e$\/.

\item \label{second}
\emph{List-like representation}. An alternative representation of
sets is based on the \emph{element insertion} constructor $\w$\/.
$\{t\,|\,s\}$ represents the set obtained by adding the element
$t$ (either a set or an individual) to the set $s$ if it is not
yet in $s$\/, that is
$$\{t\,|\,s\}  = \{ x \,:\, x \in s \vee x = t \}.$$
The empty set is represented by a distinguished constant $\e$\/.
Thus, the finite set $\{t_0,\dots,t_n\}$  is represented by a
sequence of element insertions:
$$\{t_0\,|\,\{\cdots\{t_n\,|\,\e\} \cdots \}\}$$
where
$t_0, \dots, t_n$ are either sets or individuals.
\end{enumerate}

As far as the syntactic  representation of the individuals (i.e.,
the elements of $\cal U$\/) is concerned, we can represent them
either
\begin{itemize}
\item  as constant symbols different from $\e$ (\emph{simple
individual terms}) or
\item as terms of the form $f(t_1,\dots,t_n)$,
$n>0$, $f$ different from $\cup$ and $\w$, and $t_1,\dots,t_n$
terms (\emph{general individual terms}).
\end{itemize}
Both the union-based and
the list-like representations allow the elements of the sets to be
either individual terms or other set terms. Individual and set
terms can be nested at any level.

Let us observe that the element insertion constructor $\w$ can be
represented using $\cup$\/, i.e., $\{s\,|\,t\} \defAs \{s\} \cup
t$. However, in~\cite{DPP00-FROCOS} it is  proved that, without
singleton sets, the two symbols are not mutually definable,
unless we allow the use of complex formulae involving universal
quantifiers. Observe moreover that the $\cup$ symbol allows one
to define set inclusion: $x \subseteq y$ is equivalent to $x \cup y = y$.

Furthermore, let us observe that the definition of $\cup$, being
based on membership, makes sense on sets, not on individuals. For
instance, the union of two individuals $a$ and $b$ would be a
memberless object. There is no way of stating that $a$ is equal or
different from $a \cup b$ without introducing new, non-standard,
axiomatizations. For this reason, we assume that the $\cup$
constructor is used only on sets. Similar considerations apply to
the second argument of the $\w$ operator.

\medskip

For the sake of simplicity, in the rest of the work we will make use
of a simpler and more intuitive abstract syntax to denote sets,
disregarding the concrete representation used to encode them as
terms in the language at hand.

\begin{definition}\label{abstractsetterms}
An \emph{abstract set term} is  a term of the form
$$\begin{array}{lr}
\{X_1,\dots,X_m,a_1,\dots,a_n,s_1,\dots,s_p\} \cup
Y_1\cup\cdots\cup Y_q & m, n, p,q \geq 0
\end{array}$$
where $X_i, Y_i$ are variables, $a_i$ are individual terms, and
$s_i$ are abstract set terms distinct from variables.
The $Y_i$ variables are called \emph{the set variables} of the
abstract set term.
In
particular,
\begin{itemize}
\item when $m=n=p=q=0$, the term is simply written as $\e$.
\item when $m=n=p=0$ and $q=1$, the term is the set variable $Y_1$.
\end{itemize}
The \emph{size} $\dime{s}$ of an abstract set term $s$
is the number of occurrences of symbols in $s$.
\end{definition}
As a notational convention, we will usually use $a,b,c$, possibly
subscripted, to denote individual terms, and $r,s,t$, possibly
subscripted, to denote (abstract) set terms or individual terms.
Variables are denoted by identifiers with capital letters.

When $q \leq 1$\/, the abstract set term can be rendered
concretely using both representations described above. For
example, $\{ X_1,X_2,a,b,c\} \cup Y_1$ can be seen as a shorthand
for both the concrete terms
$\{X_1\,|\,\{X_2\,|\,\{a\,|\,\{b\,|\,\{c\,|\,Y_1\}\}\}\}\}$ and
$\{ X_1\} \cup \{X_2\} \cup \{ a\} \cup \{ b \} \cup \{ c \} \cup
Y_1$\/. Conversely, when $q > 1$, the $\cup$ constructor is
required; thus, only the union-based representation is feasible.

When clear from the context we will omit the word ``abstract'',
referring to abstract set terms simply as set terms.

\medskip

Set terms may contain variables, both as individuals (the variables
$X_i$'s) and as sets (the variables $Y_j$'s). A set term containing
variables  denotes a possibly infinite collections of
sets. For instance, the term $\{a, X , b\}$ denotes all sets
containing two individuals, $a$ and $b$\/, and possibly a third
unknown element $X$\/. If $X$ takes the value $a$ or $b$ then the
set will have only 2 elements. Otherwise, e.g., $X = c$\/, the set
will contain three elements. Note that the set terms $\{a,a,b\}$,
$\{a,b,a\}$, $\{b,a,a,b\}$\/, etc.\ are accepted notations for the
same set, i.e., the (unique) set containing exactly $a$ and $b$.
Note also that variables in set terms could be implicitly forced
to assume set values using the fact that the $\cup$ constructor
requires two set arguments. Thus, for instance, the variable $Y$
in the set term $ \{a,b\} \cup Y$ can take only set values. Set
terms are called \emph{non-ground} (\emph{ground}) if they  do
(do not) contain variables.
Finally, note that general individual terms can be non-ground. For
instance, $f(X,Y)$ is a non-ground term, but the fact that the
outermost symbol is not a set constructor ensures that it is an
individual.

\begin{example}
The following are abstract set terms.
\begin{itemize}
\item $\{1,2,3\}$ ($m=0,n=3,p=0,q=0$)
\item $\{\e\}$ ($m=0,n=0,p=1,q=0$)
\item $\{X_1,X_2,a,b,c,d\}\cup Y$ ($m=2,n=4,p=0,q=1$)
\item $Y_1 \cup Y_2$ ($m=0,n=0,p=0,q=2$)
\item $\{X,a,b,c,\{1,2,3\},\{\e\}\}$ ($m=1,n=3,p=2,q=0$)
\item $\{X_1,X_2,a,f(\{a,\e\}),\e\}\cup Y_1$ ($m=2,n=2,p=1,q=1$)
\end{itemize}
\end{example}

\subsection{Set Equivalence and Set Unification}\label{classes}

The most natural decision test regarding set terms is testing
whether they represent the same set or, in the case of
non-groundness, testing whether there exists an assignment for the
variables that forces the two terms to represent the same set.

\begin{definition}
Given two terms $s$ and $t$, $s = t$ is said to be an
\emph{equation}. A conjunction $s_1=t_1 \wedge \cdots \wedge
s_n=t_n$ of equations is said to be a \emph{system of equations}.
Systems of equations are also commonly viewed as sets of
equations.
\end{definition}
If $X_1,\dots,X_n$ are the variables occurring in a system of
equations $C$, we denote with $\vec\exists C$ the formula
 $\exists X_1\cdots\exists X_n\,C$\/.
The existence of an assignment for the variables in $s$ and $t$
that forces the two terms to represent the same set will be
denoted by $\mathbb{HF} \models \vec\exists\, s=t$\/, formally
defined below.

Before defining  the interpretation of ground abstract set terms
in $\mathbb{HF}$\/, we first show how individual terms (syntax)
can be related to individuals (semantics). Let us assume that
$\mathcal{U}$ is an infinite set of individuals. Simple individual
terms denote distinct elements of $\mathcal{U}$\/. For the sake of
simplicity, in our examples, the individual terms $a,b,c,\dots$
will be interpreted as the corresponding individuals $a,b,c,\dots$
of $\cal U$\/---we use the so-called \emph{unique name assumption}.
General individual terms $f(s_1,\dots,s_m)$ and
$g(t_1,\dots,t_n)$\/, with $f$ different from $g$\/, denote
distinct elements of $\mathcal{U}$, different from all the
individuals associated to the simple individual terms. Each
function symbol $f$ of arity $n$ is interpreted as a one-to-one
function $f^{\mathbb{HF}}$ from $\mathbb{HF}$ to $\mathcal{U}$.

\begin{definition}
If $s \equiv \{a_1, \dots, a_n, s_1, \dots, s_p\}$ is a ground set
term, then its interpretation in $\mathbb{HF}$\/,
denoted by $s^{\mathbb{HF}}$\/, is the following set:
\begin{itemize}
\item if $n=0$ and $p=0$ then $s^{\mathbb{HF}}$ is the empty set

\item otherwise, $s^{\mathbb{HF}}$ is the set containing
    exactly the elements $a_1^{\mathbb{HF}}, \dots, a_n^{\mathbb{HF}}$ and
    $s_1^{\mathbb{HF}}, \dots, s_p^{\mathbb{HF}}$, where
    \begin{itemize}
    \item if $a_i$ is a simple individual term, then
    $a_i^{\mathbb{HF}}$ is simply the corresponding individual.

    \item if $a_i$ is of the form
    $f(t_1,\dots,t_n)$ then
    $a_i^{\mathbb{HF}}$ is the individual associated to
    $f^{\mathbb{HF}}(t_1^{\mathbb{HF}},\dots,t_n^{\mathbb{HF}})$.
    \end{itemize}
\end{itemize}
If $s$ and $t$ are two ground  set terms, then
$\mathbb{HF} \models (s=t)$ if and only if $s^{\mathbb{HF}}$
is the same set as $t^{\mathbb{HF}}$\/.

\noindent If $s$ and $t$ are two  set terms, and $X_1, \dots, X_n$
are all variables in $s$ and $t$\/, then $\mathbb{HF} \models
\vec\exists s=t$ if and only if there exists an assignment
$\sigma$ of ground set terms to $X_1, \dots, X_n$ such that
$\mathbb{HF} \models (s=t)\sigma$\/.
\end{definition}

\begin{definition} \label{decision}
If $s$ and $t$ are two set terms, the \emph{set unification
decision (SUD)} problem is the problem of checking whether
$\mathbb{HF} \models \vec\exists\,(s = t)$. If $s$ and $t$ are
ground, the problem is also called \emph{set equivalence}.
\end{definition}

\begin{definition}\label{solutions}
If $s$ and $t$ are two set terms and $X_1,\dots,X_n$ are the
variables occurring in them, the \emph{set unification solution
(SUS)} problem is the problem of finding an assignment $\sigma$ of
sets and/or  individuals terms to  the variables $X_1,\dots,X_n$\/,
such that $\mathbb{HF} \models (s = t)\sigma$.
\end{definition}
We give a more standard and complete definition of the unification
problem in Section~\ref{notation}.
Note that, if
two general individuals have the same outermost symbols but the
ordered list of arguments is different, then they denote two distinct
individuals (e.g., $f(a,b)$ and $f(b,a)$). However, if the two
individual terms contain set terms as their arguments, in
order to decide
whether the individuals are or not the same, one needs to compare
the sets denoted by the involved set terms. For example, the
general individual terms $f(\{a,b\},c)$ and $f(\{b,b,a\},c)$
denote the same individual since $\{a,b\}$ denotes the same set as
$\{b,b,a\}$.

\medskip

{F}rom a computational point of view, the complexity of the SUD
problem depends on the syntactic form of the two set terms $s$ and
$t$. As a matter of fact, while the set equivalence test of ground
set terms denoting flat sets, such as $\{a,b,c\}$ and $\{b,c,a\}$,
is rather easy, when the SUD problem deals with nested set terms
involving variables it becomes NP-complete (see
Section~\ref{nestedp}). Thus, in order to classify the set
unification problem, we subdivide set terms in different syntactic
classes.

\begin{definition}
For $m\geq0,n\geq0,p\geq 0,q \geq 0$, the class ${\sf
set}(m,n,p,q)$ is the collection of abstract set terms of the form:
$$\{X_1,\dots,X_{m'},
a_1,\dots,a_{n'},
s_1,\dots,s_{p'}\} \cup
Y_1\cup\cdots\cup Y_{q'}$$
where $0 \leq m' \leq m$,
 $0 \leq n' \leq n$,
 $0 \leq p' \leq p$,
 $0 \leq q' \leq q$\/, and $s_i \in {\sf set}(m,n,p,q)$\/.
\end{definition}
Observe that $\e \in {\sf set}(m,n,p,q)$ for all $m,n,p,q$\/.
Furthermore, ${\sf set}(m_1,n_1,p_1,q_1) \subseteq {\sf
set}(m_2,n_2,p_2,q_2)$ if $m_1 \leq m_2$ and $n_1 \leq n_2$ and
$p_1 \leq p_2$, and $q_1 \leq q_2$. Interesting special cases can
be obtained by setting some of these parameters to $0$:

\begin{description}
\item[${\sf ground} = \bigcup_{n\geq 0,p\geq 0}{\sf
set}(0,n,p,0)$:] the collection of set terms of the form\\
$\{a_1,\dots,a_n,s_1,\dots,s_p\}$, with $a_i$ simple individual
terms and $s_i$ ground set terms.

\item[${\sf gflat}(q)= \bigcup_{n\geq 0}{\sf set}(0,n,0,q)$:]
the collection of set terms of the form\\ $\{a_1,\dots,a_n \} \cup
Y_1 \cup \cdots \cup Y_{q'}$, with $a_i$ simple individual terms
and $Y_i$ variables ranging over {\sf gflat($q$)} sets (i.e., sets
denoted by {\sf gflat($q$)} set terms)  ($0 \leq q' \leq q$).

\item[${\sf flat}(q) = \bigcup_{m\geq 0,n\geq 0}{\sf
set}(m,n,0,q)$:] the collection of set terms of the form\\ $\{X_1
,\dots, X_m, a_1,\dots,a_n\} \cup Y_1 \cup \cdots \cup Y_{q'}$,
with $a_i$ and $X_i$ denoting simple individual terms, and $Y_i$
ranging over {\sf flat($q$)} sets  ($0 \leq q' \leq q$).

\item[${\sf nested}(q) =
\bigcup_{m\geq 0,n\geq 0,p \geq 0}{\sf set}(m,n,p,q)$:] the
collection of set terms of the form\\ $\{X_1 ,\dots, X_m,
t_1,\dots,t_n, s_1,\dots,s_p\} \cup Y_1 \cup \cdots \cup Y_{q'}$,
with $t_i$ general individual terms, $s_i$ non-variable  ${\sf
nested}(q)$ set terms, $X_i$ ranging over general individuals or
${\sf nested}(q)$ sets, and $Y_i$ ranging over ${\sf nested}(q)$
set terms ($0 \leq q' \leq q$).
\end{description}
${\sf gflat}(q)$ and  ${\sf flat}(q)$ denote flat sets only, while
{\sf ground} and ${\sf nested}(q)$ account for nested sets.
For the same $q$, we have that ${\sf gflat}(q) \subseteq
  {\sf flat}(q) \subseteq
  {\sf nested}(q).$
Moreover, {\sf ground} is included in ${\sf nested}(q)$ (namely,
in ${\sf nested}(0)$\/), but it is not included in the other
classes, since {\sf ground} accounts also for nested sets.
Actually, these classes could be further subdivided into finer
subclasses. For instance, we could further distinguish between
ground and non-ground {\sf nested} set terms, with simple or
general individuals. However, the four classes we identified above
turn out to be sufficient for our analysis.

\medskip

Observe that, in the concrete representation of sets,
 the union constructor is not required
 whenever $q \leq 1$. For these sets
we can
use the list-like representation, based on the
element insertion constructor $\w$. For $q>1$, instead, we need
the union constructor, and possibly the singleton constructor. For
instance, the abstract set term  $X \cup \{Y\} \cup Z$ can be
immediately encoded using the union-based representation  while it
has no corresponding encoding in the list-like representation.
These observations will play an important role when we will
address the various unification problems.

\section{Examples}\label{examples}

This section presents a series of instances of the set unification
problem. This allows us to give an intuitive idea of the
expressive power of the different frameworks considered in the
rest of the paper.

\begin{description}

\item[Chords:] this is the problem of
determining whether two sets of musical notes denote
the same chord (a chord is a \emph{set} of at least three
notes---i.e., order and repetitions do not matter).
We can encode the problem as a set unification problem between two
(flat) {\sf ground} set terms:
$$\{c,e,g,b\flat\} = \{g,g,e,b\flat,c,e\}$$
where $c$, $e$, $g$, $b\flat$ are constants representing musical
notes (i.e., individuals of the language).

\item[Courses covering:] this is the problem of
verifying whether two teachers are covering three courses in their
current course assignment. The problem can be encoded as a set
unification problem between a ${\sf gflat(q)}$  set term, composed
only of variables, for the teachers, and a (flat) {\sf ground} set
term for the courses:
     $$\mbox{\em Teacher}_1 \cup \: \mbox{\em Teacher}_2 =
       \{ \mbox{\em course}_1 , \mbox{\em course}_2 , \mbox{\em course}_3 \}$$
Note that, in this case, variables range over unions of elements
(i.e., subsets of $\mbox{\em course}_1 \cup \mbox{\em course}_2
\cup \mbox{\em course}_3$\/) rather than simply over individuals.
Thus we cannot use the list-like representation for its concrete
rendering.

\item[Graph Coloring:] Let us consider the graph-coloring problem
consisting of the undirected graph
$$\langle
\:\{X_1,X_2,X_3,X_4\}\:,\:\{\{X_1,X_2\},\{X_2,X_3\},\{X_3,X_4\},
          \{X_4,X_1\}\}\: \rangle$$
and a set of colors $$\{red,green,blue\}$$

This problem can be easily encoded as a single equation between
two ${\sf nested(q)}$  ($q \geq 1$) set terms in the following way:
       \[\begin{array}{l}
          \{\{X_1,X_2\},\{X_2,X_3\},\{X_3,X_4\},\{X_4,X_1\}\} \cup R   =
          \phantom{aaaaaaaaaaaaaaa}\\
          \multicolumn{1}{r}{\{ \{red,green\},\{red,blue\},\{green,blue\} \}}\\
       \end{array}\]
The right-hand side set is used to encode the set of all viable
unordered pairs of colors, and it can be a ground set term.

The solution of this equation (see Definition \ref{solutions})
provides a solution of the corresponding graph-coloring problem. A
possible solution (actually, the first one returned by the
$CLP({\cal SET})$ interpreter~\cite{TOPLAS2000}) is:
$$X_1 = red,X_2 = green,X_3 = red, X_4 = blue, R = \{\{green,blue\}\}$$
Solutions that make use of only two colors are also
computed, such as $X_1=X_3=red$, $X_2
= X_4 = green$ and $R = \{\{red,blue\},\{green,blue\}\}$\/.

\item[Finite State Automata:]
Let us consider a deterministic finite state automata on the
alphabet $\{0,1\}$, containing the set of states $Q = \{q_0, \dots,
q_{n-1},q_n\}$\/, where $q_0$ is the initial state and $q_n$ is the
unique final state. $Q_1=\{q_0, \dots, q_{n-1}\}$  denotes the set
of non-final states of the automata. We would like to ``learn''
the structure of the automata by looking at positive and negative
examples of strings that should be either accepted or rejected.
This problem can be encoded as follows. The set of transitions $D$
is represented by a ${\sf nested(q)}$  ($q \geq 0$)
set term whose elements are
triples $({\it source},{\it symbol},{\it destination})$ (where
$(\cdot\,,\,\cdot\,,\,\cdot)$ is a ternary free function symbol
used to build the triples):
\[
    D = \{ (q_0,0,X_{0,0}), (q_0,1,X_{0,1}), \dots, (q_n,0,X_{n,0}),
    (q_n,1,X_{n,1}) \}
\]
Observe that the destination of each transition is, at this point,
unknown. If $a_0 \cdots a_k$ is a string of length $k+1$
that should be accepted,
then we need to add an equation:
\[
    \{(q_0,a_0,Y_1), (Y_1, a_1, Y_2), \dots, (Y_{k},a_k,q_n) \} \cup D = D
\]
that forces the transitions $(q_0,a_0,Y_1), (Y_1, a_1, Y_2),
\dots, (Y_{k},a_k,q_n)$ to belong to $D$.
    Note that the left-hand side of the equation is a
    ${\sf nested}(q)$ set term ($q \geq 1$). Therefore, we
    can use the concrete list-like representation to encode it, based
    on the element insertion constructor $\w$, as well as the
    union-based representation.
    If $b_0 \cdots b_h$ is a string that should not be accepted, then
    we need to add the equations:
\[
    \{(q_0,b_0,Y_1), (Y_1, b_1, Y_2), \dots, (Y_{h},b_h,Y_{h+1}) \} \cup D  = D
    \:\:,\:\: \{ Y_{h+1}\} \cup Q_1 = Q_1
\]
    that force the state $Y_{h+1}$ resulting from the execution to be
    in $Q_1$\/, and hence not a final state.

    For example, if we want a four-state automata that
    accepts the strings $000$ and $001$ and rejects
    the strings $011$ and $10$, then we write the system of
    equations:
\[
\begin{array}{l}
        Q = \{q_0,q_1,q_2,q_3\},   Q_1 = \{q_0,q_1,q_2\} , \\
        D = \{(q_0,0,X_{00}), (q_0,1,X_{01}), (q_1,0,X_{10}), (q_1,1,X_{11}),\\
        \phantom{D = \{}(q_2,0,X_{20}), (q_2,1,X_{21}), (q_3, 0, X_{30}),
                  (q_3,1,X_{31}) \:\: \}, \\
        \{W_3 \}\cup Q_1 = Q_1, \{ K_2 \}\cup Q_1 = Q_1,\\
        \{(q_0,0,W_1), (W_1,1,W_2), (W_2,1,W_3) \}\cup D = D,\\
        \{(q_0,1,K_1), (K_1,0,K_2) \}\cup D   = D,\\
        \{(q_0,0,Y_1), (Y_1,0,Y_2), (Y_2,0,q_3) \}\cup D  = D,\\
        \{(q_0,0,Z_1), (Z_1,0,Z_2), (Z_2,1,q_3) \}\cup D  = D
\end{array}
\]
A possible solution (the first one returned by the $CLP({\cal
SET})$ interpreter) is (see also Figure~\ref{DFAfig}):
$$
\begin{array}{rcl}
D & = & \{(q_0,0,q_1),(q_0,1,q_2),(q_1,0,q_3),(q_1,1,q_2),\\
 && \phantom{a}      (q_2,0,q_0),(q_2,1,q_0),(q_3,0,q_3),(q_3,1,q_3)\}
\end{array}
$$

\begin{figure}
\begin{center}
\begin{picture}(200,100)(-20,-10)
\setlength{\unitlength}{1pt}
\thicklines
\put(20,20){\circle{20}}
\put(80,20){\circle{20}}
\put(80,20){\circle{15}}
\put(20,80){\circle{20}}
\put(80,80){\circle{20}}
\put(17,18){$q_2$}
\put(77,18){$q_3$}
\put(77,78){$q_1$}
\put(17,78){$q_0$}
\put(30,80){\vector(1,0){40}}
\put(48,83){0}
\put(22,70){\vector(0,-1){40}}
\put(25,50){1}
\put(12,26){\vector(0,1){48}}
\put(-5,50){0/1}
\put(73,73){\vector(-1,-1){46}}
\put(44,50){1}
\put(80,70){\vector(0,-1){40}}
\put(83,50){1}
\put(95,10){\oval(30,20)[b]}
\put(95,10){\oval(30,20)[tr]}
\put(100,20){\vector(-1,0){10}}
\put(93,25){0/1}
\end{picture}
\end{center}
\caption{The DFA computed from examples and counterexamples\label{DFAfig}}
\end{figure}
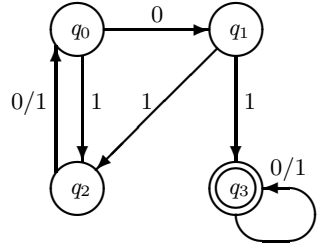

\item[Paths and Subgraphs:]
Let us represent an undirected graph $G$ as the set of all its
edges, where each edge is represented by the set of its two
constituting nodes. Let us consider the problem of computing all the
subgraphs of  $G$ with nodes
$\{c_1,\dots,c_n\}$ such that each subgraph contains at least one
path between two given nodes, e.g., $c_1,c_n$.
This problem can be immediately encoded as a set unification
problem. In fact, all the subgraphs of $G$ are given by the
solutions for $G_1$ of the equation
\begin{equation}
\label{tre}
 G_1 \cup G_2 = G
\end{equation}
The subgraphs containing the required path can be obtained by
adding the equations:
        $$G_3 = G_1 \cup \{ \{c_1,c_1\}, \dots, \{c_n,c_n\}\}
        ,\:\:\:\:
        G_3 = \{ \{c_1,X_1\},\{X_1,X_2\},\dots,\{X_{n-1},c_n\}\} \cup
        G_3$$
Observe that $G_1$ is temporarily extended to the new graph $G_3$
by introducing artificial loops, thus allowing us to recognize paths
of length less than $n$\/.
Also, observe that the equation \ref{tre} cannot be rendered
concretely using the list-like representation, since its left-hand
side set term involves more than one variable ranging over set terms
(i.e., it belongs to the ${\sf nested}(q)$ class, $q \geq 2$).
\end{description}

\section{The Set Unification Decision Problem and its Complexity}
\label{SUD}

In this section, we discuss the complexity of the Set Unification
Decision  problem for each one of the syntactic classes of set
terms listed in Section~\ref{set universe}.

\subsection{SUD for the {\sf ground} Class}\label{subground}

The set equivalence test for two ground abstract set terms  $s$
and $t$ can be solved in worst-case time
$O(\dime{s} + \dime{t})$ (see Definition \ref{abstractsetterms}).
The proof is based on a tree
representation of a well-founded set and on the  existence of a
fast algorithm for proving graph bisimulation. We first focus on
the \emph{pure} case (without individuals).

We can use a tree $G=\langle N, E\rangle$, rooted in $\nu \in N$,
where $N$ is the set of nodes and $E$ is the set of edges of $G$,
to represent a pure set.
Edges represent memberships, namely $\langle m,n \rangle$ means
that $m$ has $n$ as an element, and the nodes in the tree denote
all the sets that contribute to the construction of the
set. A node without outgoing edges represents the
empty set $\e$. It is possible to write a procedure that
translates a ground set term denoting a pure set into a tree in
linear time. An example showing two trees obtained in this way is
shown in Figure~\ref{grafi}. From the figure it is possible to
observe the sets associated to the subtrees.

\begin{figure}
\begin{center}
\footnotesize
\begin{tabular}[b]{cc}
\setlength{\unitlength}{0.5pt}
\begin{picture}(170,180)(0,40)
\thicklines
\put(50,50){\circle*{10}}
\put(100,50){\circle*{10}}
\put(150,50){\circle*{10}}
\put(50,100){\circle*{10}}
\put(150,100){\circle*{10}}
\put(100,150){\circle*{10}}
\put(50,200){\circle*{10}}
\put(50,100){\vector(0,-1){45}}
\put(150,100){\vector(0,-1){45}}
\put(100,150){\vector(1,-1){45}}
\put(100,150){\vector(0,-1){95}}
\put(50,200){\vector(1,-1){45}}
\put(50,200){\vector(0,-1){95}}
\put(60,45){$\e$}
\put(110,45){$\e$}
\put(160,45){$\e$}
\put(60,95){$\{\e\}$}
\put(160,95){$\{\e\}$}
\put(110,145){$\{\e,\{\e\}\}$}
\put(60,195){$\{\{\e\},\{\e,\{\e\}\}\}$}
\end{picture} &
\setlength{\unitlength}{0.5pt}
\begin{picture}(300,180)(0,40)
\thicklines
\put(50,50){\circle*{10}}
\put(100,50){\circle*{10}}
\put(150,50){\circle*{10}}
\put(200,50){\circle*{10}}
\put(250,50){\circle*{10}}
\put(60,45){$\e$}
\put(110,45){$\e$}
\put(160,45){$\e$}
\put(210,45){$\e$}
\put(260,45){$\e$}
\put(100,100){\circle*{10}}
\put(110,95){$\{\e\}$}
\put(200,100){\circle*{10}}
\put(210,95){$\{\e\}$}
\put(250,100){\circle*{10}}
\put(260,95){$\{\e\}$}
\put(150,150){\circle*{10}}
\put(70,145){$\{\e,\{\e\}\}$}
\put(150,200){\circle*{10}}
\put(160,195){$\{ \{\e\}, \{\e,\{\e\}\}  \}$}
\put(100,100){\vector(-1,-1){45}}
\put(100,100){\vector(0,-1){45}}
\put(200,100){\vector(0,-1){45}}
\put(250,100){\vector(0,-1){45}}
\put(150,150){\vector(-1,-1){45}}
\put(150,150){\vector(0,-1){95}}
\put(150,200){\vector(0,-1){45}}
\put(150,200){\vector(1,-2){48}}
\put(150,200){\vector(1,-1){95}}
\end{picture}
\end{tabular}
\end{center}
\caption{\label{grafi}Two bisimilar trees obtained from
$\{\{\e\},\{\e,\{\e\}\}\}$
and $\{\{\{\e,\e\},\e\},\{\e\},\{\e\}\}$ }
\end{figure}

According to  \cite{Acz88}, a \emph{bisimulation} between a graph
$G_1=\langle N_1,E_1\rangle$ and a graph $G_2=\langle
N_2,E_2\rangle$ is a relation $\bis \subseteq N_1\times N_2$ such
that:
\begin{enumerate}
\item $\forall u \in N_1 \:\:\exists v \in N_2$ such that $(u \bis v)$ and
$\forall v \in N_2\:\: \exists u \in N_1$ such that $(u \bis v)$
\item \label{c2} $(u_1 \bis u_2) \wedge \langle u_1,v_1\rangle\in E_1
      \Rightarrow
        \exists v_2\in N_2(\: (v_1 \bis v_2)\: \wedge\:
        \langle u_2,v_2\rangle\in E_2\:)$
\item \label{c3} $(u_1 \bis u_2) \wedge \langle
u_2,v_2\rangle\in E_2  \Rightarrow  \exists v_1\in N_1
(\: (v_1 \bis v_2)\:\wedge\: \langle u_1,v_1\rangle\in E_1\:).$
\end{enumerate}

We can use the notion of bisimulation on trees. Specifically, given
a tree $G_1$, rooted in node $\nu_1$\/, and a tree
$G_2$, rooted in node $\nu_2$\/, $G_1$ is
 \emph{bisimilar} to $G_2$ if and only if there exists a
bisimulation $\bis$ between $G_1$ and $G_2$ such that $\nu_1 \bis
\nu_2$\/. It is simple to verify whether the two trees of
Figure~\ref{grafi} are bisimilar. Observe that  conditions
\ref{c2}. and \ref{c3}.  resemble the extensionality axiom
(Section~\ref{universe})---in fact,
pure sets are equal if and only if their graph
representations are bisimilar~\cite{Acz88}. In~\cite{TCS2004} it
is proved that bisimilarity between \emph{acyclic} and rooted
graphs can be tested in linear time. This result is based on an
algorithm that guesses an initial partition of the
nodes---in particular,
all leaves are initially placed in the same class---and
refines it using a suitable computation strategy.

\medskip

As far as sets with individuals are concerned, the situation is
similar. Let us assume that $a_1,\dots,a_m$ are the individuals
occurring in the two terms. One can obtain the two trees as in the
previous case, but adding a label to each node: 0 for a set node
and $i$ if the node contains the individual $a_i$. Then one can
run the same graph algorithm as in the previous case with  a
single change at the beginning: the leaf nodes are split into
different classes according to their labels.

\begin{remark}\label{lineareonlogn}
In the procedure described above, for ground sets with individuals, we
need to partition leaf nodes according to their labels (individual
names). A similar problem will emerge in other procedures
presented in the paper, where constants symbols and variables must
be ordered. If we assume that the input is given as a string and
the set of constant/variable symbols used is known in advance,
then we can order them in linear time  using radix sort. If
we assume that the input terms are represented by trees using
\emph{structure sharing} (namely, there are no multiple
occurrences of nodes representing the same constant/label), we
have an implicit ordering of constants given by their memory
locations. If, otherwise,  the input is simply a string or a graph
without structure sharing, we first need to provide the ordering
of the symbols used, which requires time $O((\dime{s}+\dime{t})
\log (\dime{s}+\dime{t}))$.
\end{remark}


\subsection{SUD for the ${\sf gflat}(q)$ Class}\label{flat0dec}

Let $q$ be fixed and consider two ${\sf gflat}(q)$ set terms to be
tested: $s = \{a_1,\dots,a_n\} \cup Y_1 \cdots \cup Y_{q'}$ and $t
= \{ b_1,\dots,b_{n'}\} \cup Z_1 \cup \cdots \cup Z_{q''}$
($q'\leq q$ and $q'' \leq q$). Let:
\begin{eqnarray}
\label{vettori}
\begin{array}{rcl}
V_1 & = & \vars(s)\setminus \vars(t)\\
V_2 & = & \vars(t)\setminus \vars(s)\\
V_3 & = & \vars(s) \cap \vars(t)
\end{array} &
\begin{array}{rcl}
C_1 & = & \consts(s)\setminus \consts(t)\\
C_2 & = & \consts(t)\setminus \consts(s)\\
C_3 & = & \consts(s) \cap \consts(t)
\end{array}\end{eqnarray}
where $\vars(\alpha)$ and  $\consts(\alpha)$ denote the set of
variables and the set of simple individual terms occurring in the
term $\alpha$\/, respectively (see Remark~\ref{lineareonlogn} for
a comment on the time required to determine these sets).

If $q' = q'' = 0$ (i.e., $\vars(s) = \vars(t) = \e$), then we are
in the ground case studied in the previous section.

If $q'$ and $q''$ are both greater than 0 (i.e., $\vars(s) \neq
\e$ and $\vars(t) \neq \e$), then $s$ and $t$ are always
unifiable: a solution can be  obtained by assigning the set
$\{a_1,\dots,a_n,b_1\dots,b_{n'}\}$ to all the variables in
$vars(s)\cup vars(t)$\/.

If exactly one between $q'$ and $q''$ is 0, then we have that:
\begin{itemize}
\item if  $q' =0$, then
the problem admits a solution if and only if $C_2 = \e$\/;
\item if $q'' = 0$, then the problem admits a solution if
and only if $C_1 = \e$\/.
\end{itemize}
Thus, to solve the SUD problem for ${\sf gflat}(q)$ set terms we
simply need to compute the sets $C_i$ and $V_i$\/, a task that can
be accomplished in time $O(\dime{s}+\dime{t})$\/. The
considerations made in
 Remark \ref{lineareonlogn}
apply to this case as well.

\subsection{SUD for the ${\sf flat}(q)$ Class}\label{SUDflatq}

Let $q$ be fixed and consider two ${\sf flat}(q)$ set terms to be
tested:
$$s= \{ X_1 , \dots, X_m, a_1,\dots,a_n\} \cup Y_1 \cup
\cdots \cup Y_{q'}$$
and
$$t = \{ W_1 , \dots, W_{m'},
b_1,\dots,b_{n'}\} \cup Z_1 \cup \cdots \cup Z_{q''}$$ ($q'\leq q$
and $q'' \leq q$), and let $V_i$ and $C_i$ be the sets defined in
formula~(\ref{vettori}).

If $m = m' = 0$ we are in the case ${\sf gflat}(q)$ studied
before.
If $q'$ and $q''$ are both greater than 0, then a trivial solution
always exists, as in the ${\sf gflat}(q)$ case.

If $q'= q''=0$, then we can observe that a
necessary condition for the existence of a solution is
that:
\begin{eqnarray}\label{111111}
 |V_1| + |V_2| + |V_3 | \geq |C_1| + |C_2|,
   |V_1| +  |V_3 | \geq  |C_2|,
   |V_2| + |V_3 | \geq |C_1|
\end{eqnarray}
Condition (\ref{111111}) is also sufficient. If (\ref{111111})
holds, then we will be able to construct a solution by assigning
a different value from $C_2$ to each  variable in $V_1$\/, a
different value from $C_1$ to each variable in $V_2$\/, and by
assigning all remaining elements of $C_1$ and $C_2$ (if any) to
the variables in $V_3$. Condition (\ref{111111}) guarantees that
there are enough variables in $V_3$\/. If some variables are not
assigned by this algorithm, then the solution can be easily
completed. For example, when $|V_1| > |C_2|$, we can complete the
solution by assigning any value from $C_2$ or $C_3$ to the
remaining variables of $V_1$.

If exactly one of $q'$ or $q''$ is 0 (without loss of generality,
let us assume $q'' = 0$), then we can determine $V_i$ and $C_i$ as
in the previous cases, but without considering the variables
$Y_i,Z_i$. The  problem admits a solution if and only if $|V_2| +
|V_3| \geq |C_1|$.

\medskip

Thus, the SUD problem for {\sf flat($q$)} set terms can be
reduced to the problem of computing the sets $V_i$ and $C_i$\/. This
can be  done in time $O(\dime{s} + \dime{t})$
(again, see Remark \ref{lineareonlogn}).
As discussed more extensively  in Section~\ref{discussion2},
the class of problems {\sf flat}$(0)$
has been studied in~\cite{arni,greco2}, where these set terms
are called \emph{Bound Simple set terms}.

\subsection{SUD for the ${\sf nested}(q)$ Class} \label{nestedp}

The set unification test for nested sets with non-ground elements
(i.e., with general individuals) has been proved to be NP-hard
in~\cite{KN86} even for the simple case of ${\sf nested}(q)$
with $q=0$\/. We report here the NP-hardness proof from \cite{JLP1}.
Let us consider an instance of 3SAT, e.g.:
$$(X_1 \vee \neg X_2 \vee X_3) \wedge
  (\neg X_1 \vee X_2 \vee \neg X_3)$$
Checking its satisfiability is equivalent to testing set
unification of the two following {\sf nested(0)} set terms:
$$\{\{X_1,Y_1\},\{X_2,Y_2\},\{X_3,Y_3\},
 \{X_1,Y_2,X_3,\e\},
 \{Y_1,X_2,Y_3,\e\} \} \mbox{ and }
 \{ \{ \e,\{\e\}\}\}$$
where we interpret $\e$ as {\sf false} and $\{\e\}$ as {\sf true}.

To prove that the test is in NP, instead, one needs to prove that,
when it is satisfiable, there exists a witness for this situation
that can be verified in polynomial time. Proofs for this result
are rather complex and they can be found in~\cite{KN92,OP95}.

\subsection{Summary of Results for the SUD Problem}

Table~\ref{sum1} summarizes the complexity of the SUD problem for
the different classes of set terms we have introduced.
The \emph{Theory} row will be explained in the next sections.

\begin{table}[t]
\begin{center}
\renewcommand{\arraystretch}{1}
\begin{tabular}{r|c|c|c|c|}
          & {\sf ground}            &  {\sf gflat($q$), $q > 0$}  & {\sf flat($q$), $q = 0,1$}    \\
\cline{1-4}
\emph{SUD Complexity}  & $O(\dime{s}+\dime{t})$ &  $O(\dime{s}+\dime{t})$     & $O(\dime{s}+\dime{t})$     \\
\emph{Theory}          &   $(Ab)(C\ell)$        &  $ACI1$ with constants      & $(Ab)(C\ell)$   \\
\multicolumn{4}{c}{\phantom{aaa}}\\
& {\sf flat($q$), $q > 1$}  & {\sf nested($q$), $q = 0,1$} & {\sf nested($q$), $q > 1$} \\
\cline{1-4}
\emph{SUD Complexity}  &  $O(\dime{s}+\dime{t})$  & NP & NP \\
\emph{Theory} &  gen. $ACI1$ & $(Ab)(C\ell)$  & gen. $ACI1$  \\
\end{tabular}
\end{center}
\caption{Complexity of the SUD problem $s=t$ and $E$-theory used
to solve the SUS problem} \label{sum1}
\end{table}

\subsection{Equations vs. Systems}

We have defined the SUD and SUS problems on a single equation. The
notions can be extended to deal with systems of equations as
well: in this case we need to check whether all the equations in the
system are simultaneously unifiable.

In the {\sf ground} case nothing changes: each equation is
analyzed independently.  For ${\sf gflat}(q)$ we know
from~\cite{KN92,HK97} that the $ACI1$ with constants unification
problem for systems of equations can be reduced to propositional
Horn satisfiability and, thus, it is in $P$. In
Section~\ref{aci1-costanti} we prove the equivalence of this
problem with the ${\sf gflat}(q)$ unification problem.

As far as  the ${\sf flat}(q)$ class is concerned, we can adapt
the reduction of the 3SAT problem done for the ${\sf nested}(q)$
class, using the constant 1 instead of the set $\{\e\}$. The
instance of 3SAT is mapped to the system of equations:
$$\begin{array}{l}
  \{X_1,Y_1\} = \{ \e, 1\},
  \{X_2,Y_2\}= \{ \e, 1\},
  \{X_3,Y_3\} = \{ \e, 1\},\\
 \{X_1,Y_2,X_3,\e\} = \{ \e, 1\},
 \{Y_1,X_2,Y_3,\e\} = \{ \e, 1\}
 \end{array}$$
where all equations involve only ${\sf flat}(q)$ set terms. Thus,
while the ${\sf flat}(q)$ SUD problem for a single equation
requires linear time, the same problem for systems of equations is
NP-complete.

Regarding the ${\sf nested}(q)$ class, each system of
equations $\{s_1 = t_1,\dots,s_n = t_n\}$
can be polynomially reduced to an equisatisfiable equation as
follows:
$$ \{ (\underline{1},s_1),\dots,(\underline{n},s_n)\} =
    \{ (\underline{1},t_1),\dots,(\underline{n},t_n)\}$$
where $\underline{n}$ is a polynomial encoding of the natural number $n$
(e.g., $\underline{0} = \e$,
$\underline{n+1} =
\{ \underline{n} \}$)
and $(x,y)$ is an encoding of the ordered pair (e.g.,
$(x,y) = \{ \{x\},\{x,y\}\}$). Thus, the complexity of the
problem on systems of equations is the same as for a single
equation.

\section{$E$-Unification}\label{notation}

$E$-unification is concerned with solving term equations modulo an
equational theory $E$\/. Set unification can be seen as an
instance of the $E$-unification problem, where the underlying
equational theory contains the identities that capture the
properties of set terms---i.e., the fact that the ordering and
repetitions of elements in a set are immaterial. Different
approaches have been considered to encode sets. Accordingly,
different choices of $E$ should be considered to describe their
basic properties.

We assume the reader to be familiar with the notions of equational
theory, $E$-unification, $E$-unifier and related topics (see,
e.g., \cite{SIEK,BS01}). In this section we introduce a few basic
notations concerning $E$-unification and set unification which
will be useful in the rest of the paper.

A \emph{signature} $\Sigma$ consists of a set of function symbols.
Terms built from $\Sigma$ and from a denumerable set $\cal V$ of
variables are called \emph{$\Sigma$-terms}. ${\cal T}(\Sigma,{\cal
V})$ is the set of all the $\Sigma$-terms---and it is called the
\emph{term algebra}. Given a sequence of terms $t_1,\ldots,t_n$\/,
$\vars(t_1,\ldots,t_n)$ denotes the set of variables occurring in
the terms. $\vars$ is naturally extended to equations and sets of
equations.

A \emph{substitution} $\sigma: {\cal V} \longrightarrow {\cal
T}(\Sigma,{\cal V})$ is represented by the notation
$[X_1/t_1,\dots,X_n/t_n]$\/, where $dom(\sigma) =
\{X_1,\dots,X_n\}$ (the \emph{domain} of $\sigma$) and, for each
$i=1,\dots,n$\/, $\sigma(X_i)=t_i$\/.
A substitution is uniquely extended to a function
over $T(\Sigma,{\cal V})$ using structural induction. The
application of a substitution $\sigma$ to a term $t$ will be
denoted by $t\sigma$ (or, equivalently, by $\sigma(t)$).

An \emph{equational theory} is a finite collection of
identities $E$, where each identity is written
as $s \approx t$, and $s,t$ are terms belonging to
${\cal T}(\Sigma,{\cal V})$\/. The relation
$=_E$ is the \emph{least congruence
relation} on the term algebra ${\cal T}(\Sigma,{\cal V})$, which
contains $E$ and it is closed under substitution~\cite{BS01}.
Function symbols not occurring in $E$ are said to be \emph{free}.

\medskip

We describe now the properties of the function symbols that we use
as the set constructors. The properties that the $\cup$
constructor should have in a set theory can be described by the
following identities:
$$\begin{array}{crcll}
(A) & (X \cup Y) \cup Z & \approx &  X \cup (Y \cup Z) &
     \mbox{\emph{(Associativity)}}\\
(C) & X \cup Y & \approx & Y \cup X & \mbox{\emph{(Commutativity)}}\\
(I) & X \cup X & \approx & X & \mbox{\emph{(Idempotence)}}
\end{array}$$
Moreover, the constant symbol $\e$\/, used to denote the empty
set, is the identity element for the $\cup$ operator. This is
stated by:
$$\begin{array}{crcll}
(1) & X \cup \e  & \approx & X & \mbox{\emph{(Identity)}}
\end{array}$$
Let $E_{ACI1}$ be the equational theory consisting of identities
$(A)$\/, $(C)$\/, $(I)$\/, and $(1)$\/.

\smallskip

The $\w$ constructor, instead, should exhibit the properties
described by the following identities:
$$\begin{array}{crcll}
(Ab) &\{X\,|\,\{X\,|\,Z\}\} & \approx &
    \{X\,|\,Z\} & \mbox{\emph{(Absorption)}}\\
(C\ell) & \{X\,|\,\{Y\,|\,Z\}\} & \approx & \{Y\,|\,\{X\,|\,Z\}\}
& \mbox{\emph{(Commutativity on the left)}}
\end{array}$$

\medskip

A substitution $\sigma$ is an \emph{$E$-unifier} (or, simply, a
\emph{unifier} when the context is clear) of two terms $s,t$ if
$s{\sigma} =_E t{\sigma}$---i.e., $s{\sigma}$ and $t{\sigma}$
belong to the same $E$-congruence class.

An \emph{$E$-unification problem} over $\Sigma$ is a
{system} of equations ${\cal E} = \{s_1 = t_1,\dots,s_n =
t_n\}$ between $\Sigma$-terms. A substitution $\mu$ which is an
$E$-unifier of all the equations in ${\cal E}$ is said to be an
$E$-unifier (or an $E$-\emph{solution}) of ${\cal E}$\/. The set of
all the $E$-unifiers of $\cal E$ is denoted by ${\cal U}_E({\cal E})$\/.

Let $E$ be an equational theory and ${\cal W}$ a set of variables
(${\cal W} \subseteq {\cal V}$). ${\cal U}_E({\cal E})$ can be
sorted with respect to the pre-order $\leq_E^{\cal W}$\/: given two
substitutions $\sigma_1,\sigma_2$:
$$\begin{array}{rcl}
\sigma_1 \leq_E^{\cal W} \sigma_2 & \mbox{iff} &
\mbox{there exists a substitution $\lambda$ such that}\\
&&\mbox{$\sigma_2(X) =_E (\sigma_1 \circ \lambda)(X)$ for all $X$
in ${\cal W}$.}
\end{array}$$
In this case we say that $\sigma_1$ is \emph{more general modulo
$E$ on ${\cal W}$} than $\sigma_2$. If $\sigma_1 \leq_E^{\cal W}
\sigma_2$ and $\sigma_2 \leq_E^{\cal W} \sigma_1$\/, then we say
that $\sigma_1 =_E^{\cal W} \sigma_2$\/. Whenever ${\cal W}$ is
omitted from $\leq_E^{\cal W}$, then ${\cal W}$ is implicitly
assumed to be ${\vars}({\cal E})$.

While traditional syntactic unification problems between Herbrand
terms admit at most one most general unifier (mgu),
$E$-unification problems may not have a single most general
unifier. In this context, the role of the most general unifier is
taken on by a minimal complete set of unifiers. A \emph{complete
set of $E$-unifiers} for an $E$-unification problem $\cal E$ is a
set $\cal C$ of $E$-unifiers (i.e., a subset of ${\cal U}_E({\cal
E})$) that satisfies the additional condition:
\begin{itemize}
\item for each $E$-unifier $\sigma$
there exists an element $\theta$ in $\cal C$ such that $\theta
\leq_E \sigma$.
\end{itemize}
A complete set of $E$-unifiers $\cal C$ is called
a \emph{minimal complete set of $E$-unifiers} if it
 fulfills the
\emph{minimality} condition:
\begin{itemize}
\item  for any pair
        $\mu_1,\mu_2$ in $\cal C$\/, if
        $\mu_1 \leq_E \mu_2$, then  $\mu_1 = \mu_2$.
\end{itemize}
A substitution $\sigma$ in a minimal complete set of
$E$-unifiers $\cal C$ is called a \emph{maximal general
$E$-unifier}. When $\cal C$ is a singleton set $\{\sigma\}$ we say
that $\sigma$ is the \emph{most general $E$-unifier}.
If one minimal set of $E$-unifiers can be obtained from another
one by variable renaming and vice versa, the two sets are
equivalent and only one of them needs to be computed.

A special form of systems of equations, called the
\emph{solved form},
plays an important role in the definition of unification
algorithms. An equation $e$ of the form  $X = t$ is said to be in
\emph{solved form} with respect to a system  $\cal E$ if $X$ does
not occur neither in $t$ nor elsewhere in ${\cal E}$\/. In this
case, $X$ is said to be a \emph{solved} variable in $\cal E$\/. A
system $\cal E$ is said to be in solved form if, for all $e$ in
$\cal E$\/, $e$ is in solved form with respect to  ${\cal E}$\/.
{From} a system in solved form $\{X_1 = t_1,\dots,X_n = t_n\}$\/,
it is simple to derive the most general $E$-unifier $[X_1 /
t_1,\dots,X_n / t_n]$\/.

\medskip

$E$-unification problems can be classified according to whether
their signature $\Sigma$ contains free elements (i.e., function
symbols that do not occur in $E$). In particular, it is possible
to distinguish between:
\begin{itemize}
\item \emph{elementary unification}, where the terms
        to be unified are built only using
        the symbols appearing in the considered equational theory;
\item \emph{unification with constants}, where
        the terms to be unified are built using
        symbols in the equational theory and
        additional free constants;
\item \emph{general unification}, where the terms to be unified
        are arbitrary terms containing function symbols which
        are either free or present in the equational theory.
\end{itemize}
The unification problem studied in the next section falls in the
class of unification with constants.
The remaining sections consider general unification problems.

The SUD problem studied in Section~\ref{SUD} is an abstract case
of the \emph{$E$-unifiability problem} (namely, deciding whether
or not an $E$-unifier exists). In the next sections we deal with
the SUS problem, i.e.,the problem of determining a \emph{complete
set} of $E$-unifiers of an equation $s = t$ or of a system of
equations $\cal E$.

\section{$ACI1$ with Constants}\label{aci1-costanti}

According to the classification presented in Section
\ref{abstractsetterms} the simplest non-ground set terms we deal
with are those belonging to the ${\sf gflat}(q)$ class. In this
section we show that the SUS problem for this class can be solved
by using the solution to the $ACI1$ with constants unification
problem.

\subsection{Language and Semantics}

Let $\Sigma = \{\e,\cup,c_1,c_2,\dots \}$ be a signature composed
of the binary function symbol $\cup$\/, the constant symbol
$\e$\/, and an \emph{arbitrary} number (possibly infinite) of
free constant symbols $c_1,c_2,\dots$

\begin{definition}
An \emph{$ACI1$ with constants term} is either a variable, a
constant, or a $\Sigma$-term of the form $s_1 \cup s_2$\/, where
$s_1$ and $s_2$ are $ACI1$ with constants terms.
\end{definition}
The properties of the function symbols $\cup$\/ and $\e$\/ are
described by the identities $(A)$\/, $(C)$\/, $(I)$\/, and $(1)$
introduced in Section~\ref{notation}. Thanks to the associativity
property $(A)$\/, $ACI1$ with constants terms can be always
written as strings of the form $ \alpha_1 \cup \cdots \cup
\alpha_m$ where $\alpha_i$ is either a variable,  $\emptyset$\/,
or a constant term $c_i$\/. Moreover, using $(C),(I)$\/, and
$(1)$ we can restrict our attention to terms  without
duplications of sub-terms and without $\e$ as a sub-term (unless
the whole term is $\e$\/).

Flat set terms with variable elements (i.e., {\sf flat($q$)}
set terms) are not expressible in this language. Indeed the
language does not allow us to distinguish individuals from sets.
Variables in a set term are always interpreted as set variables.
Furthermore, nested set terms are not expressible in this
language \cite{DPP00-FROCOS}.

\subsection{Which Kind of Set Unification}\label{unificazioneACI1}

The $ACI1$ with constants language allows us to describe the set
unification problem for ${\sf gflat}(q)$ set terms.
The SUS problem for this class can be solved  using the solution
to the corresponding  $ACI1$ with constants
unification problem (defined below).
As an example, let us consider the ${\sf gflat}(q)$ unification
problem:
$$\{a,b\}\cup Y_1 \cup Y_2  =
  \{a,b,c,d \}$$
The solutions for this problem
are those mapping $Y_1$ and $Y_2$ to subsets of $\{a,b,c,d\}$ such
that $c$ and $d$ are in the image of $Y_1$ or $Y_2$. For instance,
$[Y_1/\{a,c\},Y_2/\{a,b,d\}]$ is a solution. Let us consider now
the related $ACI1$ with constants unification problem:
$$a \cup b \cup Y_1 \cup Y_2  =
  a\cup b \cup c \cup d$$
In this case, $a$, $b$, $c$, $d$ are not interpreted as set
elements. However, thanks to the properties of the $\cup$
operator, the solutions for this problem are closely related to
those for the ${\sf gflat}(q)$ unification problem.
The solutions for the $ACI1$ with constants unification
problem are those mapping $Y_1$ and $Y_2$ to unions of elements of
$\{a,b,c,d\}$ such that $c$ and $d$ are in the image of $Y_1$ or
$Y_2$. For instance, $[Y_1/ a\cup c, Y_2/ a \cup b \cup d]$.

We formalize this idea by defining a function $(\cdot)^*$ that
translates ${\sf gflat}(q)$ set terms into $ACI1$ with constants
terms as follows:
$$(\{a_1,\dots,a_n\}\cup Y_1 \cup \cdots \cup Y_q)^* \defAs
   a_1 \cup \cdots \cup a_n \cup Y_1 \cup \cdots \cup Y_q$$
$(\cdot)^*$ admits an inverse function.
The function can also be extended to substitutions:
$\sigma^*(X) = (\sigma(X))^*$.

\begin{lemma} \label{equivaleACI1}
$\sigma$ is a solution of the ${\sf gflat}(q)$ SUS problem $s = t $
if and only if
$\sigma^*$ is a $ACI1$ unifier of $s^*=t^*$.
\end{lemma}
For the proof, see  \ref{auxproofs}.

\begin{example}
The following are set terms and set unification problems which are
allowed in $ACI1$ with constants:
\begin{itemize}
\item $X_1 \cup X_2 \cup X_3 = X_4 \cup X_1$
\item $a \cup b  \cup X_1 \cup X_2 = c \cup X_3$\/---that is
$(\{a,b\} \cup X_1 \cup X_2)^* = (\{c\} \cup X_3)^*$\/
\item the first problem of Section \ref{examples} (the
\emph{Chords} problem) can be encoded as the $ACI1$ with constants
problem $ c \cup e \cup g \cup b\flat = g \cup g \cup e \cup
b\flat \cup c \cup e$.
\end{itemize}
\end{example}

\subsection{Unification Algorithm}   \label{baader-algo}

A general algorithm capable of computing a minimal complete set of
$ACI1$-unifiers for $ACI1$ with constants unification problems has
been presented in~\cite{BB88}.

Given two $\Sigma$-terms  $s$ and $t$ the algorithm computes a
complete set $S$ of $ACI1$-unifiers for $ s=t $. Without loss of
generality, we assume that if only one of the terms is ground,
then it is $t$\/. The set $S$ can be extracted from a schema of
Boolean $ACI$-matrices. Each column of the matrix is associated to
a variable in $s=t$. Each row, instead, is associated to new
variables that will enter in the solutions. The matrix is composed
of identity matrices, by matrices of 0 with exactly one column set
to 1, and by 0 matrices.

\begin{example}\label{var_con_shar}
Let us consider the problem:
$$S_1 \cup S_2 \cup X = T_1 \cup T_2 \cup X$$
The sets $V_1,V_2,V_3,C_1,C_2,C_3$ are computed as in formula
(\ref{vettori}) of Section~\ref{flat0dec}:
 $V_1 = \{S_1,S_2\}$, $V_2
        = \{T_1,T_2\}$, $V_3 = \{X\}$\/, and $ C_1 = C_2 = C_3 = \e$.
\begin{figure}[h]
$$\begin{array}[b]{rl}
\renewcommand{\arraystretch}{1}
\begin{array}[b]{|ccccc|}
\multicolumn{2}{c}{\overbrace{\phantom{aaaaaa}}^{ V_1}}
& \multicolumn{2}{c}{\overbrace{\phantom{aaaaaa}}^{ V_2}}
& \multicolumn{1}{c}{\overbrace{\phantom{aa}}^{ V_3}}\\
\multicolumn{1}{c}{S_1} &
S_2 &  T_1 & T_2 &
\multicolumn{1}{c}{X} \\
\cline{1-5}
1 & 0 & 1 & 0 & 0  \\
1 & 0 & 0 & 1 & 0   \\
\cline{1-5}
0 & 1 & 1 & 0 & 0   \\
0 & 1 & 0 & 1 & 0   \\
\cline{1-5}
1 & 0 & 0 & 0 & 1  \\
0 & 1 & 0 & 0 & 1   \\
\cline{1-5}
0 & 0 & 1 & 0 & 1  \\
0 & 0 & 0 & 1 & 1   \\
\cline{1-5}
0 & 0 & 0 & 0 & 1  \\
\cline{1-5}
\end{array} &
\renewcommand{\arraystretch}{1.01}
\hspace{-.5cm}\begin{array}[b]{l}
\left.\begin{array}{l}
R_1 \\ R_2 \end{array}\right.
\\
\left.\begin{array}{l}
R_3 \\ R_4 \end{array}\right.
\\
\left.\begin{array}{l}
R_5  \end{array}\right.
\\
\left.\begin{array}{l}
R_6 \end{array}\right.
\\
\left.\begin{array}{l}
R_7 \\ R_8 \end{array}\right.
\\
\left.\begin{array}{l}
R_9 \end{array}\right.
\end{array}
\end{array}
$$
\caption{The $ACI$-matrix for the problem
$S_1 \cup S_2 \cup X = T_1 \cup T_2 \cup X$}
\label{figesempio}
\end{figure}
Since the given problem does not involve constants, the matrix is
unique (see Figure~\ref{figesempio}). $R_1,\dots,R_9$ are new
variables that allow to compactly represent the unique mgu:
$$
    \left[\begin{array}{lcl}
       S_1 & / & R_1 \cup R_2 \cup R_5,\\
           S_2 & / & R_3 \cup R_4 \cup R_6, \\
           T_1 & / & R_1 \cup R_3 \cup R_7,\\
           T_2 & / & R_2 \cup R_4 \cup R_8,\\
           X &  / & R_5 \cup R_6 \cup R_7 \cup R_8 \cup R_9
      \end{array}
\right]$$
The two 1's in a row state that the two variables should have a part
in common in each solution. For instance, in the first row it is
stated that $R_1$ is a part of $S_1$ and of $T_1$ (in other words,
$R_1 = S_1\cap T_1$).
\end{example}

When the problem involves constants, the matrices have also rows
for $C_1,C_2,C_3$. In this case several matrices are
non-deterministically generated. Each of them describes a
solution; their union covers the whole solution space.

\begin{example}\label{3su2}
Let us consider the problem:
$$X_1 \cup X_2 \cup X_3 = a \cup b$$
where $ V_1 = \{ X_1,X_2,X_3 \},
        C_2 = \{ a,b \}$\/,
$V_2 = V_3 = C_1 = C_3 = \e$.
There are 49 $ACI$-matrices for
this problem. Two of them are:
$$\begin{array}[b]{clccl}
\begin{array}[b]{|ccc|}
\multicolumn{3}{c}{\overbrace{\phantom{aaaassssaaaa}}^{V_1}}\\
\multicolumn{1}{c}{X_1} & X_2 &
\multicolumn{1}{c}{X_3} \\
  \cline{1-3}
  0 & 1 & 1 \\
  \cline{1-3}
  1 & 0 & 1  \\
  \cline{1-3}
\end{array} &
\hspace{-.5cm}\begin{array}[b]{l}
\left.\begin{array}{l}
a \\ b \end{array}\right\} C_2 \\
\vspace{-0.5cm}
\end{array} &
\phantom{aaa} &
\begin{array}[b]{|ccc|}
\multicolumn{3}{c}{\overbrace{\phantom{aaaassssaaaa}}^{V_1}}\\
\multicolumn{1}{c}{X_1} & X_2 & \multicolumn{1}{c}{X_3} \\
  \cline{1-3}
  0 & 0 & 1 \\
  \cline{1-3}
  0 & 1 & 0 \\
  \cline{1-3}
  \end{array}  &
\hspace{-.5cm}
\begin{array}[b]{l}
\left.\begin{array}{l}
a \\ b \end{array}\right\} C_2 \\
\vspace{-0.5cm}
\end{array}
\end{array}$$
yielding the unifiers: $\begin{array}{cc}
[X_1 / b, X_2 /a, X_3 / a \cup b], &
[X_1 / \e, X_2 / b, X_3 / a].
\end{array}$
\end{example}

The number of $ACI$-matrices to be computed for a given $ACI1$
unification problem is $(2^{| V_2|} - 1 + | V_3|)^{| C_1|}
  (2^{| V_1|} - 1 + | V_3| )^{| C_2|}
(2^{| V_1|} + 2^{| V_2|} - 1)^{|  C_3|} $
which is $O(2^{(\dime{s}+\dime{t})^2})$\/~\cite{BB88}.

\medskip

The detection of a solution of a unification problem (i.e.,
solving the SUS problem) clearly implies solving the related
decision problem. Thus,  the complexity of finding a solution can
be no better than the complexity of solving the corresponding
decision problem. In this case, both the problems can be solved
in linear time (with the assumption in
Remark~\ref{lineareonlogn}). This can be achieved as follows.
First verify that the decision problem $s=t$ has a positive
answer; this can be done in linear time thanks to the results in
Lemma~\ref{equivaleACI1} and Section~\ref{flat0dec}. If the test
succeeds, then a solution can be constructed by assigning to each
variable $X$ in $s=t$ a term composed of the union of all the
constants present in $s = t$\/.  For further  details the reader
is referred to~\cite{BB88}.

\subsection{Discussion}\label{discussion1}

A simpler unification problem---called the \emph{elementary}
$ACI1$ unification problem---has been considered in the
literature. This problem involves terms which are constructed
using only variables, the constant $\emptyset$\/, and the binary
constructor $\cup$ (i.e., a subcase of {\sf gflat($q$)} with $n =
0$ and $q \geq 0$).
This problem is simpler in the sense that the decision problem
has always a positive answer---i.e., each unification problem $s=t$
has a solution.
Therefore, the complexity of finding an arbitrary
solution is $O(1)$.
Furthermore, each elementary $ACI1$ unification
problem admits a single most general unifier.
In~\ref{matrix} we show a variant of the
$ACI$-matrices for this simplified problem.

As a final remark, \cite{HK97,KN92} show how the result presented
in this section can be extended to provide a polynomial time
solution to systems of  $ACI1$ with constants unification
problems.

\section{General $(Ab)(C\ell)$ Unification}\label{setlog}

Set terms involving variable elements and/or nested sets are not
expressible in the language of $ACI1$ with constants (see
Section~\ref{unificazioneACI1}). The proposal we describe in this
section is intended to enlarge the domain of discourse to the
more general class of ${\sf nested}(q)$ set terms with $q\leq1$.
As already observed at the end of Section~\ref{classes}, in this
case we can rely on the element insertion operator $\w$ as the set
constructor for the concrete implementation of sets. This choice
allows the presence of at most one set variable in each set term,
while $ACI1$ with constants does not place any restriction on the
number of set variables which can occur in each set term. On the
other hand, it allows us to represent \emph{nested} sets---which
is not possible using $ACI1$ with constants unification.
Moreover, it allows sets to be viewed and manipulated in a
fashion similar to \emph{lists}. As a matter of fact, this
approach has been adopted in a number of  logic and
functional-logic programming languages (e.g., $CLP({\cal
SET})$~\cite{DR93,TOPLAS2000}, SEL~\cite{Jay92},
SETA~\cite{SETA}).

The unification algorithm we propose here is similar to the one
presented in~\cite{JLP1}---but with a considerably simpler
termination proof. The underlying equational theory contains the
two identities $(Ab)$ and $(C\ell)$ shown in
Section~\ref{notation}, stating the fundamental properties of the
set constructor $\w$\/.

\subsection{Language and Semantics}

$\Sigma$ is a signature containing the binary function symbol
$\{\cdot\,|\,\cdot\}$\/, the empty set constant symbol $\e$\/, and
an arbitrary number (possibly infinite) of free function symbols
with arbitrary arities.

\begin{definition}
An \emph{$(Ab)(C\ell)$ set term} is either a variable, or the
constant $\e$, or a $\Sigma$-term of the form $\{t\,|\,s\}$, where
$t$ is a $\Sigma$-term and $s$ is an $(Ab)(C\ell)$ set term. An
\emph{individual term} is either a variable or a $\Sigma$-term of
the form $f(t_1,\dots,t_n)$ with $f \not\equiv \w,
f \not\equiv \e$ and
$t_1,\dots,t_n$ are $\Sigma$-terms
(if $n=0$ it is a constant term).
\end{definition}
The function symbol $\{\cdot\,|\,\cdot\}$ has the properties
described by the identities $(Ab)$ and $(C\ell)$ introduced in
Section~\ref{notation}. Hence, set terms denote hereditarily
finite sets based on $\cal U$, while individual terms denote
arbitrary elements of the universe $\cal U$. As a notational
convenience $\{\,s_1\,|\,\{\,s_2\,|\,\cdots\,\{\,s_n\,|\,t\,\}
\cdots \}\}$ will be written as $\{s_1,\dots,s_n\,|\,t\}$ or
simply as $\{s_1,\dots,s_n\}$ when $t$ is $\e$\/.

\subsection{Which Kind of Set Unification}

The $(Ab)(C\ell)$ language allows us to describe the SUD and SUS
problems for {\sf nested(1)} set terms---i.e.,
arbitrary nested sets with at most one set variable per
set term. In particular, the language allows us to
deal with all classes
of set terms that are included in {\sf nested(1)}, namely {\sf
ground}, {\sf gflat(1)}, and {\sf flat(1)}.

\begin{example} The following are set terms and set unification
problems which are allowed in $(Ab)(C\ell)$:
\begin{itemize}
\item $\{X,\{Y\}\} =  \{Z,\e\}$
\item $\{\{X_1,a\}\,|\,Y_1\} =  \{X_3\,|\,Y_2\}$ (i.e., in abstract
syntax---cf. Section \ref{abstractsetterms}---$\{\{X_1,a\}\} \cup
Y_1 =  \{X_3\} \cup Y_2$)
\item the \emph{Graph coloring} problem of Section \ref{examples}
can be encoded as an $(Ab)(C\ell)$ problem:
\[\begin{array}{c}
 \hspace{-15pt}  \{\{X_1,X_2\},\{X_2,X_3\},\{X_3,X_4\},\{X_4,X_1\}\,|\,
                  R \} =
          \{ \{c_1,c_2\},\{c_1,c_3\},\{c_2,c_3\} \}\\
       \end{array}\]
\end{itemize}
On the other hand, the problem $A \cup B \cup C = \{a\} \cup D$
cannot be expressed in this framework.
\end{example}

\subsection{Unification Algorithm}\label{sets.unification}

The algorithm consists of three parts. The first part ({\fp
AbCl\_unify}---see Figure~\ref{algobags.a}) chooses one equation
at a time using a semi-deterministic strategy. The
second part ({\fp AbCl\_unify\_actions}---see
Figure~\ref{genalgo}) performs the rewriting of the selected
equation. A final processing of \emph{membership equations}, i.e.,
equations of the form $X = \{t_0, \dots, t_n \,|\, X\}$ with $X
\not\in \vars(t_0,\dots,t_n)$\/, ({\fp AbCl\_unify\_final}---see
Figure~\ref{membership}) constitutes the third and final part of
the algorithm.

The system ${\cal E}$ is split into three parts: ${\cal E}_{\fp
s}$ is the solved form part (initially set to empty), ${\cal
E}_{\fp ns}$\/ is a system of equations (initially set to the
input system ${\cal E}_{\fp in}$), and ${\cal E}_{\fp aux}$ is a
system of equations dealt with as a stack. For ${\cal E}_{\fp
aux}$ we assume the existence of a {\fp push} operation that puts
an equation on the top of the stack and of a {\fp pop} operation
that returns and removes the equation on the top of the stack.
Given a system of equations ${\cal E}_{\fp in}$, the algorithm
non-deterministically returns either {\fp fail} or a collection of
systems in solved form.

\begin{figure}[htb]
\small
$$\begin{array}{l}
\mbox{\fp AbCl\_unify}({\cal E}_{\fp in}):\\
\phantom{aaa}\mbox{\fp ${\cal E}_{\fp s} := \e$\/;
                 ${\cal E}_{\fp ns} := {\cal E}_{\fp in}$\/;
                 ${\cal E}_{\fp aux} := \e$\/;}\\
\phantom{aaa}\mbox{\fp  ${\cal E} := \langle {\cal E}_{\fp s}$,
${\cal E}_{\fp ns}$,
  ${\cal E}_{\fp aux}\rangle$;} \\
\phantom{aaa}\mbox{\fp while ${\cal E}_{\fp ns}$ $\neq$ $\e$ or
                        ${\cal E}_{\fp aux}$ $\neq$ $\e$ do}\\
\phantom{aaaaaa}\mbox{\fp if ${\cal E}_{\fp aux}$ $\neq$ $\e$
then }
                \mbox{\fp $e :=$ {\fp pop}(${\cal E}_{\fp aux}$) }\\
\phantom{aaaaaa}\mbox{\fp else }
                \mbox{\fp select arbitrarily an equation $e$
                                from ${\cal E}_{\fp ns}$ and remove it;}\\
\phantom{aaaaaa}\mbox{\fp AbCl\_unify\_actions(${\cal E}$,$e$);}\\
\phantom{aaa}\mbox{\fp AbCl\_unify\_final}({\cal E})\\
\hline\hline
\end{array}$$
\caption{General $(Ab)(C\ell)$ Unification Algorithm (main)}
\label{algobags.a}
\end{figure}

In the algorithm we make use of the function {\fp tail}, defined
as follows:
$$\begin{array}{rcl}
  {\fp tail}(t) & = & \left\{
\begin{array}{ll}
t & \mbox{if $t$ is a variable or a term $f(t_1,\dots,t_n)$\/, $f
\not\equiv
\{\cdot\,|\,\cdot\}$}\\
{\fp tail}(t_2) & \mbox{if $t$ is $\{t_1\,|\,t_2\}$}
\end{array}
\right.
\end{array}$$
For instance, if $s = \{a,b\}$, namely $s =
\{a\,|\,\{b\,|\,\e\}\}$, then ${\fp tail}(s)= \e$\/.

\begin{figure}[htbp]
\small
$$
\renewcommand{\arraystretch}{0.6}
\begin{array}{crcl}
\multicolumn{4}{l}{\mbox{\fp AbCl\_unify\_actions(${\cal E},e$):}}\\
\multicolumn{4}{l}{\phantom{aaa}\mbox{\fp case $e$ of}}\\
\hline
(1) & \begin{array}{r}
X = X
\end{array} & \mapsto &  {\cal E}_{\fp ns} := {\cal E}_{\fp ns}\\
\hline
(2) & \left.\begin{array}{r}
t = X \\
\mbox{$t$ is not a variable}
\end{array} \right\}
& \mapsto & {\cal E}_{\fp ns} := {\cal E}_{\fp ns} \wedge ( X = t ) \\
\hline
(3) & \left.\begin{array}{r}
X = f(t_1,\dots,t_n)\\
f \not\equiv \w \mbox{ and $X$ occurs in } f(t_1,\dots,t_n)
\end{array} \right\}
& \mapsto & {\fp fail} \\
\hline
(4) & \left.\begin{array}{r}
X = \{t_0,\dots,t_n \,|\,t\}\\
t \not\equiv \{\cdots\} \mbox{ and $X$ occurs in $t$ ($X \not\equiv t$),}\\
\mbox{or $X$ occurs in $t_0,\dots,t_n$}
\end{array} \right\}
& \mapsto & {\fp fail} \\
\hline
(5) &  \left.\begin{array}{r}
X = t\\
\mbox{$X$ does not occur in $t$} \\
\end{array} \right\}
& \mapsto & \begin{array}{l}
          {\cal E}_{\fp s} := {\cal E}_{\fp s}[X/t] \wedge ( X = t );\\
          {\cal E}_{\fp ns} := {\cal E}_{\fp ns}[X/t];
          {\cal E}_{\fp aux} := {\cal E}_{\fp aux}[X/t]
                    \end{array}\\

\hline
(6) & \left.\begin{array}{r}
f(s_1,\dots,s_m) = g(t_1 ,\dots ,t_n) \\
f \not\equiv g \end{array} \right\}
& \mapsto &  {\fp fail}\\
\hline
(7) & \left.\begin{array}{r}
f(s_1,\dots ,s_n) = f(t_1,\dots,t_n)\\
f \not\equiv \w
\end{array} \right\}
& \mapsto & {\cal E}_{\fp ns} := {\cal E}_{\fp ns} \wedge
              ( s_1 = t_1 \wedge \dots \wedge s_n = t_n )\\
\hline
(8) & \begin{array}{r}
\{t\,|\,s\} = \{t'\,|\,s'\}\\
\end{array}
& \mapsto & {\fp AbCl\_{step}}({\cal E},\{t\,|\,s\} = \{t'\,|\,s'\})\\
\hline
\hline
\end{array}$$

$$\begin{array}{crcl}
\multicolumn{4}{l}{\mbox{{\fp AbCl\_step}}({\cal E},
\{ t\,|\,s \}  =  \{ t'\,|\,s' \}) : }\\
\multicolumn{4}{l}{\phantom{aaa}
   \mbox{\fp if tail($s$) and tail($s'$) are not the same variable  then choose one among:} }\\
\multicolumn{4}{l}{\phantom{aaa}\begin{array}{cl}
            (i) & {\cal E}_{\fp ns} := {\cal E}_{\fp ns} \wedge (t  =  t');
               {\fp push}(s  =  s'\:,\: {\cal E}_{\fp aux}) \\
            (ii) & {\cal E}_{\fp ns} := {\cal E}_{\fp ns} \wedge (t  =  t');
               {\fp push}(\{t\,|\,s\}  =  s'\:,\: {\cal E}_{\fp aux}) \\
            (iii) & {\cal E}_{\fp ns} := {\cal E}_{\fp ns} \wedge (t  =  t');
               {\fp push}(s  =  \{t'\,|\,s'\}\:,\: {\cal E}_{\fp aux}) \\
            (iv) & {\fp push}(s  =  \{t'\,|\,N\}\:,\: {\cal E}_{\fp aux});
                   {\fp push}(\{t\,|\,N\}  =  s'\:,\: {\cal E}_{\fp aux}) \\
                             & \mbox{$N$ new variable}
          \end{array} }\\
\multicolumn{4}{l}{\phantom{aaa}
   \mbox{\fp else
   let $\{t\,|\, s\} \equiv \{t_0,\dots,t_m\,|\,X\}$ and
               $\{t'\,|\, s'\}\equiv \{t_0',\dots,t_n'\,|\,X\}$,
   $X$ variable;}}\\
\multicolumn{4}{l}{\phantom{aaa}
   \mbox{\fp select arbitrarily $i$ in $\{0,\dots,n\}$\/; }
   \mbox{\fp choose one among: } }\\
\multicolumn{4}{l}{\phantom{aaa}
\begin{array}{cl}
         (i) & {\cal E}_{\fp ns} := {\cal E}_{\fp ns} \wedge (t_0  =  t_{i}');
              {\fp push}(\{t_1,\dots,t_m\,|\,X\}  =
\{t_0',\dots,t_{i-1}',t_{i+1}',\dots,t_n'\,|\,X\},
                            {\cal E}_{\fp aux}) \\
         (ii) & {\cal E}_{\fp ns} := {\cal E}_{\fp ns} \wedge (t_0  =  t_{i}');
               {\fp push}(\{t_0,\dots,t_m\,|\,X\}  =
                        \{t_0',\dots,t_{i-1}',t_{i+1}',\dots,t_n'\,|\,X\},
                            {\cal E}_{\fp aux}) \\
        (iii) & {\cal E}_{\fp ns} := {\cal E}_{\fp ns} \wedge (t_0  =  t_{i}');
                {\fp push}(\{t_1,\dots,t_m\,|\,X\}  =
                 \{t_0',\dots,t_n'\,|\,X\}, {\cal E}_{\fp aux})\\
         (iv) & {\fp push}(X  =  \{t_0\,|\,X\}, {\cal E}_{\fp aux});
                {\fp push}(\{t_1,\dots,t_m\,|\,X \} =
                           \{t_0',\dots,t_n'\,|\,X \}, {\cal E}_{\fp aux}) \\
         \end{array} }\\
\hline
\hline
\end{array}$$

\small
$$\begin{array}{ crcl }
\multicolumn{4}{l}{\mbox{\fp AbCl\_unify\_final}({\cal E}):}\\
\multicolumn{4}{l}{\phantom{aaa}
      \mbox{\fp Repeatedly perform any of the following actions;}}\\
\multicolumn{4}{l}{\phantom{aaa}
      \mbox{\fp  if neither applies then stop with success;}}\\
  \hline
  (1) & \left.\begin{array}{r}
  X = \{t_0^0,\dots,t_{n_0}^0\,|\,X\} \wedge
  \dots \wedge
  X = \{t_0^k,\dots,t_{n_k}^k\,|\,X\} \wedge {\cal E}\\
  \mbox{$k > 0$, the number of all membership equations involving $X$} \\
  \mbox{$X$ does not occur in $t_0^0,\dots,t_{n_0}^0,
  \dots,t_0^k,\dots,t_{n_k}^k$}
  \end{array}  \right\}
  & \mapsto &
  \phantom{aaaaaaaaaa}\\
  &  \multicolumn{3}{r}{
  X = \{ t_0^0,\dots,t_{n_0}^0,
  \dots,t_0^k,\dots,t_{n_k}^k\,|\,N\} \wedge
  {\cal E}[X/\{ t_0^0,\dots,t_{n_0}^0,
  \dots,t_0^k,\dots,t_{n_k}^k\,|\,N\}]}\\
  \hline
  (2) & \left.\begin{array}{r}
  X = t \wedge {\cal E}\\
  \mbox{$X$ occurs in $t$}
  \end{array} \right\}
  & \mapsto & {\fp fail}\\
  \hline\hline
  \end{array}$$

\caption{General $(Ab)(C\ell)$ Unification Rewriting Rules}
\label{genalgo} \label{membership}
\end{figure}

The core of the unification algorithm (Figure~\ref{genalgo}) is
very similar in structure to the traditional unification
algorithms for standard Herbrand terms (e.g.,~\cite{MM82}). In
particular, rule (1) is also known as the \emph{Trivial} rule,
rule (2) as the \emph{Orient} rule, rules (3) and (4) are the
\emph{Occurs Check} rules, rule (5) is known as the \emph{Variable
Elimination} rule, rule (6) as the \emph{Symbol Clash} rule, and
rule (7) as the \emph{Term Decomposition} rule~\cite{JOU91}. The
main difference is represented by the presence of rule (8)\/,
whose aim is the reduction of equations between two set terms.
Reduction of this kind of equations is performed by the procedure
{\fp AbCl\_step} (see Figure~\ref{membership}) that implements the
two identities $(Ab)$ and $(C\ell)$\/. $(Ab)$ and $(C\ell)$ are
equivalent, for terms denoting sets, to the following
axiom~\cite{DPR98funda}:
$$\begin{array}{clll}
(E^s_k)  &
   \forall  Y_1 Y_2 V_1 V_2 &
  \multicolumn{2}{l}{ \left(\begin{array}{l}
   \{ Y_1\,|\,V_1 \}  =  \{ Y_2\,|\,V_2 \} \:\:
   \leftrightarrow \\
   \phantom{aaaaaaaa}
   (Y_1  =  Y_2 \wedge V_1  =  V_2) \vee\\
   \phantom{aaaaaaaa}
   (Y_1  =  Y_2 \wedge V_1  =  \{ Y_2\,|\,V_2 \}) \vee\\
   \phantom{aaaaaaaa}
   (Y_1  =  Y_2 \wedge \{ Y_1\,|\,V_1 \}  =  V_2) \vee\\
   \phantom{aaaaa}\exists K \:( V_1  =  \{Y_2\,|\,K\} \wedge
                 V_2  =  \{Y_1 \,|\, K \} )
   \end{array}\right) }
\end{array}$$

\noindent which can be easily converted into a rewriting rule to
be used in the unification algorithm.
 $(E^s_k)$ is in a sense a ``syntactic version'' of the
extensionality axiom, which allows the extensionality property to
be expressed in terms of only equations, without having to resort
to any membership, universal quantifiers, or inclusion operation.
$(E^s_k)$ allows also to account for equations of the form
$$\{t_0,\dots,t_m\,|\,X\}  =
\{t_0',\dots,t_n'\,|\,X\} \,,$$ where the two sides are set terms
with the same variable as tail element. Unfortunately, a blind
application of the rewriting rule obtained from $(E^s_k)$ would
lead to non-termination in this situation. This is the reason why
this case has been isolated and dealt with as special in the
algorithm (within the procedure {\fp AbCl\_step}), actually
splitting the rewriting rule obtained from $(E^s_k)$ into two
distinct rules.

A call to {\fp AbCl\_step} introduces equations in the stack $\fp
{\cal E}_{aux}$ that are immediately processed. This generates a
deterministic sequence of actions. We refer to the sequence of
actions performed until the stack becomes empty as the
\emph{global effect} of {\fp AbCl\_step}.

Membership equations, i.e., equations of the form $X  =
\{t_0,\dots,t_n \,|\,X\}$, with $X \not\in
\vars(t_0,\dots,t_n)$\/, are not dealt with by any rule of {\fp
AbCl\_unify\_actions}. This kind of equations turns out to be
satisfiable for any $X$ containing $t_0,\dots,t_n$ since
duplicates are immaterial in a set thanks to $(Ab)$ and
$(C\ell)$\/ (this justifies the name \emph{membership equations}).
These equations are processed at the end of {\fp AbCl\_unify} by
the procedure {\fp AbCl\_unify\_final}. Also, observe that the
occur-check test performed by the standard unification algorithm
is modified accordingly, so as to distinguish this special case
from others (rules $( 3)$ and $( 4)$).

Correctness and completeness of the algorithm presented in this
paper derive immediately from the similar algorithm
of~\cite{JLP1}. The termination proof for this algorithm,
however, turns out to be simpler than that in~\cite{JLP1}, since
here we rely on a more deterministic strategy, and we provide a
separate treatment of membership equations. Basically, in the
algorithm of this paper we avoid the repeated application of the
rewriting rule:
$$X = \{t_0,\dots,t_n \,|\,X\} \mapsto
  X = \{t_0,\dots,t_n \,|\,N\}$$
that increases the number of variables in the algorithms
in~\cite{JLP1,TOPLAS2000}. This change
 allows the number of variables in the system to be
kept under control. The simpler termination proof can be found
in~\ref{auxproofs}.

\begin{example} Let us consider the unification problem
$$\{X_1,X_2,X_3 \} = \{a,b,c\}$$
(i.e., $\{ X_1 \,|\, \{ X_2
\,|\, \{ X_3 \,|\, \e \} \} \}
  =
  \{ a \,|\, \{ b \,|\, \{ c \,|\, \e \} \} \}$).
The algorithm {\fp AbCl\_unify} returns the following six
independent solutions that constitute the minimal complete set of
$E$-unifiers for the given unification problem:

$$\begin{array}{ccc}
\begin{array}{l}
X_1 = a, X_2 = b, X_3 = c \\
X_1 = a, X_2 = c, X_3 = b \\
X_1 = b, X_2 = a, X_3 = c
\end{array} & \phantom{aaa}&
\begin{array}{l}
X_1 = c, X_2 = a, X_3 = b \\
X_1 = b, X_2 = c, X_3 = a \\
X_1 = c, X_2 = b, X_3 = a
\end{array}
\end{array}$$

\end{example}
In general, the algorithm {\fp AbCl\_unify} may open a
large---though finite---number of alternatives, possibly leading
to redundant solutions. \cite{AD97,Sto99} show how to improve the
algorithm to minimize the number of redundant unifiers.

\subsection{Discussion}\label{discussion2}

The problem of finding solutions we tackle here extends the
satisfiability problem for set unification (i.e., the SUD
problem), shown to be NP-complete (c.f. Sect \ref{nestedp}). To be
precise, we mean that there exists an algorithm on a
non-deterministic machine that can also find the answer (the
correct class is FNP). \cite{OP95} proposes a methodology to guess
a solution of a conjunction of literals built using variables, the
constant symbol $\e$, the function symbol $\w$ and the predicate
symbols $ =,\in,\cup,\cap$, and $\setminus$. The unification
problem is the particular case where only positive literals based
on the equality predicate $=$ are used. A {\em guess} is
represented by a graph containing a number of nodes polynomially
bounded by the number of variables in the original problem.
Verification of whether a guess is a solution of the problem can
be done in polynomial time. \cite{OP95} also shows how this
technique can be extended to the general problem with free
function symbols---the one we deal with in this paper. A
non-deterministic algorithm based on a ``guess-and-verify''
technique has also been proposed in~\cite{KN86}.

The algorithm presented here, as well as those
in~\cite{JLP1,AD97}, have the common drawback that, due to the
explicit application of substitutions during the solving process
they have a computational complexity which falls outside of the
FNP class. Nevertheless, it is possible to encode this algorithm
using well-known techniques---such as multi-equations or graphs
with structure sharing~\cite{MM82,patersonwegman}---that allow us
to maintain a polynomial time complexity along each
non-deterministic branch of the computation. For instance,
in~\cite{ADR99} a goal driven algorithm in FNP for
non-well-founded and hybrid sets has been presented. In that paper
it is also shown how to use the algorithm for well-founded sets,
to solve the problem dealt with in this section. A similar result
is presented in~\cite{voronkov}. A detailed discussion of such
kinds of enhancements, however, is outside the scope of this
paper.

As far as the size of the computed complete set of unifiers is
concerned, we can observe that the algorithm opens, for each level
of nesting, a number of alternatives equivalent to the number of
solutions returned by the global effect of {\fp AbCl\_step}. This
number is no greater than the size of the minimal complete set of
$(Ab)(C\ell)$-unifiers for the problem:
$$\{X_1,\dots,X_h\,|\,M\} =
  \{X_{h+1},\dots,X_n\,|\,N\}$$
This value has a rough upper bound equal to
$O(2^{n\lg n})\,$~\cite{AD97}.
 Since this process can be repeated once for each nesting, a rough
upper bound to the number of solutions is $O(2^{n^2\lg n})$.

\medskip

Various authors have considered simplified versions of the
$(Ab)(C\ell)$ problem obtained by imposing restrictions on the
form of the set terms. Most notable is the use of sets in the
context of relational and deductive
databases~\cite{relationlog,AG91,ldl,ng}. Typical restrictions
which have been considered are flat and completely specified set
terms, i.e., elements either of the {\sf gflat}$(q)$ or {\sf
flat}$(0)$ classes. Specialized algorithms have been provided for
some of these cases. In particular, various works have been
proposed to study the simpler case of matching and unification of
\emph{Bound Simple} set terms \cite{greco2}, i.e., elements of
{\sf flat}$(0)$\/. These restrictions are sufficient to make the
task of deciding unifiability between set terms very simple---as
also discussed in Section~\ref{SUD}.

Let us illustrate the results in the simple case of
matching~\cite{greco1}
 (the approach has been generalized to sequential
 unification in~\cite{greco2} and to parallel unification in~\cite{ng}).
In the case of matching, the two set terms $s$ and $t$
to be unified can be written as:
$$\begin{array}{rclcrcl}
s & \equiv & \{ c_1, \dots, c_{r}, X_1, \dots, X_{h}\} & \phantom{a} &
t & \equiv & \{ b_1, \dots, b_{k}, c_1, \dots, c_{r}\}
\end{array}$$
where, according to our notation (see Section~\ref{flat0dec}),
$b_i \in C_2$, $c_i \in C_3$\/, and $X_i \in V_1$ ($C_1 = \e$
otherwise the problem has no solutions). The two terms unify iff
$h \geq k$ (see Section~\ref{SUDflatq}). {F}rom~\cite{greco1} we
know that the number of solutions is $$\sum_{i=0}^{k} (-1)^i
\left(
\begin{array}{c}
                                k\\
                                i
                             \end{array}
                  \right) (k + r - i)^{h}$$
The set of substitutions representing the correct solutions of
the matching problem $s = t$ can be
obtained by:
\begin{itemize}
\item computing all the $h$-multisets of
        $\{b_1, \dots, b_{k}, c_1, \dots, c_{r}\}$
        that  contain all the elements of the set $\{b_1, \dots, b_{k}\}$
\item computing all the distinct permutations of each multiset.
\end{itemize}
An algorithm based on this approach is optimal, in the sense that
it computes exactly a complete and minimal set of unifiers, with a
complexity that is linear in the size of such set of unifiers.

\section{General $ACI1$ Unification}
\label{ACI_gen}\label{unificazione-general}

The unification problem considered in Section~\ref{aci1-costanti}
is capable of dealing with flat set terms containing an arbitrary
number of set variables. On the other hand, the unification
problem of Section~\ref{setlog} allows unification between
possibly nested set terms with at most one set variable per set
term. The goal of this section is to provide a solution to
unification problems which do not fall in any of the two above
categories, namely, unification problems in presence of set terms
which can be nested at any depth and which may contain an
arbitrary number of set variables.
 We will refer to this kind of problems as
\emph{general $ACI1$ unification problems}.

We propose a novel solution that combines the algorithms of
Sections~\ref{aci1-costanti} and~\ref{setlog} developed for
solving $ACI1$ unification with constants and general
$(Ab)(C\ell)$ unification. The result is a new goal-driven
algorithm for general $ACI1$ unification.

\subsection{Language and Semantics}
\label{lansem} We consider a language whose signature $\Sigma$
contains the constant $\e$\/, the binary function symbol $\cup$\/,
and  a (possibly infinite) collection of free function symbols
with arbitrary arities.

\begin{definition}
An \emph{$ACI1$ set term} is either a variable, or the constant
$\e$, or a $\Sigma$-term of the form $t \cup s$\/,
 where
$t$  and $s$ are $\Sigma$-terms. An
\emph{individual term} is either a variable or a $\Sigma$-term of
the form $f(t_1,\dots,t_n)$ with
$f \not\equiv \cup$ and
$f \not\equiv \e$ and
$t_1,\dots,t_n$ are $\Sigma$-terms
(if $n=0$ it is a constant term).
\end{definition}
The function symbols $\cup$ and $\e$ have the properties described
by the identities $(A)$\/, $(C)$\/, $(I)$ and $(1)$ introduced in
Section~\ref{notation}. Hence, set terms denote hereditarily
finite sets based on $\cal U$, while individual terms denote
arbitrary elements of the universe $\cal U$.

In the rest of the discussion we will assume the existence of at
least one function symbol $f \in \Sigma$ of arity greater than
zero---note that if such symbol does not exist, then we are in the
case discussed in Section~\ref{aci1-costanti}. Intuitively, terms
based on such symbol will be used to encode singleton sets.
Without loss of generality we assume to use the unary function
symbol $\{\cdot\}$ to represent singleton sets (more generally, if
the chosen symbol $f$ is of arity $n$, $n \geq 0$, we could assume
that the term $f(s,\e,\dots,\e)$ is used to denote the singleton
set containing the element $s$). In this way, it will be possible,
for instance, to distinguish the individual element $a$ from the
set containing $a$ (i.e.,  $\{a\}$). Moreover, as a notational
convenience, we will denote the term $\{s_1\} \cup \cdots \cup \{
s_n\}$ with  $\{s_1, \dots, s_n\}$\/.

\subsection{Which Kind of Set Unification}

The general $ACI1$ language allows us to describe the SUD and SUS
problems for any abstract set terms in {\sf set}$(m,n,p,q)$\/. In
particular, the cases {\sf flat}$(q)$ and {\sf nested}$(q)$ with
$q \geq 2$ are handled in this framework (and not in any of the
previous ones).

\begin{example}
The following are set terms and set unification problems which
are allowed in general  $ACI1$\/:

\begin{itemize}
\item $\{ \{ A , B\} \cup C \cup D \} \cup E \cup F
    = \{\{X,1\}\} \cup E \cup G$
\item $\{ \{g(a)\} \cup X  \} \cup Z = \{b\} \cup T \cup S$
\end{itemize}
\end{example}

\subsection{Unification Algorithm}

In this section, we propose a novel algorithm to directly solve the
general $ACI1$ unification problem.  The algorithm is
composed of a main procedure ({\fp general\_aci}) and a
rewriting function ({\fp aci\_{step}}), which deals with equations
between set terms (see Figure~\ref{nuovatavola}).

The structure of the main procedure is very similar to the
structure of standard unification algorithms for the Herbrand
case. The algorithm maintains two separate collections of
equations, ${\cal E}_{\fp s}$ and ${\cal E}_{\fp ns}$\/: the first
collects the equations in solved form while the second contains
the equations that require further processing. As in the case of
$(Ab)(C\ell)$ unification, the main changes with respect to
standard Herbrand unification are concerned with the two rules
dealing with set terms (i.e., terms containing occurrences of
$\cup$ at the outermost level):
\begin{itemize}
\item rule $( 5)$ which is aimed at dealing with equations of the form
        $X = \cdots \cup X$
     which are satisfiable in the case of $ACI1$ theory, whereas they
     were not satisfiable if the $\cup$ symbol would be uninterpreted;
\item rule $( 8)$ which is used to solve equations
         between two set terms.
\end{itemize}
We will use the notation $\bar s$ to denote the list of terms
$s_1, \dots, s_n$\/, and
 $\bar s = \bar t$ to denote
  $s_1 = t_1 ,\dots, s_n = t_n$\/.

{\fp aci\_{step}} receives as input the equation between set terms
to be solved and non-deterministically produces as result two
systems of equations (corresponding to the ${\cal E}_{\fp s}$ and
${\cal E}_{\fp ns}$ of the main unification procedure) and a
substitution.
{\fp aci\_{step}} performs its task in four successive steps, as
shown in Figure~\ref{nuovatavola}. \emph{Term Propagation} is the
only (don't know) non-deterministic step of the whole algorithm.
Both
 \emph{Term Propagation} and  \emph{Variables Removal} can lead to a
failure for some of the non-deterministic choices performed within
 \emph{Term Propagation}. Let us analyze these steps in more
detail.

\begin{description}
\item[Normalization:] \hspace{1cm}\\
\noindent
\emph{input:} A system consisting of the single equation
\begin{equation}
f_1(\bar l_1) \cup \cdots \cup
f_{k_1}(\bar l_{k_1}) \cup L_1 \cup \cdots \cup L_{k_2} =
g_1(\bar r_1) \cup \cdots \cup g_{h_1}(\bar r_{h_1}) \cup
R_1 \cup \cdots \cup R_{h_2}
\end{equation}
\noindent where $L_i$\/, $R_j$ ($0 \leq i \leq k_2$, $0 \leq j \leq h_2$)
are variables and  $f_i$\/, $g_j$ ($0 \leq i \leq k_1$, $0 \leq j \leq h_1$)
are function symbols different from $\cup$\/.\\
\noindent
\emph{output:} A system
        $$\begin{array}{rcll}
        \vspace{.2cm}{\cal E}^n  & = &
         & N^L_1 = f_1(\bar l_1) \wedge \cdots \wedge
         N^L_{k_1} = f_{k_1}(\bar l_{k_1}) \;\wedge\\
        \vspace{.2cm} &&& N^R_1 = g_1(\bar r_1)\wedge \cdots \wedge
        N^R_{h_1} = g_{h_1}(\bar
             r_{h_1})\;\wedge  \\
             &&&  N^L_1 \cup \cdots \cup N^L_{k_1} \cup
                 L_1 \cup \cdots \cup L_{k_2} =
                 N^R_1 \cup \cdots \cup N^R_{h_1}
                 \cup R_1 \cup \cdots \cup R_{h_2}
        \end{array}$$
        where $N^L_i$ and $N^R_j$ are new distinct variables.

This step, following the idea used in~\cite{adventure,BS96},
performs a normalization of the problem $\cal E$ into the problem
${\cal E}^n$\/---producing an equation between set terms that
contains only variables.

\item[Elementary $ACI1$ Solution:] \hspace{1cm}\\
\emph{input:} The system ${\cal E}^n$ produced by the
\emph{Normalization} step;\\
\emph{output:} A pair of systems ${\cal E}^{ACI}_1$ and ${\cal
  E}^{ACI}_2$
obtained by solving the elementary $ACI1$ unification problem
\begin{equation}
\label{simplereq}
N^L_1 \cup \cdots \cup N^L_{k_1} \cup L_1 \cup \cdots \cup L_{k_2} =
N^R_1 \cup \cdots \cup N^R_{h_1} \cup R_1 \cup \cdots \cup R_{h_2}
\end{equation}
of ${\cal E}^n$\/. This problem can be directly solved by using
the techniques seen in Section~\ref{aci1-costanti} (see also
Example~\ref{var_con_shar}). The result of the computation is a
collection of equations of the form $V = A_{a_1,b_1} \cup
A_{a_2,b_2} \cup \dots$ where $V$ is a variable occurring in the
two terms to be unified and $A_{a_1,b_1}, A_{a_2,b_2}, \dots$ are
new variables generated by the unification algorithm. The solved
form equations associated to $L_j$ and $R_i$ form the set ${\cal
E}^{ACI}_1$. ${\cal E}^{ACI}_2$ is composed of  the equations
concerning the variables  $N^L_j$ and $N^R_i$. These variables are
immediately  replaced by the terms they have been set equal to
during the  \emph{Normalization} step.

\item[Term Propagation:] \hspace{1cm}\\
\emph{input:} The pair of systems ${\cal E}^{ACI}_1$ and
${\cal E}^{ACI}_2$ produced in the previous step;\\
\emph{output:} A pair of systems ${\cal E}_1$ and ${\cal E}_2$\/.

The equations in ${\cal E}^{ACI}_2$ can be simplified using the
semantic properties of $\e$ and $\cup$\/. As a matter of fact,
the equations in  ${\cal E}^{ACI}_2$\/ can be immediately
satisfied by binding each $A_{i,j}$ appearing in the right-hand
side of an equation either to $\e$ or to a term which unifies
with the left-hand side of the equation. Observe, however, that
each $A_{i,j}$ can occur in the right-hand side of more than one
equation; thus, it should receive a consistent binding in order
to satisfy  ${\cal E}^{ACI}_2$\/.

More precisely, a substitution $\lambda$ describing the solution
of the equations in ${\cal E}^{ACI}_2$ can be build as follows.
Let us assume that an ordering has been fixed on the equations in
${\cal E}^{ACI}_2$ and on the variables $A_{i,j}$\/. Thus, for
each $A_{i,j}$ occurring in ${\cal E}^{ACI}_2$ we can identify an
equation $e_{A_{i,j}}$ which contains the ``first'' occurrence of
such variable in its right-hand side. If $f(\bar s)$ is the
left-hand side of such equation, then $\lambda(A_{i,j})$ is
\emph{non-deterministically} defined to be either
\begin{itemize}
\item $\lambda(A_{i,j}) = \e$ or
\item $\lambda(A_{i,j}) = f(\bar s)$\/.
\end{itemize}
As soon as the value of $\lambda(A_{i,j})$ has been determined,
the substitution is immediately applied to ${\cal E}^{ACI}_2$\/.
Once all the $A_{i,j}$ occurring in ${\cal E}^{ACI}_2$ have been
processed, the system is reduced to a collection of equations of
the form:
$$f(\bar s) = f_1 (\bar s_1) \cup \cdots \cup f_h(\bar s_h)$$
with $h \geq 1$ (without loss of generality, we may assume that
all the occurrences of $\e$ in the union have been removed, as
well as repetitions of the same term). The above result also
relies on the assumption that at least one $A_{i,j}$ per equation
is assigned a term different from $\e$\/.

If some of the $f_i$ is different from $f$ for some equation, then
another guess for $\lambda$ must be chosen; if no choice leading
to the satisfaction of this condition can be made, then the system
does not admit solutions. Otherwise, let the output system ${\cal
E}_2$ consist of all equations of the form:
$$\bar s = \bar s_1 \wedge \dots \wedge \bar s = \bar s_h$$
for each equation in ${\cal E}^{ACI}_2$\/.
\label{termpropagation}

The other output system, ${\cal E}_1$\/, is obtained by applying
$\lambda$ to the input system ${\cal E}^{ACI}_1$\/, with the usual
elimination of $\e$ and repetitions in the unions. Thus,

        \[ \renewcommand{\arraystretch}{1.8}
        \begin{array}{rcl}
         {\cal E}_1 & = &
         \bigwedge_{1 \leq j \leq k_2 }
         L_j = \bigcup_{i=1}^{h_1} \lambda(A_{i,k_1+j}) \cup
                \bigcup_{i=h_1+1}^{h_1+h_2} A_{i,k_1+j}
                \;\wedge \\
&&   \bigwedge_{1 \leq i \leq h_2  }
 R_i = \bigcup_{j=1}^{k_1} \lambda(A_{h_1+i,j}) \cup
                \bigcup_{j=k_1+1}^{k_1+k_2} A_{h_1+i,j} \end{array}
        \]

\item[Variables Removal:] \hspace{1cm}\\
\emph{input:} The pair ${\cal E}_1$ and ${\cal E}_2$ computed in
the previous step;\\
 \emph{output:} The substitution $\rho$\/.

{From} ${\cal E}_1$ we can directly produce a substitution which
allows all  variables $L_j$ and $R_i$ to be removed. More
precisely, this is obtained as follows. Let $\rho_{L_j}$ and
$\rho_{R_i}$ denote the substitutions that respectively replace
$L_j$ ($1 \leq j \leq k_1$) and $R_i$ ($1 \leq i \leq h_1$). In
order to guarantee that ${\cal E}_1 \cup {\cal E}_2$  admits
solutions we need to make sure that no cyclic conditions occur.

Let us define the relation $\Rightarrow$ as follows:
$$\begin{array}{rcl}
X \Rightarrow Y& \mbox{ iff } & Y \in \vars(X \rho_X)
\end{array}$$
and let us denote with $\Rightarrow^*$ the transitive
closure of $\Rightarrow$.

A necessary condition for the solvability
of the  set of equations ${\cal E}_1$ is that $$(\forall X \in
\{L_1,\dots,L_{k_1},R_1,\dots,R_{h_1}\})
                        (X \not\Rightarrow^* X).$$

If this test is satisfied, then we can construct a
global substitution
        $$\rho = \rho_{L_1} \circ \cdots \circ \rho_{L_{k_1}} \circ
\rho_{R_1} \circ \cdots \circ \rho_{R_{h_1}}$$ which allows all
variables $\{L_1,\dots,L_{k_1},R_1,\dots,R_{h_1}\}$ to be removed.
\end{description}
A detailed description of the algorithms for the
\emph{Elementary $ACI1$ Solution} step and the \emph{Term
Propagation} step is reported in~\ref{matrix}.

\begin{figure}[htp]
\small
$$\begin{array}{crcl}
\multicolumn{4}{l}{\mbox{\fp general\_aci}({\cal E}):}\\
\multicolumn{4}{l}{\phantom{aaa}{\cal E}_{\fp s} := \e;
       {\cal E}_{\fp ns} := {\cal E}\,\,\,\mbox{\fp (i.e., the
       initial system of equations)}\/;}\\
\multicolumn{4}{l}{\phantom{aaa}
      \mbox{\fp while ${\cal E}_{\fp ns} \neq \e$ do}}\\
\multicolumn{4}{l}{\phantom{aaaaaa}
      \mbox{\fp select arbitrarily an equation $e$ from ${\cal E}_{\fp ns}$
      and remove it;}}\\
\multicolumn{4}{l}{\phantom{aaaaaa}\mbox{\fp case $e$ of}}\\
\hline
(1) & \begin{array}{r}
 X=X
\end{array} & \mapsto &  {\cal E}_{\fp ns} := {\cal E}_{\fp ns}\\
\hline
(2) & \left.\begin{array}{r}
t = X \\
\mbox{$t$ is not a variable}
\end{array} \right\}
& \mapsto & {\cal E}_{\fp ns} := {\cal E}_{\fp ns} \wedge ( X = t ) \\
\hline
(3) & \left.\begin{array}{r}
X = t\\
\mbox{$t$ can be re-ordered as}\\
f_1(\bar s_1) \cup \cdots \cup f_n(\bar s_n) \cup
    V_1 \cup \cdots \cup V_m  \\
\mbox{$n \geq 0$\/, $f_i  \not\equiv \cup$\/, $m \geq 0$\/, }
\mbox{and $X \in \vars(\bar s_1,\dots,\bar s_n)$}
\end{array} \right\}
& \mapsto & {\fp fail} \\
\hline
(4) & \left.\begin{array}{r}
X = t \\
\mbox{$X$ does not occur in $t$}
\end{array} \right\}
& \mapsto & \\
& \multicolumn{3}{r}{
{\cal E}_{\fp s} := {\cal E}_{\fp s}[X/t] \wedge ( X
= t );
          {\cal E}_{\fp ns} := {\cal E}_{\fp ns}[X/t]}\\
\hline
(5) &  \left.\begin{array}{r}
X = t  \\
\mbox{$t$ can be re-ordered as
$t' \cup X \cup \cdots \cup X$\/,}\\
\mbox{$t'= f_1(\bar s_1) \cup \cdots \cup f_n(\bar s_n) \cup
    V_1 \cup \cdots \cup V_m$\/,}\\
\mbox{$f_i  \not\equiv \cup$\/, $m \geq 0$\/,}\\
\mbox{$X \not\in \vars(t')$}
\end{array}\right\}
& \mapsto & \begin{array}{l}
   {\cal E}_{\fp ns} := {\cal E}_{\fp ns} \wedge ( X = t' \cup N )\\
   \mbox{$N$ new variable}
                \end{array}\\
\hline
(6) & \left.\begin{array}{r}
f(s_1,\dots,s_m) = g(t_1 ,\dots ,t_n) \\
f \not\equiv g
\end{array} \right\}
& \mapsto &  {\fp fail}\\
\hline
(7) & \left.\begin{array}{r}
f(s_1,\dots ,s_n) = f(t_1,\dots,t_n)\\
f \not\equiv \cup
\end{array}\right\}
& \mapsto & \\
& \multicolumn{3}{r}{
{\cal E}_{\fp ns} := {\cal E}_{\fp ns} \wedge
( s_1 = t_1 \wedge \dots \wedge s_n = t_n )}\\
\hline
(8) &  s_1 \cup s_2 = t_1 \cup t_2
& \mapsto & \\
& \multicolumn{3}{r}{
\begin{array}{l}
\mbox{Let $\langle {\cal E}'_1,{\cal E}'_2,\theta \rangle$ be a result of}\\
{\fp aci\_{step}}(s_1 \cup s_2 = t_1 \cup t_2);\\
{\cal E}_{\fp s} := {\cal E}_{\fp s}\theta \wedge {\cal E}'_1;
{\cal E}_{\fp ns} := {\cal E}_{\fp ns}\theta \wedge {\cal E}'_2
\end{array}}\\
\hline\hline
\end{array}$$
{\fp
\begin{tabular}{l}
{\fp aci\_{step}}$( e )$\/:\\
\phantom{aaa}${\cal E}^n :=$ Normalization($e$);\\
\phantom{aaa}$\langle {\cal E}_1^{ACI}, {\cal E}_2^{ACI} \rangle :=$
      Elementary\_$ACI1$\_Solution(${\cal E}^n$) ;\\
\phantom{aaa}$\langle {\cal E}_1, {\cal E}_2 \rangle :=$
      Term\_Propagation(${\cal E}_1^{ACI}, {\cal E}_2^{ACI}$) ;\\
\phantom{aaa}$\rho :=$
      Variables\_Removal(${\cal E}_1, {\cal E}_2 $) ;
\\
\phantom{aaa}return $\langle {\cal E}_1,
         {\cal E}_2\rho,
         \rho \rangle$\\
\hline\hline
\end{tabular}
}
\caption{General $ACI1$ Unification Procedure and the function {\fp aci\_step}}
\label{generalaciproc}
\label{nuovatavola}
\end{figure}

\newpage
\begin{example}
Let us consider the unification problem:
$$\{\{a\}\} \cup \{b\} \cup X = \{\{W\}\} \cup Y \cup Z$$
The  \emph{Normalization} step leads to the system
$${\cal E}^n   \equiv  N^L_1 = \{\{a\}\} \wedge N^L_2 = \{b\} \wedge
         N^R_1 = \{\{W\}\}  \wedge
              N^L_1 \cup N^L_2 \cup X  =
                 N^R_1 \cup Y \cup Z  $$
The equation $N^L_1 \cup N^L_2 \cup X  =
                 N^R_1 \cup Y \cup Z$ can be solved and its solution
applied to the rest of the system (\emph{Elementary $ACI1$
Solution} step), leading to:
$$
\begin{array}{cc}
\begin{array}{rcl}
{\cal E}^{ACI}_1  &  \equiv &   X = A_{1,3} \cup A_{2,3}
       \cup A_{3,3}  \;\wedge \\
        &&  Y = A_{2,1} \cup A_{2,2} \cup A_{2,3}  \;\wedge \\
          &&  Z = A_{3,1} \cup A_{3,2} \cup A_{3,3}
\end{array} &
\begin{array}{rcl}
{\cal E}^{ACI}_2  &  \equiv &   \{\{a\}\} = A_{1,1} \cup
         A_{2,1} \cup A_{3,1}  \;\wedge \\
         && \{b\} = A_{1,2} \cup A_{2,2} \cup A_{3,2}  \;\wedge \\
         && \{\{W\}\} = A_{1,1} \cup A_{1,2} \cup A_{1,3}  \\
\end{array}
\end{array}
$$
A possible substitution $\lambda$ produced by the  \emph{Term
Propagation} step is the following:
\begin{center}
\begin{tabular}{|ccccccc|}
$A_{1,1}$ & $A_{2,1}$ & $A_{3,1}$ & $A_{1,2}$ & $A_{2,2}$ & $A_{3,2}$ &
$A_{1,3}$\\
\cline{1-7}
\multicolumn{1}{|c}{$\{\{a\}\}$} &
\multicolumn{1}{|c}{$\{\{a\}\}$} &
\multicolumn{1}{|c}{$\e$} &
\multicolumn{1}{|c}{$\e$} &
\multicolumn{1}{|c}{$\{b\}$} &
\multicolumn{1}{|c}{$\e$} &
\multicolumn{1}{|c|}{$\e$}
\end{tabular}
\end{center}
This produces the systems
$$
\begin{array}{lcr}
{\cal E}_2  \equiv   \{W\} = \{a\} & &
{\cal E}_1  \equiv  X = A_{2,3} \cup A_{3,3}  \wedge \, Y = \{\{a\}\}
\cup \{b\}
\cup A_{2,3} \wedge  Z = A_{3,3}
\end{array}
$$
and the substitution
$\rho = [X/A_{2,3} \cup A_{3,3}, \,\, Y/\{\{a\}\} \cup \{b\} \cup
A_{2,3}, \,\, Z/A_{3,3} ].$
{F}rom ${\cal E}_2$ it is then computed $[W/a]$.
\end{example}

\subsection{Results for the General $ACI1$ Unification Algorithm}

\subsubsection{Soundness and Completeness}
The soundness and completeness results can be derived as follows.
\begin{lemma} \label{corr_aci}
Given an equation $e$ of the form
\[f_1(\bar l_1) \cup \cdots \cup
f_{k_1}(\bar l_{k_1}) \cup L_1 \cup \cdots \cup L_{k_2} =
g_1(\bar r_1) \cup \cdots \cup g_{h_1}(\bar r_{h_1}) \cup
R_1 \cup \cdots \cup R_{h_2}\]
let  $\langle {\cal E}^i_1, {\cal E}^i_2, \rho^i \rangle$\/, for  $i =
1,\dots,k$\/,
be the collection of all the distinct solutions non-deterministically
produced by the call ${\fp aci\_{step}}( e )$\/. Then:
\begin{itemize}
\item if $\sigma$ is a unifier of ${\cal E}^i_1 \cup  {\cal E}^i_2$
then $\sigma$ is a unifier of $e$ and $\sigma \leq \rho^i$\/
\item if $\sigma$ is a unifier of $e$ then there exists $1 \leq i \leq
k$ and a substitution $\gamma$ such that $\sigma \cup
  \gamma$ is a unifier of ${\cal E}^i_1 \cup  {\cal E}^i_2$\/.
\end{itemize}
\end{lemma}
For the proof, see  \ref{auxproofs}.

\begin{theorem}\label{correctnessandcompl}
The unification procedure {\fp general\_aci} is correct and complete
with respect to the general $ACI1$ theory\/.
\end{theorem}
\begin{proof}
Immediate from the  above Lemma~\ref{corr_aci} concerning the
auxiliary function ${\fp aci\_{step}}$, and from the classical
results regarding Herbrand unification  for the remaining rules.
\end{proof}

\subsubsection{Termination of {\fp general\_aci}}

The development of a termination proof for general unification
algorithms for theories obtained using some or all of the $(A)$,
$(C)$, $(I)$, and $(1)$ axioms is a well-known challenging
task~\cite{BS96}. In~\cite{Fages} Fages proposed a termination
proof for general $AC$ unification. The complexity measure
developed by Fages to prove termination, however, is not
applicable to our problem---mainly due to the need, in our
algorithm, to introduce new variables to handle cases such as $X =
Y \cup X$\/, that are unsatisfiable in $AC$ but admit solutions in
$ACI1$.

The detailed termination proof (Theorem~\ref{aciterm}) is reported
in~\ref{auxproofs}. We give here the main ideas behind
that proof. First of all, ${\fp aci\_{step}}$ replaces an
 equation between two sets with equations between members of the
sets, thus with equations of a ``lower level''. The process cannot
enter in a loop thanks to the occur-check test which avoids the
possibility of generating infinitely-nested sets. To formalize
this idea we define the notion of \emph{$p$-level}
(Def.~\ref{pilevel}). Terms can be naturally viewed as trees. We
use two kinds of edges in these trees, edges connecting
$\cup$-nodes to their children and edges linking all other types
of nodes. We show how the unification algorithm operates on this
tree representation of terms, and we determine some properties
related to cycles involving edges of the second type (+1-edges).
Finally, we define a complexity measure built from the notion of
$p$-level of the terms occurring in the system of equations. We
show that this measure is well-ordered and that any given sequence
of applications of rules either decreases it, or an occur-check
failure is detected.

\subsection{Discussion}\label{discussion3}

A non-deterministic algorithm for general $ACI$ is presented by
Kapur and Narendran in~\cite{KN92} that can be adapted to general
$ACI1$.
Another algorithm for general $ACI1$ unification can be obtained
as an instance of the general technique of Baader and
Schulz~\cite{BS96} for combining unification algorithms. Combining
unification procedures for different unification problems has been
a major topic of investigation for years~\cite{siek84}. Various
proposals have been put forward to allow combination of
unification procedures under different conditions on the
equational
theories~\cite{yelick85,herold86,tiden86,kirchner89,ss87}.
In~\cite{BS96} Baader and Schulz proposed a general technique for
combining unification procedures over disjoint theories under very
simple restrictions---i.e., \emph{constants restriction}. In the
context of general $ACI1$ unification, we need to combine two
theories: the theory $ACI1$ for $\e$ and $\cup$\/, and the empty
equational theory for all the other function symbols. The
technique proposed by Baader and Schulz can thus be used to
integrate the unification procedure for $ACI1$ with constants and
a standard Herbrand unification algorithm to obtain a unification
procedure for general $ACI1$.

Let us briefly compare these two proposals with the unification
algorithm for general $ACI1$ presented in this paper. All three
unification procedures start with a  \emph{Normalization} step
(implicit in~\cite{KN92}). New variables are introduced for
subterms.
As an example, the problem
\begin{eqnarray}
\{X\}\cup\{Y\} &=& \{a\}\cup\{b\}  \label{formula.esempio}
\end{eqnarray}
is rewritten as
\begin{eqnarray}\label{formula.esempio.piatta}
N_1^L \cup N_2^L = N_1^R \cup N_2^R, N_1^L  = \{X\}, N_2^L =
\{Y\}, N_1^R = \{a\}, N_2^R = \{b\}
\end{eqnarray}
All three procedures introduce \emph{don't know} non-determinism.
In particular, \cite{BS96} introduces non-determinism in steps~3
and~4, where
\begin{itemize}
\item step~3 computes an arbitrary partition of
     the variables in independent sets
    (all the variables in the same component of the partition will
    be aliased to each other in the final solution);
\item step~4 imposes an arbitrary order over the elements of the previously
    computed partition.
\end{itemize}
In the formula \ref{formula.esempio.piatta}, for instance, there
are 6 variables. Therefore, there are $\sum_{i=1}^6 \stir{6}{i} =
203$ possible partitions of the set of
variables\footnote{$\stir{n}{i}$ is the number of partitions of
$n$ elements into $i$ classes, known as \emph{Stirling number of
the second type}~\cite{Knuth}.}, and $6!=720$ possible strict
orderings among the 6 variables. Actually, the
problem~(\ref{formula.esempio}) has only two independent solutions
$X=a,Y=b$ and $X=b,Y=a$ that suggests the need of only 2
non-deterministic choices. The high number of choices
in~\cite{BS96} derives from the generality of the combination
procedure (which is not specifically tied to the problem of set
unification). On the other hand, it is unclear whether the
instantiation of that framework to the problem at hand would
actually reduce the number of alternatives compared to the
algorithm we propose in this paper.

The unification procedure presented in~\cite{KN92} is rather
different. It performs a series of non-deterministic guesses for
the variables in order to find ground substitutions. It has two
main practical drawbacks. The first is that the number of choices
does not depend on the structure of the problem but rather on the
signature. The second drawback is that the algorithm always
returns ground substitutions. The number of ground substitutions
of a general $ACI$ problem can be infinite. Let us consider, for
instance, the problem
\begin{eqnarray}
\{ \e \}\cup Y =  Y \label{esempioinfinito}
\end{eqnarray}
$Y = \{\e\}$, $Y = \{\e\} \cup \{\{\e\}\}$, $Y = \{\e\} \cup
\{\{\e\}\} \cup \{\{\{\e\}\}\},\dots\:\:$  are all the ground
solutions for (\ref{esempioinfinito}). However, a unique most
general unifiers, $Y = \{\e\} \cup N$ is sufficient to finitely
describe all solutions (this is exactly what our algorithm
returns). Even for problems where only ground unifiers are
present, our algorithm has the advantage of using the symbols in
the problem to drive the construction of the solution, instead of
performing a blind enumeration based on the language signature.

As concerns the non-determinism introduced by our algorithm, first
observe that the  \emph{Normalization} step allows us to call the
elementary $ACI1$ unification  step with terms containing only
variables. In this case it is known that the unification problem
admits a unique mgu. So, we are not exploiting the possibility of
the $ACI1$ with constants unification algorithm to return
non-deterministically all the mgus and we perform that choice
later. The rationale behind this is that the non-variable terms
$s_1,\dots,s_\ell, t_1,\dots,t_r$ in an equation $X_1 \cup \cdots
\cup X_m \cup s_1\cup \cdots \cup s_\ell = Y_1 \cup  \cdots \cup
Y_n \cup t_1\cup \cdots \cup t_r$ can be compound terms. We do not
know (yet) if some of them can be unified, and thus we cannot
consider them as equal or different constants when calling the
$ACI1$ with constants algorithm. Possible optimizations of our
algorithm include the use of $ACI1$ with constants in those cases
where a simple preprocessing allows us to quickly determine what
individuals in the equations are equal or distinct. If $V_1, V_2,
V_3$ are the set of variables in the elementary $ACI1$ unification
problem as defined in Section~\ref{flat0dec}, then the Boolean
$ACI$ matrix \cite{BB88} is of size $(|V_1|+|V_2|+|V_3|)
(|V_1||V_2|+ |V_1||V_3|+
 |V_2||V_3| + |V_3|)$
and the new variables introduced are $|V_1||V_2|+ |V_1||V_3|+
 |V_2||V_3| + |V_3|$.
Our \emph{elementary $ACI$ matrix} (see~\ref{matrix})
introduces the same number of variables, but its size is
$(|V_1|+|V_3|)( |V_2||V_3|)$. For instance, if
$|V_1|=|V_2|=|V_3|=v$ we need space $4v^2$ against space
$9v^3+3v^2$.

All non-deterministic choices are performed in the \emph{Term
Propagation} step. If $k$ is the number of variables introduced by
the matrix, this would potentially open $2^k$ non-deterministic
choices. However, using the auxiliary Boolean matrix
(see~\ref{matrix}) we do not try all these choices, since for each
column and each row of the matrix for \emph{Term Propagation}
there must be at least one variable which is different from $\e$.
This decreases the number of choices. In the case of the system of
equations (\ref{formula.esempio.piatta}) we have only 8
non-deterministic choices instead of the $2^4$ expected (and the
$203 \times 720$ of the naive application of the Baader-Schulz
procedure).

As far as the difference in non-determinism between the general
$ACI1$ and the general $(Ab)(C\ell)$ unification is concerned, we
can observe that the $ACI1$ algorithm opens, for each level of
nesting, a number of alternatives equivalent to the resolution of
an $ACI1$ with constants problem; this leads to $O(2^{n^2})$
solutions (see Section~\ref{baader-algo}). Since this process can
be repeated once for each nesting, a rough upper bound to the
number of solutions is $O(2^{n^3})$\/. Observe that this number
of solutions is greater than those computed by the $(Ab)(C\ell)$,
namely $O(2^{n^2\log n})$. This fact suggests that the general
$ACI1$ unification should be used only when the problem is really
not expressible using the general $(Ab)(C\ell)$ unification and
the full range of solutions is required.

\section{Related Work}\label{related}

Most of the related proposals have already been discussed
throughout the paper. In
this section we provide a  brief overview of other related
contributions.

\begin{description}

\item[Boolean unification.]
Boolean unification is a very powerful framework that allows one,
in particular, to mimic the $ACI1$ with constants unification
problems. The richer language of Boolean unification, however,
allows the various solutions of a given $ACI1$
problem to be encoded in a very compact
way, as a single complex solution---instead of using multiple
$ACI$-matrices as in Section~\ref{aci1-costanti}.
A fundamental work in this area
is~\cite{MN89}, which surveys both the \emph{Boole's method} and
the \emph{L\"owenheim's method}. The former has been originally
described in~\cite{BS87} while the second  has been initially
described in~\cite{MN88}. All these approaches deal with Boolean
unification with constants, where the signature $\Sigma$ contains
a possibly infinite collection of constants, which intuitively
represent the \emph{elements} of the universe $\cal U$\/. The
class of terms allowed in this framework extends the one
considered in this paper by allowing a variety of different
operators to be used in the construction of sets, such as
intersection $\cap$ and complementation $\bar{(\cdot)}$.

The complexity of the decision problem of Boolean unification has
been studied in~\cite{Baa98}. In the elementary case, i.e.,
without constants, the problem is NP-complete, while in the case
with constants the problem becomes PSPACE-complete. However, if
the input is of the form admitted by $ACI1$ unification, the test
between two ground terms can be performed in linear time. The
computation of the unifier for a given Boolean unification
problem $s = t$ is based on the fact that $\mu$ is a unifier of
$s = t$ if and only if $\mu$ is a unifier of $s \dif t = \e$\/,
where $\dif$ is a function symbol which is interpreted as the
symmetric difference. Thus, to solve a unification problem, it is
sufficient to solve a \emph{matching problem}. The work
in~\cite{BS87} shows that a \emph{unique} most general unifier is
sufficient to cover all the solutions. The generality of this
scheme and the power of this unification procedure are balanced
by the complexity of the answers produced---sets built using
$\dif$ are arguably more complex and less intuitive than those
constructed using~$\cup$\/.

\item[Computable Set Theory.]
The work on Computable Set Theory \cite{CST2} has been
mainly developed at the New
York University, with the objective of  enhancing the expressive
power of inference engines for automated theorem provers,
and for the implementation of the imperative set-based programming
language
SETL~\cite{SDDS86}. The general problem is to  identify computable
classes of formulae of suitable sub-theories of Zermelo-Fraenkel
set theory. In this context, the set unification problem is seen
as a special case of the satisfiability problem for the
$\exists^*\forall$-class of formulae. As a matter of fact, thanks
to the extensionality axiom, testing whether two terms $s$ and $t$
with variables $X_1,\dots,X_n$ are unifiable is equivalent to
testing whether the following holds:
$$\mathbb{HF} \models
\exists X_1 \cdots \exists X_n\,\forall Z ( Z \in s
\leftrightarrow Z \in t).$$ Unification algorithms can be obtained
by instantiating the general (and complex) techniques for testing
satisfiability of $\exists^*\forall$-formulas~\cite{DFO04}.

\item[Set constraints.]
Set constraints \cite{kozen,aiken} are conjunctions of literals
of the form $e_1 \subseteq e_2$ where $e_1$ and $e_2$ are
\emph{set expressions}, constructed using variables, constant and
function symbols, and the union, intersection, and complement of
set expressions. Set expressions denote sets of Herbrand terms.
An expression identifies a subset of the Herbrand universe. A
unification problem of the type $s=t$ can be expressed in this
framework as the constraint $s \subseteq t \wedge t \subseteq
s$\/.

The framework is sufficiently powerful to solve $ACI1$ unification
problems with constants; nevertheless, the peculiar interpretation
given to terms in the language is such to prevent to encode large
classes of set unification problems. In particular, to represent
nested sets in set constraints we need to make use of a
distinguished functional symbol $\{\cdot\}$ (as described also in
Section~\ref{lansem}); on the other hand, using the set constraint
interpretation of expressions, the two expressions $\{\{s,t\}\}$
and $\{\{s\},\{t\}\}$ would be mapped to the same set.

\item[Alternative representations of sets.]
Other syntactic representations of sets  are also feasible. For
instance a set of $n$ elements can be represented by
$\{\}_n(t_0,\dots,t_n)$\/, where $\{\}_n$ is a function symbol of
arity $n$\/. This solution requires the introduction of an
infinite signature, with a different set constructor for each
possible finite set cardinality. This approach has been adopted,
for example, in~\cite{STZ92}. In order to use this solution it is
necessary
 to introduce a complex
infinite equational theory, capable of specifying the
unifiability of set terms with different main functors---as in
the case
$\{\}_3(X,Y,Z) = \{\}_2(a,b)$.

This representation scheme allows one to express only set terms
with a known upper bound on their cardinality. Namely,
$|\,\{\}_n(t_1,\dots,t_n)\,| \leq n\;.$

\end{description}

\section{Conclusions}\label{conclusioni}

In this paper we have presented a survey of the problem of solving
unification in the context of algebras for sets. We have
abstractly defined the set unification problem and developed the
corresponding equational theories, starting from the simpler case
of $ACI1$ with constants and proceeding to the most comprehensive
case of general $ACI1$ unification. We have presented decision and
unification procedures for the different classes of unification
problems and analyzed their complexity. Complexity results, as
well as the suitable equational theory for a given set unification
problem, are summarized in Table~\ref{sum1}. The algorithms
presented are either drawn from the literature or are brand new
algorithms developed by the authors.

We believe this work fills a gap in the literature on this topic,
by providing a uniform and complete presentation of this problem,
and by presenting a comparative study of the different solutions
proposed.

\subsection*{Acknowledgments}

We thank the anonymous referees that helped us to improve the
quality of presentation of the paper. The research presented in
this paper has benefited from discussions with A.~Formisano,
E.~G.~Omodeo, C.~Piazza, A.~Policriti, and D.~Ranjan, all
of whom we would like to thank.

\bibliographystyle{acmtrans}

\appendix

\section{Proofs}\label{auxproofs}

\subsection{Termination of {\fp AbCl\_unify}}\label{terminasemplice}

To prove the following theorem, we will use the notions of solved
variable and solved equation. Given a system $\mathcal{E}$ an
equation in $\mathcal{E}$ is \emph{solved} if it is of the form $X
= t$ and $X$ does not occur neither in $t$ nor elsewhere in
$\mathcal{E}$. If $X$ is the r.h.s.\ of a solved equation then it
is a solved variable. Moreover, $size$ is the function returning
the number of occurrences of constant and functional symbols in a
term ($size(X)=0,size(f(t_1,\dots,t_n))=1+\sum_{i=1}^n
size(t_i)$).

\begin{theorem}[{\fp AbCl\_unify} termination]\label{termunif}
For any Herbrand system ${\cal E}$\/, and for any possible
sequence of non-deterministic choices, ${\fp AbCl\_unify}({\cal E}$)
terminates.
\end{theorem}
\begin{proof}
To start, do not consider the final call to $\fp AbCl\_unify\_final$.
We associate the complexity pair
$\langle A,B \rangle$
to a system  $\cal E$, where:
\begin{itemize}
\item $A$ is the number of non-solved variables in $\cal E$\/
\item let $p = \max \{ size(\ell) : \ell = r \mathrel{in} {\cal E} \}$.
For $i=0,\dots,p$, let $\eta(i)$ be the number of non-solved
equations $\ell = r$ in $\cal E$ s.t. $size(\ell) = i$. Then $B$
is the list: $[\eta(p),\eta(p-1),\dots,\eta(0)]$. We define the
ordering among lists as follows:
$$\begin{array}{rcl}
  x <_{list} y & \mbox{iff} & ({\it length}(x) < {\it length}(y))
\mathrel{or}\\
&&  ({\it length}(x) = {\it length}(y) \mathrel{and}
    {\it head}(x) < {\it head}(y)) \mathrel{or} \\
&&  ({\it length}(x) = {\it length}(y),
    {\it head}(x) = {\it head}(y), \mathrel{and}
    {\it tail}(x) <_{list} {\it tail}(y))
\end{array}$$
\end{itemize}
where ${\it length},
{\it head}$, and ${\it tail}$ are
three functions on lists returning the length of the list,
its first element, and the list deprived of its first element,
respectively.

The ordering between two complexity pairs is the lexicographic
ordering in which usual $<$ is used for the integer numbers of the
first argument and  $<_{list}$ for the second.
It is immediate to prove that this ordering is well-founded.

\medskip

We show that each non-failing call to
{\fp AbCl\_unify\_actions}
causes the decreasing of the complexity. Well-foundedness
of the ordering implies termination.
By case analysis, we note that:
\begin{itemize}
\item rules 1, 2, and 7 cannot increase $A$\/, while $B$ always
      decreases
\item rule  5 decreases $A$
\item rule 8 is more complicated to analyze, since it calls
${\fp AbCl\_{step}}
  (\mathcal{E}, \{t\,|\,s\}=\{t'\,|\,s'\} )$.
In this case,
equations are added on the part of the system dealt as a stack,
driving the following rule applications.
These sequences of rule applications always allow to
empty the stack. We consider these operations as a unique step
that removes $ \{t\,|\,s\}=\{t'\,|\,s'\} $ and introduces other
equations in the system. Four cases must be distinguished:
\begin{enumerate}
\item ${\fp tail}(s)$ and
      ${\sf tail}(s')$ are not variables:
in this case $A$ cannot increase and $B$ decreases, since the
equation is replaced by a certain number of equations between the
elements of the two sets and between their tails, but all of
fewer (leftmost) size;

\item exactly one of them is a variable. Assume ${\fp tail}(s)$
is a variable: a
substitution for it is computed and applied: $A$ decreases. The
situation when ${\fp tail}(s')$ is a variable is perfectly symmetrical.

\item ${\fp tail}(s)$ and ${\fp tail}(s')$
are the different variables $X$ and $Y$, respectively.
One of the following cases
happens:
\begin{enumerate}
\item a substitution $X = \{ \dots \,|\, Y \}$ is computed,
\item a substitution $Y = \{ \dots \,|\, X \}$ is computed,
\item a substitution $X = \{ \dots \,|\, N \}$ and
                     $Y = \{ \dots \,|\, N \}$ ($N$ a new variable,
the same for the two equations) is computed.
\end{enumerate}
In all the three cases the application of the substitution cause $A$
to decrease.
\item ${\fp tail}(s)$ and ${\fp tail}(s')$
are the same variable $X$. In this case one
      equation $X =  \{ \dots \,|\, X \}$ is added to $\cal E$
together with a certain number of equations between elements
of the two sets $\{t\,|\,s\}$ and $\{t'\,|\,s'\}$. All those
equations have  (leftmost) size smaller than
$\{t\,|\,s\}$.
\end{enumerate}
\end{itemize}

To conclude the proof, let us observe that the termination of {\fp
AbCl\_unify\_final} is evident. For any variable $X$ occurring in
a equation $X =  \{ \dots \,|\, X\}$ we  perform at most one
rewriting and application of substitution. $X$ occurs elsewhere in
the system only as l.h.s.\ Equations  in solved form remains in
solved form and do not fire any new action.
\end{proof}

\subsection{Correspondence between $ACI1$ with
Constants and ${\sf gflat}(q)$ Unification}

\noindent {\bf Lemma \ref{equivaleACI1}}.~~{\it $\sigma$ is a
solution of the SUS problem $s = t $ if and only if $\sigma^*$ is
a $ACI1$ unifier of $s^*=t^*$.}

\begin{proof}
Without loss of generality, we assume that symbols in $s$
and $t$ are sorted, so as they are of the form
$$\begin{array}{rcl}
s & = & \underbrace{\{a_1 , \dots , a_m\}}_{C_1}
\cup
\underbrace{\{b_1  , \dots  ,  b_n\}}_{C_3}
\cup
\underbrace{Y_1 \cup \cdots \cup Y_p}_{V_1}
\cup
\underbrace{W_1 \cup \cdots \cup W_q}_{V_3} \\
t & = & \underbrace{ \{ d_1  , \dots  ,  d_{m'}\}}_{C_2}
\cup
\underbrace{\{ b_1  , \dots , b_n\}}_{C_3}
\cup
\underbrace{Z_1 \cup \cdots \cup Z_{p'}}_{V_2}
\cup
\underbrace{W_1 \cup \cdots \cup W_q}_{V_3} \\
\end{array}$$
where $C_i$ and $V_i$ are determined according to formula
(\ref{vettori})---Section~\ref{flat0dec}. The corresponding
 $(s)^*$ and $(t)^*$ are:
$$\begin{array}{rcl}
s & = & \underbrace{a_1 \cup \cdots \cup a_m}_{C_1}
\cup
\underbrace{b_1 \cup \cdots \cup b_n}_{C_3}
\cup
\underbrace{Y_1 \cup \cdots \cup Y_p}_{V_1}
\cup
\underbrace{W_1 \cup \cdots \cup W_q}_{V_3} \\
t & = & \underbrace{d_1 \cup \cdots \cup d_{m'}}_{C_2}
\cup
\underbrace{b_1 \cup \cdots \cup b_n}_{C_3}
\cup
\underbrace{Z_1 \cup \cdots \cup Z_{p'}}_{V_2}
\cup
\underbrace{W_1 \cup \cdots \cup W_q}_{V_3} \\
\end{array}$$

\noindent$\sigma$ is a solution of $s=t$ if and only if
\begin{itemize}
\item for each $a_i$ in  $C_1$ there is $X$ in $V_2 \cup V_3$ such that
$\sigma(X) = \{a_i , \cdots\}$ and
\item for each $b_j$ in  $C_2$ there is $X$ in $V_2 \cup V_3$ such that
$\sigma(X) = \{b_j, \cdots\}$ and
\item each variable in $V_1\cup V_2 \cup V_3$ is mapped on a
set of constants in $C_1 \cup C_2 \cup C_3$ plus, possibly, other
constants.
\end{itemize}
$\mu$ is a solution of $s=t$ if and only if
\begin{itemize}
\item for each $a_i$ in  $C_1$ there is $X$ in $V_2 \cup V_3$ such that
$\mu(X) = a_i \cup \cdots$ and
\item for each $b_j$ in  $C_2$ there is $X$ in $V_2 \cup V_3$ such that
$\mu(X) = b_j \cup \cdots$ and
\item each variable in $V_1\cup V_2 \cup V_3$ is mapped on a
union of constants in $C_1 \cup C_2 \cup C_3$ plus, possibly, other
constants.
\end{itemize}
Clearly, $\mu = \sigma^*$.
\end{proof}

\subsection{Soundness and Completeness of General $ACI1$ Unification Algorithm}

\noindent{\bf Lemma \ref{corr_aci}.}~~{\em Given an equation $e$
of the form \[f_1(\bar l_1) \cup \cdots \cup f_{k_1}(\bar l_{k_1})
\cup L_1 \cup \cdots \cup L_{k_2} = g_1(\bar r_1) \cup \cdots \cup
g_{h_1}(\bar r_{h_1}) \cup R_1 \cup \cdots \cup R_{h_2}\] let
$\langle {\cal E}^i_1, {\cal E}^i_2, \rho^i \rangle$\/, for  $i =
1,\dots,k$\/, be the collection of all the distinct solutions
non-deterministically produced by the call ${\fp aci\_{step}}( e
)$\/. Then:
\begin{itemize}
\item if $\sigma$ is a unifier of ${\cal E}^i_1 \cup  {\cal E}^i_2$
then $\sigma$ is a unifier of $e$ and $\sigma \leq \rho^i$\/
\item if $\sigma$ is a unifier of $e$ then there exists $1 \leq i \leq
k$ and a substitution $\gamma$ such that $\sigma \cup
  \gamma$ is a unifier of ${\cal E}^i_1 \cup  {\cal E}^i_2$\/.
\end{itemize}}
\begin{proof}
Let us prove the lemma by showing that the conditions hold at each step
of the construction of each solution.

\begin{itemize}
\item For the \emph{Normalization} step, it is trivial to show that
$\sigma$ is a unifier
        for $e$ if and only if $\sigma \cup \gamma$ is a unifier for ${\cal
E}^n$\/, where
        $\gamma$ possibly binds the new variables $N^L_i, N^R_j$\/. In
this case $k$ is equal to $1$\/. The substitution
        $\gamma$ is $[ N^L_i/f_i(\bar l_i)\sigma \,|\, 1 \leq i \leq
k_1 ] \cup [ N^R_j/g_j(\bar r_j)\sigma \,|\, 1 \leq j \leq
h_1 ]$\/.

\item
For the \emph{Elementary $ACI1$ Solution} step the result follows
from the results for elementary $ACI1$ unification~\cite{BB88}. In
this case we
        have that $\sigma$ is a unifier of ${\cal E}^n$ if
        and only if $\sigma \cup
\gamma$
        is a unifier for ${\cal E}^{ACI}$\/, where $dom(\gamma)
\subseteq
\{A_{i,j} \, | \,1 \leq i \leq h_1+h_2 \wedge 1 \leq j \leq k_1 + k_2
\}$\/.

\item
Let us consider  the \emph{Term Propagation} step. We prove that
        $\sigma'=\sigma \cup \gamma$ is a unifier for ${\cal E}^{ACI}$
(where
        $dom(\sigma) \cap dom(\gamma) = \e$ and $dom(\gamma) =
\{A_{i,j}
\, | \,
                1\leq i \leq h_1+h_2 \wedge 1 \leq j \leq k_1 + k_2
\}$\/) iff
        $\sigma \gamma'$ is a unifier for ${\cal E}_1 \cup {\cal E}_2$
        (where $\gamma'$ is the restriction of $\gamma$ to
        $\{A_{i,j} \, | \, h_1+1 \leq i \leq h_1+h_2 \wedge k_1+1 \leq
j\leq k_1+k_2\}$\/).

        Let $\sigma'=\sigma\cup\gamma$ be a unifier for ${\cal E}^{ACI}$
and
        let us consider the equations  $e$ in ${\cal E}^{ACI}_2$ in the
same arbitrary
        order used to build ${\cal E}_2$\/.
         Such equations have the
        form $f(\bar s) = \bigcup A_{i} \cup \bigcup B_{j}$\/. If
$\sigma'$
is a unifier
        for ${\cal E}^{ACI}$\/, then (from the ACI properties and
Clark's
        Equational Theory) each $A_i\gamma$ and $B_j\gamma$ must be
either
$\e$ or
        $f(\bar s)\sigma$\/; furthermore, at least one of $A_i,B_j$ must
be
assigned
        $f(\bar s)\sigma$\/. Let $I$ and $J$ be the collection of
indices
for which
        respectively $A_i$ and $B_j$ receive $f(\bar s)\sigma$ in
$\sigma'$\/.
        We can use $I$ and $J$ to select a certain
        ${\cal E}_1 \cup {\cal E}_2$\/---the
        one in which the $\lambda$ is constructed by taking
    $\lambda(A_i) = f(\bar s)$
        ($\lambda(B_j) = f(\bar s)$) for $i\in I$ ($j\in J$), and $\e$
    for the
        remaining variables in the equation. The process can be repeated
    for the remaining equations, until all the variables have received an
    assignment in $\lambda$\/. The consistency of $\sigma'$ guarantees
    that this construction
        will provide a consistent ${\cal E}_2$\/. It is straightforward
    to observe
        that  $\sigma'$ is a unifier for ${\cal E}_2$\/. Observe also
    that
        $\sigma' \leq \lambda$\/, i.e., $\sigma' = \lambda \circ \theta$\/.
    This last
        fact, together with the fact that $\sigma$ is a unifier for
        ${\cal E}^{ACI}_1$\/, is sufficient to conclude that $\sigma'$
    is a unifier for ${\cal E}_1$\/.

        Vice versa, let  $\sigma'$ be a unifier for a certain
        ${\cal E}_1 \cup {\cal E}_2$ produced by the algorithm.
        Since the construction was possible, then there is a
substitution
$\lambda$ which
        has been used to convert ${\cal E}^{ACI}_2$ into ${\cal E}_2$\/.
If
$\sigma'$
        is a solution of  the equations $\bar s = \bar s_1, \dots, \bar s
=
\bar s_h$
        present in ${\cal E}_2$\/, then $\sigma'$ is also a
unifier for the equation
        $f(\bar s) = (\bigcup A_i \cup \bigcup B_j)\lambda$ which
produced such elements
        of ${\cal E}_2$\/. Thus, $\sigma' \cup [A/A\lambda\circ\sigma'
\,|\, A \in dom(\lambda)]$
        is a unifier for ${\cal E}^{ACI}_2$\/. The result for ${\cal
E}^{ACI}_1$ is
        obvious.

\item
Correctness for the \emph{Variables Removal} step follows from the
fact that we are not interested in solutions over infinite
terms.
\end{itemize}
\end{proof}

\subsection{Termination of General $ACI1$ Unification Algorithm}

\begin{definition}  \label{pilevel}
Let ${\cal E}$ be a set of equations, and let us consider a
function $\lev:{\vars}({\cal E}) \longrightarrow \nat$. This
function can be extended over elements of $T(\Sigma,{\cal V})$ as
follows:
$$\begin{array}{rcll}
  \lev(f(t_0,\dots,t_n)) & = &
     1 + \max\{\lev(t_0),\dots,
           \lev(t_n) \} & \mbox{$f \in \Sigma$\/,
          $f \not\equiv \cup$}\\
  \lev(s \cup t) & = &
      \max\{\lev(s), \lev(t)\}
 \end{array}$$
The function $\lev$ is said to be a \emph{partial $p$-level} if
it satisfies the condition:
\begin{itemize}
\item[$(*)$] $\lev(\ell) , \lev(r)  \leq  p$\/,
for any equation $\ell = r$ in $\cal E$\/.
\end{itemize}
Any partial $p$-level $\lev$ is said to be a \emph{(complete)
$p$-level} if it satisfies also the condition:
\begin{itemize}
\item[$(**)$] $\lev(\ell)  =  \lev(r)$ for
any equation $\ell = r$ in $\cal E$.
\end{itemize}
\end{definition}

\begin{lemma}\label{level-1}
Let us consider  a system of equations $\cal E$\/, and
let  $p$ be the number of
occurrences of elements of $\Sigma$ in $\cal E$\/; then,
exactly one of the following conditions holds:
\begin{itemize}
\item there exists a complete $p$-level for $\cal E$
\item for any natural number $q$, there are no complete $q$-levels
        for $\cal E$\/.
\end{itemize}
\end{lemma}
\begin{proof}
Given the system $\cal E$, it is possible to obtain, by adding a
suitable number of new variables,  an equivalent
system $\cal E'$ in \emph{flat form}, i.e., each equation in $\cal E$ is
in one of the following forms:
\begin{enumerate}
\item $X = Y$
\item \label{case2} $X = f(Y_1, \dots, Y_n)$,
      $f \in \Sigma$ and $f \not\equiv \cup$
\item $X = Y_1 \cup Y_2$
\end{enumerate}
Observe that at most $p$ equations of type (\ref{case2}) can appear
in $\cal E'$\/.

The goal is to map $\cal E'$ to a set of linear integer constraint
systems. Each possible complete $p$-level for $\cal E'$ (and thus
for $\cal E$) is a solution of at least one of such systems of
constraints. Vice versa, each solution of one of these systems
can be used to generate a complete $q$-level for $\cal E'$, for a
suitable $q$. Such mapping is realized as follows: for each
(term) variable $X$ in $\cal E'$ we introduce a corresponding
(integer) variable $x$; then we add equations and disequations
according to the following rules:

\[\begin{array}{rcl}
    X=X  &  \mapsto  &  \mbox{\em if $X$ does not occur elsewhere,
    then add $x=0$}\\
    X=Y  &  \mapsto  &  x = y\\
    X=f(Y_1,\dots,Y_n), n > 0 & \mapsto  &
       \bigvee_{i=1}^n (x = y_i+1
\wedge \bigwedge_{j=1,j\neq i}^n y_j \leq y_i)\\
        X=a  &   \mapsto  & x=1\\
        X=Y_1 \cup Y_2 & \mapsto & (x=y_1 \wedge y_2 \leq y_1) \vee
(x=y_2 \wedge y_1 \leq y_2)
  \end{array}
\]

Through simplifications (e.g., distributivity) it is possible
to obtain a disjunction of systems $S_1 \vee \cdots \vee S_k$, where each
system $S_i$
contains
only equations of the form:
\[
x=y \:\:\:\:\:\:\:\:\:x=y+1\:\:\:\:\:\:\:\:\:x=0\:\:\:\:\:\:\:\:\:x=1\:\:\:\:\:\:\:\:\:x \leq y
\]

Furthermore, in each system there can be at most $p$ occurrences
of equations of the type $x=y+1$ and $x=1$. Our aim is to show that, if
one of the systems $S_i$ is satisfiable, then there will be one solution
$\sigma$ of  $S_i$
such that for each variable $x$ we have $\sigma(x) \leq p$.

Each system $S_i$ can be further simplified using the following
observations:
\begin{itemize}
\item All equations of the form $x=1$ can be eliminated and replaced
        with the equations $w=0$ and $x=w+1$, where $w$ is a new
        variable. Note that the total number of equations $x=y+1$ is
        still at most $p$ even after this simplification.
\item  The equations of the form $x=y$ induce an equivalence relation
        on the variables. We can remove all these equations and replace
        each occurrence of each variable in $S_i$
         with a selected representative from its equivalence
        class.
\end{itemize}

For each $S_i$ we can construct a labeled graph $G_{S_i} = \langle
\vars(S_i), E_i\rangle$ as follows (see example in
Figure~\ref{nuovo esempio}; thick lines are used for $+1$ edges
and thin lines for $0$ edges):
\begin{itemize}
\item for each equation of the form $x=y+1$ in $S_i$, generate
        an edge $(y,x)$ with label $+1$
\item for each equation of the form $x \leq y$ in $S_i$, generate
        an edge $(x,y)$ with label $0$,
          unless there is already
        an edge $(x,y)$ with label $+1$\/.
\end{itemize}

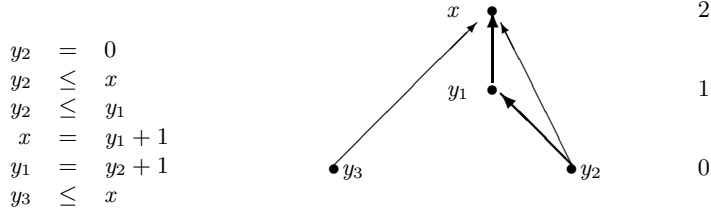
\begin{figure}
$$\begin{array}{ccc}
\begin{array}[b]{rcl}
y_2 & = & 0\\
y_2 & \leq & x\\
y_2 & \leq & y_1\\
x & = & y_1 + 1\\
y_1 & = & y_2 + 1\\
y_3 & \leq & x
\end{array} &
\begin{array}[t]{c}
 \setlength{\unitlength}{1pt}
 \begin{picture}(200,70)(0,50)
 \thinlines    \put(100,120){$\bullet$}
               \put(85,120){$x$}
                    \put(100,90){$\bullet $}
                    \put(85,90){$ y_1 $}
                    \put(130,60){$\bullet~y_2$}
                    \put(40,60){$\bullet~y_3$}
\thicklines    \put(102.5,95){\vector(0,1){27}}
\thicklines    \put(132.5,64){\vector(-1,1){27}}
\thinlines     \put(42.5,64){\vector(1,1){54}}
\thinlines     \put(132.5,64){\vector(-1,2){27}}
\thinlines   \put(180,120){$2$}
\thinlines   \put(180,90){$1$}
\thinlines   \put(180,60){$0$}
\end{picture}
\end{array}
\end{array}$$
\caption{\label{nuovo esempio}A simplified integer system,
the corresponding graph, and a solution}
\end{figure}

If $G_{S_i}$ contains a cycle with at least one edge labeled
$+1$, then $S_i$ will not admit solutions. Cycles in $G_{S_i}$
composed only of edges of type $0$ denote an implicit equality
between the nodes in the cycle---thus we can collapse the
cyclic component.
These observations allow us to focus only on directed acyclic
graphs.

A solution of $S_i$ can be described as a labeling of the
nodes of the graph. A \emph{consistent} labeling $\sigma$ of the
nodes of the graph representing a solution should fulfill the
following conditions:
\begin{itemize}
\item if $(x,y)$ is an edge of type $+1$, then $\sigma(y) = \sigma(x)+1$
\item if $(x,y)$ is an edge of type $0$, then $\sigma(x) \leq \sigma(y)$
\item if there is an equation $x=0$ in $S_i$, then $\sigma(x) = 0$\/.
\end{itemize}

We claim that if the $G_{S_i}$ admits a labeling with the
above properties, then $G_{S_i}$ also admits a labeling  $\phi$ of the
nodes such that for each node $X$ we have $\phi(X)\leq r$, where $r$ is
the number of $+1$ edges in $G_{S_i}$---in particular $r \leq p$.

Let us develop a proof by lexicographical induction over the measure
$\langle A,B \rangle$, where $A$ is the number of $+1$ edges
and $B$ is the number of $0$ edges in the graph.

\begin{description}
\item [$\langle 0,0 \rangle$] In this case the graph is composed
        only of disconnected nodes, and the original system $S_i$
        contains only equations of the form $x=0$\/;
        the solution $\sigma$ such that $\sigma(x)=0$ for each node
        $x$ is a consistent 0-labeling.

\item [$\langle m,n \rangle$] Let $x$ be an arbitrary node of
        $G_{S_i}$ with no outcoming edges, $(v_1,x), \dots, (v_h,x)$
        incoming edges of type $+1$, and $(w_1,x),\dots,(w_k,x)$
incoming
        edges of type $0$. With no loss of generality we assume $h + k
\geq 1$.
         Let us distinguish
        the following cases:

        \begin{enumerate}
        \item $h=0$: consider
        the graph $G_{S_i}'$ obtained by removing node $x$ and all
        its incoming edges (all of type $0$). The measure for the
        graph $G_{S_i}'$ is $\langle m, n-k \rangle$. By inductive
        hypothesis, there is a consistent $m$-labeling $\sigma$ of
$G_{S_i}'$.
        $\sigma$ can be extended to a consistent $m$-labeling of
        $G_{S_i}$ by assigning $\sigma(x) =
\max\{\sigma(w_1),\dots,\sigma(w_k)\}$.

        \item $h > 1$ and $k \geq 0$: in each consistent labeling of
$G_{S_i}$ we
        must have that $\sigma(v_1) = \cdots = \sigma(v_h) =
\sigma(x)-1$.
Let us consider
        the graph $G_{S_i}'$ obtained by collapsing nodes $v_1, \dots,
v_h$ into
         a single node $v$. The measure of $G_{S_i}'$ is $\langle m-h+1,
n
\rangle$;
        thus, by inductive hypothesis, it is possible to determine a
        consistent $(m-k+1)$-labeling $\sigma$ of $G_{S_i}'$. $\sigma$
can be
        extended into a consistent $(m-k+1)$-labeling of $G_{S_i}$ by
defining
        $\sigma(v_1) = \cdots = \sigma(v_h) = \sigma(v)$. By definition
$\sigma$
        is also a consistent $m$-labeling of the graph.

        \item $k=0$ and $h=1$: consider the graph $G_{S_i}'$ obtained by
        removing $X$ and its incoming edge. The measure of $G_{S_i}'$ is
        $\langle m-1,n \rangle$, thus, by inductive hypothesis, there
        is a consistent $(m-1)$-labeling $\sigma$ of such graph. This
labeling
        can be extended to a consistent $m$-labeling of $G_{S_i}$ by
defining $\sigma(x) = \sigma(v_1)+1$.

        \item $k>0$ and $h=1$:     in each consistent labeling of
$G_{S_i}$ we
        must have that:
        \begin{enumerate}
        \item \label{aaa}$\sigma(v_1) = \sigma(x) - 1$
        \item \label{bbb}$\sigma(w_i) \leq \sigma(x)$ for $i=1,\dots,k$,
thus
                $\sigma(w_i) = \sigma(x)$ or $\sigma(w_i) \leq
\sigma(v_1)$
        \end{enumerate}
        Let us consider the following class of simplified graphs:
        we arbitrarily partition $\{w_1,\dots,w_k\}$ into two subsets
                $B_1,B_2$ and we consider the graph obtained by:
                \begin{itemize}
                \item removing all edges $(w_i,x)$
                \item   collapsing all nodes in $B_1 \cup \{x\}$
                \item adding the edges $(w_i,v_1)$ for each
                        $w_i \in B_2$
                \item if $B_1 = \emptyset$, then the node $x$ and
                        the edge $(v_1,x)$ are removed.
\end{itemize}

        The two properties (\ref{aaa}) and (\ref{bbb}) guarantee that
        each consistent labeling of $G_{S_i}$ is a consistent labeling
        of at least one of the simplified graphs, and each consistent
        labeling of a simplified graph can be extended (see below) to
        a consistent labeling of $G_{S_i}$.
        Since we are under the assumption that $G_{S_i}$ admits
consistent
        labelings, at least one of the simplified graph admits
consistent
        labelings.
                The measure of each simplified graph is
                $\langle m, n - |B_1| \rangle$ if $B_1 \neq \emptyset$,
                $\langle m-1, n \rangle$ otherwise. By inductive
hypothesis
                we can build a consistent $m$-labeling (or
$(m-1)$-labeling
                in the last case) $\sigma$ for such graph.
                If $B_1 \neq \e$, then $\sigma$ can be extended to
                a consistent $m$-labeling of $G_{S_i}$ by defining
                $\sigma(w_i) = \sigma(x)$ for each $w_i \in B_1$.
Otherwise,
                a consistent $m$-labeling of $G_{S_i}$ is obtained by
                defining $\sigma(x) = \sigma(v_1) + 1$.

        \end{enumerate}
\end{description}
\end{proof}

The notion of $p$-level has a direct interpretation
on a graph-encoding of the system of equations. The unification
algorithm itself can be mapped on a collection of graph manipulation
operations. The mapping of the unification algorithm on graphs
allows us to intuitively demonstrate that for each intermediate
system of equations during the unification process it is possible
to determine a  partial $p$-level (where
$p$ is the number of occurrences of elements of $\Sigma$ in the
initial system).

Given the initial system ${\cal E}_0$ we define the directed labeled
graph $G_0$ as follows:
\begin{itemize}
\item $G_0$ contains a node for each occurrence of a function
        symbol in ${\cal E}_0$; without loss of generality, we
        assume that each occurrence of a constant $c$ has
    been replaced with a term
        $c(B)$\/, where $B$ is a fixed variable, and $c$ is a
        new unary function symbol.

\item   $G_0$ contains a node for each variable in ${\cal E}_0$\/.

\item For each term $f(t_1,\dots,t_n)$ ($f$ different from $\cup$)
        in ${\cal E}_0$, if
        $\mu$ is the node created for the specific occurrence of
        $f$, and $\nu_i$ is the node created for the main functor
        of $t_i$ (or for the variable $t_i$), then the edge
        $(\mu,\nu_i)$ with label $+1$ is added to $G_0$

\item  let $t$ be a term $t_1 \cup \cdots \cup t_n$ such that:
       $n>1$,
        the main functor of each $t_i$ is different from $\cup$,
        and either
        \begin{itemize}
        \item the term $t$ is the left-hand side or the right-hand
                side of an equation in ${\cal E}_0$; or
        \item there exists a  term $f(t_1,\dots,t_n)$ in ${\cal E}_0$ such
                that $t \equiv t_i$ and $f$ is different from $\cup$\/.
        \end{itemize}

     Let $\mu$ be the node introduced for the first occurrence of
     $\cup$ in $t$, i.e.,
     $$t_1
     \underbrace{\cup}_{\Uparrow} t_2 \cup  \cdots \cup t_n$$
     and let $\nu_i$ be the node created for the
     main functor of $t_i$ (or for the variable $t_i$); the graph
     $G_0$ contains the edges
        $(\mu,\nu_i)$ with label $0$ for $i = 1,\dots,n$\/.

\item remove from $G_0$ all the nodes created for occurrences of
        $\cup$ which do not have any outgoing edges.
\end{itemize}

\begin{example}
Let ${\cal E}_0$ be the system $f(X) \cup f(g(Y)) = f(g(Z) \cup
h(Z,V)) \cup V \cup W$. Then $G_0$ is the graph (thick lines are
used for $+1$ edges while thin lines are used for $0$ edges):
$$ \setlength{\unitlength}{1pt}
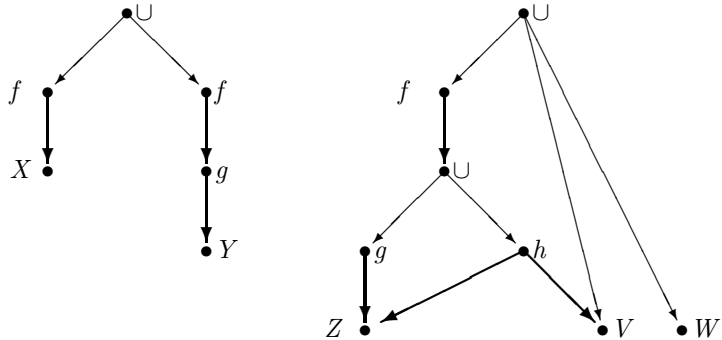
 \begin{picture}(300,130)(0,0)
 \thinlines    \put(50,120){$\bullet~\cup$}
                    \put(20,90){$\bullet$}
                    \put(8,90){$f$}
                    \put(20,60){$\bullet$}
                    \put(8,60){$X$}
                    \put(80,90){$\bullet~f$}
                    \put(80,60){$\bullet~g$}
                    \put(80,30){$\bullet~Y$}
\thinlines     \put(52.5,122.5){\vector(-1,-1){27}}
\thinlines     \put(52.5,122.5){\vector(1,-1){27}}
\thicklines    \put(22.5,92.5){\vector(0,-1){27}}
\thicklines    \put(82.5,92.5){\vector(0,-1){27}}
\thicklines    \put(82.5,62.5){\vector(0,-1){27}}


 \thinlines    \put(200,120){$\bullet~\cup$}
                    \put(170,90){$\bullet$}
                    \put(155,90){$f$}
                    \put(170,60){$\bullet~\cup$}
                    \put(140,30){$\bullet~g$}
                    \put(200,30){$\bullet~h$}
                    \put(140,0){$\bullet$}
                    \put(127,0){$Z$}
                    \put(230,0){$\bullet~V$}
                    \put(260,0){$\bullet~W$}
\thinlines     \put(202.5,122.5){\vector(-1,-1){27}}
\thinlines     \put(202.5,122.5){\vector(1,-4){29}}
\thinlines     \put(202.5,122.5){\vector(1,-2){58}}

\thinlines     \put(172.5,62.5){\vector(-1,-1){27}}
\thinlines     \put(172.5,62.5){\vector(1,-1){27}}

\thicklines    \put(172.5,92.5){\vector(0,-1){27}}

\thicklines    \put(142.5,32.5){\vector(0,-1){27}}
\thicklines    \put(202.5,32.5){\vector(1,-1){27}}
\thicklines    \put(202.5,32.5){\vector(-2,-1){54}}

\end{picture}$$
\end{example}

Let us define an \emph{iteration} to be a single application of a
rule of the procedure {\fp general\_aci}. Each rule of the
unification algorithm can be mapped onto an operation on the
graph. If ${\cal E}_i$ is the system obtained after $i$ iterations
of the unification algorithm, then we denote with $G_i$ the
corresponding graph. The graph operations corresponding to the
different non-failing unification rules are the following:
\begin{itemize}
\item [1.] if ${\cal E}_{i+1}$ is obtained by removing an equation
        $X=X$ from ${\cal E}_i$, then $G_{i+1} = G_i$
\item [2.] if ${\cal E}_{i+1}$ is obtained by replacing $t=X$ with
        $X=t$ in ${\cal E}_i$, then $G_{i+1} = G_i$
\item [4.] if ${\cal E}_{i+1}$ is obtained by replacing each occurrence
        of $X$ with $t$ in ${\cal E}_i$, then $G_{i+1}$ is obtained
        by adding the edge $(\mu,\nu)$ with label $0$, where $\mu$
        is the node associated with the variable $X$ and $\nu$ is
        the node created for the main functor of term $t$ (or for the
        variable $t$)
\item [5.] if ${\cal E}_{i+1}$ is obtained by replacing the equation
        $X = t$ where $t \equiv f_1(\bar{s}_1) \cup \cdots \cup
f_n(\bar{s}_n) \cup V_1 \cup \cdots
                \cup V_m \cup X$ (assumed in this ordered form as
explained
        in the unification algorithm---note that this simplification is
not
        needed in the graph representation) with the equation
        $X = f_1(\bar{s}_1) \cup \cdots \cup f_n(\bar{s}_n) \cup V_1 \cup
\cdots
                \cup V_m \cup N$, $N$ new variable, then $G_{i+1}$ is
        obtained by adding a new node $\nu$ for $N$, by removing the
edge
$(\mu,\xi)$
        where $\mu$ is the node for the functor of $t$ and $\xi$ the
node
for $X$, and
        by adding the new edge $(\mu,\nu)$
\item [7.] if ${\cal E}_{i+1}$ is obtained by replacing the equation
        $f(t_1,\dots,t_n)=f(s_1,\dots,s_n)$ in ${\cal E}_i$\/, then
        $G_{i+1} = G_{i}$
\item [8.] let us assume that ${\cal E}_{i+1}$ is obtained by replacing
the equation
        $$f_1(\bar{s}_1) \cup \cdots \cup f_n(\bar s_m) \cup X_1 \cup
\cdots
\cup X_h = g_1(\bar{t}_1) \cup \cdots \cup g_n(\bar t_n) \cup Y_1
\cup
\cdots \cup   Y_k$$
        in ${\cal E}_i$ with a family of equations:
$$f_i(\bar s_i) = g_j(\bar t_j)\phantom{aaa}\mbox{ for some $i,j$}$$
        and by substituting $X_i$ ($Y_i$) with terms of the form:
$$X_i = g_{i_1}(\bar{t}_{i_1}) \cup \cdots \cup g_{i_r}(\bar t_{i_r})
\cup
                                N_1 \cup \cdots \cup N_s$$
        (similarly for $Y_i$).
        $G_{i+1}$ is obtained from $G_i$ as follows:
        \begin{itemize}
        \item introducing a new node $\nu_i$ for each new variable $N_i$
        \item if $\mu$ is the node for $X_i$ ($Y_i$) and $\eta_j$ is the
                node for the main functor of term $g_j(\bar t_j)$
($f_j(\bar s_j)$),
                then add the edge $(\mu,\nu_w)$ and $(\mu,\eta_j)$ with
label $0$
                for each $g_j(\bar t_j)$ ($f_j(\bar s_j)$) and for
                each $N_w$ present in the substitution
                for $X_i$ ($Y_i$).
        \end{itemize}
\end{itemize}

\begin{lemma}\label{noneri}
Let $One(G)$ be the set of $+1$ edges present in the graph $G$.
Then for each $G_i$ obtained from the above transformations we
have $One(G_i) = One(G_0)$. Furthermore, $G_i$ does not contain
any cycles which  include edges labeled~$+1$.
\end{lemma}
\begin{proof}
The first property is obvious from the definition of the
transformations.

The second property is straightforward for the cases $(1)$,
$(2)$, and $(7)$ of the unification algorithm, since they do not
add edges---and thus cannot generate cycles. Case $(5)$ adds a new
edge, but the destination of the edge is a new variable which has
no outgoing edges.

Case $(4)$ can be seen as follows: let us assume, by
contradiction, that the addition of the edge from the node of $X$
to the node of $t$ generates a cycle with $+1$ edges. This means
that, before this operation, there exists a path from the root of
$t$ to the node of $X$ (with at least one $+1$ edge). This path
can be only the result of a sequence of edge additions leading
from a node reachable from the root of $t$ to the node of $X$.
Each of these edges has been introduced during previous variable
substitutions---and each of the nodes reachable using this path
identifies a sub-term of $t$. Thus, $X$ is a sub-term of $t$.
This contradicts the possibility of applying case $(4)$, since
this situation is explicitly handled by case $(3)$ and leads to a
failure.

Case $(8)$ can be seen as a combination of cases $(7)$ (new
equations of the type $f_i(\bar s_i) = g_j(\bar t_j)$ which do
not modify the graph), $(5)$ for the new variables $N_i$, and
$(4)$ for the substitution of existing variables.
\end{proof}

\begin{lemma}\label{corneri}
Let us assume that there is a non-failing sequence
of $k$ non-deterministic
choices, such that ${\fp general\_aci}({\cal E})$ generates (one
per each successive iteration) the systems ${\cal E}={\cal
E}^{(0)},{\cal E}^{(1)}, {\cal E}^{(2)},\dots, {\cal E}^{(k)} $.
Let $p$ be the number of occurrences of function symbols in
${\cal E}^{(0)}$\/. Then, there exists
$$\lev : {\vars}\left(
\bigcup_{j = 0}^k {\cal E}^{(j)} \right) \longrightarrow \nat$$
 such that:
\begin{itemize}
\item it fulfills condition $(*)$ of Def.~\ref{pilevel}
      for all systems of equations ${\cal E}^{(j)}$ (i.e.,
      $\lev(\ell)  \leq p$ and $\lev(r) \leq p$
        for all the equations $\ell = r$ in ${\cal E}^{(j)}$),  and
\item any time a substitution $[X/t]$ has been applied,
      then $\lev(X) = \lev(t)$\/.
\end{itemize}
\end{lemma}
\begin{proof}
Let us consider the graphs $G_j$ associated to the systems ${\cal
E}^{(j)}$. First of all, observe that if there is a function
fulfilling the requirements for the system ${\cal E}^{(j)}$, then
the same function works for all graphs ${\cal E}^{(i)}$ with $i <
j$. This allows us to concentrate on ${\cal E}^{(k)}$. By
Lemma~\ref{noneri}, we know that $G_k$ is acyclic and it contains
the same number ($p$) of $+1$ edges as ${\cal E}^{(0)}$. {F}rom
this fact, starting from leaf nodes and going back on edges,
augmenting a value only if a $+1$ edge is encountered, it is
natural to find a function $\lev$ fulfilling the required
property. \end{proof}

\begin{theorem}[termination]\label{aciterm}
Given a system of equations ${\cal E}$\/,  all
the non-deterministic branches of the computation of
$\mbox{{\fp general\_aci}}({\cal E})$ terminate in
a finite amount of time.
\end{theorem}
\begin{proof}
Assume that there is a non-failing sequence of
non-deterministic choices ${\cal E}^{(0)},{\cal E}^{(1)}$,
${\cal E}^{(2)},\dots, {\cal E}^{(k)}$ (they are the values of $\cal E$
at the $0^{th},1^{st},2^{nd},\dots,k^{th}$ iteration,
respectively), and let $p$ be the number of occurrences of
function symbols in ${\cal E}^{(0)}$\/. We know from
Lemma~\ref{corneri} that there exists a function $\lev :
{\vars}\left( \bigcup_{j \geq 0} {\cal E}^{(j)}\right)
\longrightarrow \nat$
 such that
\begin{itemize}
\item it fulfills condition $(*)$ for all the systems of equations ${\cal
E}^{(j)}$\/,
  and
\item any time a substitution $[X/t]$ has been applied,
      then $\lev(X) = \lev(t)$\/,
\end{itemize}
We call this property condition $(\alpha)$.

Picking such a $\lev$\/, we define a measure of complexity ${\cal
L}_{\cal E}$ for the system of equations $\cal E$\/:

$$\begin{array}{rcl} {\cal L}^{(\lev)}_{\cal E} & = & [\#(2p),\#(2p -
1),\#(2p
- 2), \dots , \#(1) , \#(0)] \end{array}$$
\noindent
where $\#(j)$ returns the number of equations \emph{not in solved
  form}
$\ell = r$ in ${\cal E}$ such that $\lev(\ell) + \lev(r) = j$\/.
The ordering between two lists of this form is the usual well-founded
lexicographical ordering.

Let $h$ be the number of equations in the initial system. The
initial tuple ${\cal L}^{(\lev)}_{{\cal E}^{(0)}}$ is necessarily
less than or equal to $[h,0,\dots,0]$. Let us consider how the
various rules in Figure~\ref{generalaciproc} modify the complexity
measure tuple:
\begin{itemize}
\item rule $( 1)$ clearly reduces the complexity by removing one
        equation
\item rule $( 2)$ does not affect the complexity but can be
        safely ignored (we could easily rewrite the algorithm
        without it by adding explicit cases for equations
        $t=X$ wherever we analyze $X=t$\/)
\item rule $( 4)$ reduces the complexity: in fact one equation
        of complexity $2\lev(X)$ is removed, while the rest
        of the system is unaffected, since $X$ is replaced by
        a term with the same level
\item rule $( 5)$ will lead in one additional iteration to an
        rule $( 4)$, which means that the complexity of
        the original equation must be $2\lev(X)$\/; by assigning
        $\lev(N) = \lev(X)$ we have that after two reductions the
        complexity will decrease
\item rule $( 7)$ replaces an equation of complexity
        $2+l_1+r_1$ with a collection of equations each  having
        complexity $l+r \leq l_1+r_1$\/, leading to a  smaller
        total complexity (thanks to lexicographical  ordering)
\item rule $( 8)$ is a complex rule which leads to the execution
        of the ${\fp aci\_{step}}$ function. Let $e$ be the equation
        communicated to ${\fp aci\_{step}}$.
        The only equations in non-solved form
        that are generated by ${\fp aci\_{step}}$ are
        the equations $\bar s = \bar s'$ present in ${\cal E}_{\fp ns}$\/. Such
        an equation $\bar s = \bar s'$ originates
        from simplifying an equation $f(\bar s) = f(\bar s') \cup
        \cdots$\/.
        Observe that in this equation $f(\bar s)$ and $f(\bar s')$
        originally
        appeared on distinct sides of the equation $e$---in the general
        structure of the equation, one of the two is a $f_j(\bar l_j)$
        and the other is a $g_i(\bar r_i)$\/. Thus, the equation
        $f(\bar s) = f(\bar s')$ has a complexity which is less or equal
        than that of $e$\/, which implies also that the complexity of
        $\bar s = \bar s'$ is strictly lower than that of $e$\/.
        Thus the original equation is replaced by a collection of
        equations
        of smaller complexity (assuming, as stated earlier, that the
        equations
        of the form $L_i = t_i$ and $R_j = s_j\cdots$ that lead to
        $\rho$
        are all such that $\lev(L_i) = \lev(t_i)$ and $\lev(R_j) =
       \lev(s_j)$\/).
\end{itemize}

Thus, every rule application decreases the complexity measure
${\cal L}^{(\lev)}_{\cal E}$. The lexicographical ordering on
constant-length lists of non-negative integers is a well-founded
ordering, and thus this activity cannot be done indefinitely.

However, this is not sufficient for termination, since we are not
sure that the complexity measure tuple reaches the value
$[0,\dots,0]$ within $k$ rule applications. Moreover, we do not
know if the function $\lev$ fulfills condition $(\alpha)$ for the
successive systems ${\cal E}^{(k+1)}, {\cal E}^{(k+2)}$, \dots .

To prove termination, a further measure is needed:
let
$${\cal M}_{\cal E} = \mo {\cal L}^{(\ell)}_{\cal E}  :
  \mbox{$\ell$ is a function from $\vars({\cal E})$ to $\{0,\dots,p\}$
        that fulfills condition $(\alpha)$} \mc$$
Multisets of tuples are governed by (well-founded) multiset ordering.

${\cal M}_{{\cal E}^{(0)}}$ is finite. All the initial tuples are
less than or equal to $[r,0,\dots,0]$; each of them is associated
to a function from $\vars({\cal E}^{(0)})$ to $\{0,\dots,p\}$ that
fulfills condition $(\alpha)$.

Let us consider this multiset and the effects of an iteration
over each of its tuples. After one iteration it holds that:

\begin{itemize}
\item The function $\ell$
  fulfills condition $(\alpha)$ for the successive systems. In this
  case $t$ is replaced by a fewer tuple (see the proof above).

\item The function $\ell$
  does not assign values to new variables. However, it is possible to extend
  $\ell$ into $\ell'$ in order to assign values for these variables.
  In this case the tuple $t$ is replaced by a certain (finite) number of
  tuples fewer than $t$ (the new variables $N$ are introduced
  in equations of the form $X = \cdots \cup N$ and thus,
  $\ell'(N) \leq \ell(X)$).

\item The function $\ell$
  does not fulfill condition $(\alpha)$ for the new system and, moreover,
  it is not possible to extend
  $\ell$ into $\ell'$ in order to assign values for these variables
  to fulfill condition $(\alpha)$.
  In this case the tuple $t$ is simply removed from the multiset.

\end{itemize}
Since multiset ordering is well-founded, this ensures termination.
\end{proof}

\section{Matrix for  \emph{Term Propagation}}\label{matrix}

In this section we briefly show how it is possible to compute
automatically the output equations of the \emph{Term Propagation}
phase of the General ACI unification algorithm
(Section~\ref{termpropagation}). The method we propose builds on
the solution of the $ACI$ unification with constants problem based
on $ACI$-matrices; the novelty is the use of a simplified form of
$ACI$-matrix that takes advantage of the format of the equations
to be dealt with in this context---i.e., elementary $ACI1$
equations.

Given an elementary $ACI1$ unification problem
\[ S_1 \cup \cdots \cup S_n \cup X_1 \cup \cdots \cup X_p =
        T_1 \cup \cdots \cup T_m \cup X_1 \cup \cdots \cup X_p
\]
the \emph{elementary $ACI$-matrix} is as follows:
\[
\begin{array}[b]{rl}
\renewcommand{\arraystretch}{1}
\begin{array}[t]{|ccc|ccc|}
\multicolumn{1}{c}{S_1} & {\dots} & \multicolumn{1}{c}{S_n} &
        \multicolumn{1}{c}{X_1} & \dots & \multicolumn{1}{c}{X_p} \\
\cline{1-6}
A_{1,1} & \dots & A_{1,n} & A_{1,n+1} & \dots & A_{1,n+p} \\
\vdots   & \ddots & \vdots   & \vdots     & \ddots & \vdots     \\
A_{m,1} & \dots & A_{m,n} & A_{m,n+1} & \dots & A_{m,n+p} \\
\cline{1-6}
A_{m+1,1} & \dots & A_{m+1,n} & A_{m+1,n+1} & \dots & A_{m+1,n+p} \\
\vdots   & \ddots & \vdots   & \vdots     & \ddots & \vdots     \\
A_{m+p,1} & \dots & A_{m+p,n} & A_{m+p,n+1} & \dots & A_{m+p,n+p} \\
\cline{1-6}
\end{array} &
\renewcommand{\arraystretch}{1.01}
\hspace{-.3cm}\begin{array}[t]{l}
\\
T_1\\
\vdots\\
T_m\\
X_1\\
\vdots\\
X_p
\end{array}
\end{array}
\]
However, variables $A_{m+i,n+j}$ with $i>0,j>0,i\neq j$ are not
used and thus we can avoid to introduce them. The most general
unifier for the elementary problem can be obtained as follows:
\[
\renewcommand{\arraystretch}{1.4}
\begin{array}{lclclcl}
S_j & = & \displaystyle \bigcup_{i=1}^{m+p} A_{i,j} &
\phantom{aaa} &
T_i & = & \displaystyle \bigcup_{j=1}^{n+p} A_{i,j}\\
X_j &=& \multicolumn{5}{l}{ \displaystyle \bigcup_{i=1}^m
A_{i,n+j} \cup \bigcup_{k=1}^{n} A_{m+j,k} \cup  A_{m+j,n+j}}
\end{array}
\]
One can easily prove that this method provides the same solution
as the $ACI$ unification with constants algorithm based on Boolean
$ACI$ matrices of \cite{BB88}, briefly recalled in Section
\ref{aci1-costanti}.

\begin{example}
Let us consider the same unification problem $S_1 \cup S_2 \cup X
= T_1 \cup T_2 \cup X$ as in Example~\ref{var_con_shar}; the
elementary $ACI$-matrix is
\[
\begin{array}[b]{rl}
\renewcommand{\arraystretch}{1}
\begin{array}[t]{|cc|c|}
\multicolumn{1}{c}{S_1} & \multicolumn{1}{c}{S_2} &
\multicolumn{1}{c}{X} \\
\cline{1-3}
R_1 & R_3 & R_7\\
R_2 & R_4 & R_8\\
\cline{1-3}
R_5 & R_6 & R_9\\
\cline{1-3}
\end{array} &
\renewcommand{\arraystretch}{1.01}
\hspace{-.3cm}\begin{array}[t]{l}
\\
T_1\\
T_2\\
X
\end{array}
\end{array}
\]
Let us observe that the variables in the matrix have been named to
show the correspondence with the new variables used in
Example~\ref{var_con_shar}.
\end{example}

Given the unification problem:
$$\renewcommand{\arraystretch}{1.8}
\begin{array}{rcl}
{\cal E}^n  & \equiv &  N^L_1 = f_1(\bar l_1) \wedge \cdots \wedge
 N^L_{k_1} =
f_{k_1}(\bar l_{k_1})\;\wedge \\
&& N^R_1 = g_1(\bar r_1)\wedge \cdots \wedge N^R_{h_1} =
g_{h_1}(\bar
             r_{h_1})\;\wedge  \\
&&  N^L_1 \cup \cdots \cup N^L_{k_1} \cup
                 L_1 \cup \cdots \cup L_{k_2} =
                 N^R_1 \cup \cdots \cup N^R_{h_1}
                 \cup R_1 \cup \cdots \cup N^R_{h_2}
        \end{array}$$
we solve the elementary $ACI1$ problem on the equation:
$$
N^L_1 \cup \cdots \cup N^L_{k_1} \cup L_1 \cup \cdots \cup L_{k_2}
= N^R_1 \cup \cdots \cup N^R_{h_1} \cup R_1 \cup \cdots \cup
R_{h_2}
$$
We build an auxiliary Boolean matrix $B$ that allows us to reduce
the non-determinism. We deal with two cases:

\begin{itemize}
\item
If $ \{L_1, \dots, L_{k_2} \} \cap \{ R_1, \dots, R_{h_2} \} = \e
$ any (non-deterministic) solution can be described using a
$(h_1+h_2)\times(k_1+k_2)$ matrix $B$ such that
\begin{itemize}
        \item for $h_1+1 \leq i \leq h_1+h_2$ and $k_1+1 \leq j \leq
k_1+k_2$ we have
                        $B[i,j] = \bot$
        \item all the other components of $B$ have a value taken from
$\{0,1\}$
        \item for each $1\leq i \leq h_1$ $\sum_{j=1}^{k_1+k_2} B[i,j]
\geq 1$ and
                for each $1\leq j \leq k_1$ $\sum_{i=1}^{h_1+h_2} B[i,j]
\geq 1$.
        \end{itemize}

Thus, $B$ is a boolean matrix with the exception of the fourth
quadrant, where
 the matrix contains only the value $\bot$.
  The matrix $B$ can be used to describe the substitution $\lambda$:
\[ \lambda(A_{i,j}) = \left\{
                        \begin{array}{ll}
                                A_{i,j} & \mbox{if }B[i,j] = \bot\\
                                \e & \mbox{if } B[i,j] = 0 \\
                                h(r_i) & \mbox{if } B[i,j]=1 \wedge j >
h_1\\
                                h(l_j) & \mbox{if } B[i,j]=1 \wedge j
\leq h_1
                        \end{array}
                        \right. \]
        Additionally, $B$ generates the new set of equations:
\[ E^{conf} = \bigwedge_{ B[i,j]=1 \wedge
          1\leq i \leq k_1 \wedge
          1 \leq j \leq h_1} h(l_j) = h(r_i)
\]

\item Assume now that the
two sides of the equation share some variables. I.e., let us
assume that the problem at hand is
\[\begin{array}{l}
N^L_1 \cup \cdots \cup N^L_{k_1} \cup L_1 \cup \cdots \cup L_{k_2}
        \cup Com_1 \cup \cdots \cup Com_c =\phantom{aaaaaaaaa}\\
\multicolumn{1}{r}{    N^R_1 \cup \cdots \cup N^R_{h_1} \cup R_1 \cup \cdots \cup R_{h_2}
        \cup Com_1 \cup \cdots \cup Com_c}
   \end{array}
\]
The solution of the problem in this case can be built around the
elementary $ACI$-matrix shown in Figure~\ref{extendedmatrix1}.
The table in Figure~\ref{extendedmatrix1} assumes $h=h_1+h_2$ and
$k=k_1+k_2$\/. The solution of  the ACI problem, in this case,
will be composed of equations of the form:
\[\renewcommand{\arraystretch}{2.5}
\begin{array}{rcl}
  L_j  & = & \displaystyle{\bigcup_{i=1}^{h_1}} A_{i,k_1+j} \cup  \bigcup_{i=h_1+1}^{h_1+h_2}
A_{i,k_1+j} \cup
        \bigcup_{i=h_1+h_2+1}^{h_1+h_2+c} A_{i,k_1+j}\\
R_i  &  =   & \displaystyle \bigcup_{j=1}^{k_1} A_{h_1+i,j} \cup
            \bigcup_{j=k_1+1}^{k_1+k_2} A_{h_1+i,j} \cup
            \bigcup_{j=k_1+k_2+1}^{k_1+k_2+c} A_{h_1+i,j}\\
Com_v & =& \displaystyle\bigcup_{i=1}^{h_1+h_2+c} A_{i,k_1+k_2+v}
\cup
                \bigcup_{j=1}^{k_1+k_2+c} A_{h_1+h_2+v,j}
\end{array}
\]
In Figure~\ref{extendedmatrix2}, we depict the boolean matrix
$B$ which will be used in this case. The matrix $B$ should satisfy
the following properties:
\begin{itemize}
\item quadrant 5, 6, and 8 are filled with $\bot$\/;
\item the non-zero entries in quadrant 9 are assigned $\bot$\/; observe
        that the quadrant 9 is a diagonal matrix with non-zero elements
        only along the main diagonal;
\item quadrant 1, 2, 3, 4, and 7 are boolean matrices;
\item for $1 \leq j \leq k_1$ we have
$\sum_{i=1}^{h_1+h_2} B[i,j] + \sum_{i=1}^{c} B[i,j] \geq 1 $
\item for $1 \leq i \leq h_1$ we have
$ \sum_{j=1}^{k_1+k_2} B[i,j] + \sum_{j=1}^{c} B[i,j] \geq 1 $
\end{itemize}
The substitution $\lambda$ and the collection of new equations
$E^{\mbox{\em conf}}$ are defined exactly as above.

\begin{figure}[htb]
{\footnotesize
\[
\renewcommand{\arraystretch}{1.1}
\begin{array}{|c|c|c|c|c||c|c|c|c|c||c|c|c|c|c||c}
\multicolumn{1}{c}{N^L_1} &
\multicolumn{3}{c}{\dots} &
\multicolumn{1}{c}{N^L_{k_1}} &
\multicolumn{1}{c}{L_1} &
\multicolumn{3}{c}{\dots} &
\multicolumn{1}{c}{L_{k_2}} &
\multicolumn{1}{c}{Com_1} &
\multicolumn{3}{c}{\dots} &
\multicolumn{1}{c}{Com_c} & \\
\cline{1-15}
A_{1,1} & \multicolumn{3}{|c|}{\dots} &
A_{1,k_1} &  A_{1,k_1+1}  & \multicolumn{3}{|c|}{\dots} &
A_{1,k}  &  A_{1,k+1}  & \multicolumn{3}{|c|}{\dots} &
A_{1,k+c} &  N^R_1\\
\cline{1-15}
\multicolumn{5}{|c||}{\dots}  &
\multicolumn{5}{|c||}{\dots} & \multicolumn{5}{|c||}{\dots} &
\vdots\\
\cline{1-15}
A_{h_1,1} & \multicolumn{3}{|c|}{\dots} & A_{h_1,k_1}
& A_{h_1,k_1+1} & \multicolumn{3}{|c|}{\dots} & A_{h_1,k} &
A_{h_1,k+1} & \multicolumn{3}{|c|}{\dots} & A_{h_1,k+c} &
N^R_{h_1}\\
\cline{1-15}
\cline{1-15}
 A_{h_1+1,1} & \multicolumn{3}{|c|}{\dots} &
A_{h_1+1,k_1} & A_{h_1+1,k_1+1} &
\multicolumn{3}{|c|}{\dots} & A_{h_1+1,k} & A_{h_1+1,k+1} &
\multicolumn{3}{|c|}{\dots} & A_{h_1+1,k+c} &
 R_1\\
\cline{1-15}
\multicolumn{5}{|c||}{\dots}  & \multicolumn{5}{|c||}{\dots}  &
\multicolumn{5}{|c||}{\dots} &
\vdots\\
\cline{1-15}
A_{h,1} & \multicolumn{3}{|c|}{\dots} & A_{h,k_1} &
A_{h,k_1+1} & \multicolumn{3}{|c|}{\dots} & A_{h,k} &
A_{h,k+1}& \multicolumn{3}{|c|}{\dots} & A_{h,k+c} &
 R_{h_2}\\
\cline{1-15}
\cline{1-15}
 A_{h+1,1} & \multicolumn{3}{|c|}{\dots} &
A_{h+1,k_1} & A_{h+1,k_1+1} & \multicolumn{3}{|c|}{\dots} &
A_{h+1,k} & A_{h+1,k+1} & \multicolumn{3}{|c|}{\dots} &
A_{h+1,k+c} &
 Com_1\\
\cline{1-15}
\multicolumn{5}{|c||}{\dots}  & \multicolumn{5}{|c||}{\dots}  &
\multicolumn{5}{|c||}{\dots} &
\vdots\\
\cline{1-15}
A_{h+c,1} & \multicolumn{3}{|c|}{\dots} & A_{h+c,k_1}
& A_{h+c,k_1+1} & \multicolumn{3}{|c|}{\dots} & A_{h+c,k} &
A_{h+c,k+1}& \multicolumn{3}{|c|}{\dots} & A_{h+c,k+c} &
 Com_c\\
\cline{1-15}
\cline{1-15}
\end{array}
\]}
\caption{Elementary $ACI$-matrix} \label{extendedmatrix1}
\end{figure}

\begin{figure}[htb]
\begin{center}
{\footnotesize
\begin{tabular}{|c c c c c|c c c c c|c c c c c |c}
 \multicolumn{1}{c}{$N^L_1$} & \multicolumn{3}{c}{$\dots$} &
 \multicolumn{1}{c}{$N^L_{k_1}$} &
 \multicolumn{1}{c}{$L_1$} &
\multicolumn{3}{c}{$\dots$} &
\multicolumn{1}{c}{$L_{k_2}$} &
\multicolumn{1}{c}{$Com_1$} & \multicolumn{3}{c}{$\dots$} &
\multicolumn{1}{c}{$Com_c$} & \\
\cline{1-15}
 \multicolumn{5}{|c|}{ }  & \multicolumn{5}{c|}{ } &
\multicolumn{5}{c|}{ }  & $N^R_1$\\
 \multicolumn{5}{|c|}{\sf Quad 1} &  \multicolumn{5}{c|}{\sf Quad 2} &
\multicolumn{5}{c|}{\sf Quad 3} & $\vdots$\\
 \multicolumn{5}{|c|}{ }  & \multicolumn{5}{c|}{ } &
\multicolumn{5}{c|}{ }  & $N^R_{h_1}$\\
\cline{1-15}
 \multicolumn{5}{|c|}{ }  & \multicolumn{5}{c|}{ } &
\multicolumn{5}{c|}{ }  & $R_1$\\
 \multicolumn{5}{|c|}{\sf Quad 4} &  \multicolumn{5}{c|}{\sf Quad 5} &
\multicolumn{5}{c|}{\sf Quad 6} & $\vdots$\\
 \multicolumn{5}{|c|}{ }  & \multicolumn{5}{c|}{ } &
\multicolumn{5}{c|}{ }  & $R_{h_2}$\\
\cline{1-15}
 \multicolumn{5}{|c|}{ }  & \multicolumn{5}{c|}{ } &
\multicolumn{5}{c|}{\sf Quad 9}  & $Com_1$\\
 \multicolumn{5}{|c|}{\sf Quad 7} &  \multicolumn{5}{c|}{\sf Quad 8} &
\multicolumn{5}{c|}{$I_{c,c}$} & $\vdots$\\
 \multicolumn{5}{|c|}{ }  & \multicolumn{5}{c|}{ } &
\multicolumn{5}{c|}{ }  & $Com_c$\\
\cline{1-15}
\end{tabular}}
\end{center}
\caption{Extended Boolean Matrix $B$} \label{extendedmatrix2}
\end{figure}
\end{itemize}

\end{document}